\documentclass[useAMS, usenatbib]{mn2e}
\usepackage{color}
\usepackage{epsfig}
\usepackage{epstopdf}
\usepackage{amsmath, amssymb}
\usepackage{hyperref}
\usepackage{wasysym} 
\usepackage{subfigure}
\usepackage{multirow}
\usepackage{lipsum}
\usepackage{ulem}
\usepackage{mn2e-breakabs}
\usepackage{setspace}  

\newlength{\plotwidth}
\newlength{\fullwidth}
\newcommand{\llan}{\left\langle}
\newcommand{\rran}{\right\rangle}
\newcommand{\ihMpc}{h{\rm\;Mpc^{-1}}}
\newcommand{\hMpc}{h^{-1}{\rm\;Mpc}}

\newcommand{\itMpc}{h^3{\rm\;Mpc^{-3}}}

\newcommand{\zeff}{z_{\rm eff}}
\def\vc#1{{\bf#1}}

\pagerange{\pageref{firstpage}--\pageref{lastpage}} \pubyear{2013}

\def\LaTeX{L\kern-.36em\raise.3ex\hbox{a}\kern-.15em
    T\kern-.1667em\lower.7ex\hbox{E}\kern-.125emX}

\title[BOSS: Anisotropic Fourier-space analysis]{The clustering of galaxies in the completed SDSS-III Baryon Oscillation Spectroscopic Survey: Anisotropic galaxy clustering in Fourier-space}
\author[Florian Beutler et al.]
{\parbox{\textwidth}{Florian Beutler$^{1,2}$\thanks{E-mail: \texttt{florian.beutler@port.ac.uk}}, Hee-Jong Seo$^{3}$, Shun Saito$^{4,5}$, Chia-Hsun Chuang$^{6, 7}$, Antonio J. Cuesta$^{8}$, Daniel J. Eisenstein$^{9}$, H\'ector Gil-Mar\'in$^{10,11}$, Jan Niklas Grieb$^{12, 13}$, Nick Hand$^{14}$, Francisco-Shu Kitaura$^{7}$, Chirag Modi$^{15}$, Robert C. Nichol$^{1}$, Matthew D. Olmstead$^{16}$, Will J. Percival$^{1}$, Francisco Prada$^{6,17,18}$, Ariel G. S{\'a}nchez$^{12}$, Sergio Rodriguez-Torres$^{6,17}$,  Ashley J. Ross$^{10, 1}$, Nicholas P. Ross$^{20}$, Donald P. Schneider$^{21, 22}$, Jeremy Tinker$^{23}$, Rita Tojeiro$^{24}$, Mariana Vargas-Maga\~na$^{25}$}\vspace{0.4cm}\\
\parbox{\textwidth}{
$^{1}$Institute of Cosmology \& Gravitation, Dennis Sciama Building, University of Portsmouth, Portsmouth, PO1 3FX, UK\\
$^{2}$Lawrence Berkeley National Lab, 1 Cyclotron Rd, Berkeley CA 94720, USA\\
$^{3}$Department of Physics and Astronomy, Ohio University, 251B Clippinger Labs, Athens, OH 45701, USA\\
$^{4}$Kavli Institute for the Physics and Mathematics of the Universe (WPI),The University of Tokyo Institutes for Advanced Study, The University of Tokyo, Kashiwa, Chiba 277-8583, Japan\\
$^{5}$Max-Planck-Institut fŸ\"ur Astrophysik, Karl-Schwarzschild-Strasse 1, D-85740 Garching bei MŸnchen, Germany\\
$^{6}$Instituto de F\'{\i}sica Te\'orica, (UAM/CSIC), Universidad Aut\'onoma de Madrid, Cantoblanco, E-28049 Madrid, Spain \\
$^{7}$Leibniz-Institut f\"ur Astrophysik Potsdam (AIP), An der Sternwarte 16, D-14482 Potsdam, Germany\\
$^{8}$Institut de Ci{\`e}ncies del Cosmos (ICCUB), Universitat de Barcelona (IEEC-UB), Mart{\'\i} i Franqu{\`e}s 1, E08028 Barcelona, Spain\\
$^{9} $Harvard-Smithsonian Center for Astrophysics, 60 Garden St., Cambridge, MA 02138, USA\\
$^{10}$ Sorbonne Universit\'es, Institut Lagrange de Paris (ILP), 98 bis Boulevard Arago, 75014 Paris, France \\
$^{11}$ Laboratoire de Physique NuclŽaire et de Hautes Energies, Universit\'e Pierre et Marie Curie, 4 Place Jussieu, 75005 Paris, France
$^{12}$Max-Planck-Institut f\"ur extraterrestrische Physik, Postfach 1312, Giessenbachstr., 85741 Garching, Germany\\
$^{13}$Universit\"ats-Sternwarte M\"unchen, Ludwig-Maximilians-Universit\"at M\"unchen, Scheinerstra\ss{}e 1, 81679 M\"unchen, Germany\\
$^{14}$Department of Astronomy, University of California Berkeley, CA 94720, USA\\
$^{15}$Department of Physics, University of California Berkeley, CA 94720, USA\\
$^{16}$Institute for Gravitation and the Cosmos, The Pennsylvania State University, University Park, PA 16802\\
$^{17}$Campus of International Excellence UAM+CSIC, Cantoblanco, E-28049 Madrid, Spain \\
$^{18}$Instituto de Astrof\'{\i}sica de Andaluc\'{\i}a (CSIC), Glorieta de la Astronom\'{\i}a, E-18080 Granada, Spain \\
$^{19}$Department of Physics, Ohio State University, 140 West 18th Avenue, Columbus, OH 43210, USA\\
$^{20}$Institute for Astronomy, University of Edinburgh, Royal Observatory, Edinburgh EH9 3HJ, UK\\\
$^{21}$Department of Astronomy and Astrophysics, The Pennsylvania State University, University Park, PA 16802\\
$^{22}$Institute for Gravitation and the Cosmos, The Pennsylvania State University, University Park, PA 16802\\
$^{23}$Center for Cosmology and Particle Physics, Department of Physics, New York University, 4 Washington Place, New York, NY 10003, USA\\
$^{24}$School of Physics and Astronomy, University of St. Andrews, St Andrews, Fife, KY16 9SS, UK\\
$^{25}$Instituto de Fisica, Universidad Nacional Autonoma de Mexico, Apdo. Postal 20-364, Mexico.\vspace{0cm}}}
 
\begin{document}
 
\label{firstpage}
\maketitle

\begin{abstract}
We investigate the anisotropic clustering of the Baryon Oscillation Spectroscopic Survey (BOSS) Data Release 12 (DR12) sample, which consists of $1\,198\,006$ galaxies in the redshift range $0.2 < z < 0.75$ and a sky coverage of $10\,252\,$deg$^2$. We analyse this dataset in Fourier space, using the power spectrum multipoles to measure Redshift-Space Distortions (RSD) simultaneously with the Alcock-Paczynski (AP) effect and the Baryon Acoustic Oscillation (BAO) scale. We include the power spectrum monopole, quadrupole and hexadecapole in our analysis and compare our measurements with a perturbation theory based model, while properly accounting for the survey window function.
To evaluate the reliability of our analysis pipeline we participate in a mock challenge, which resulted in systematic uncertainties significantly smaller than the statistical uncertainties. While the high-redshift constraint on $f\sigma_8$ at $z_{\rm eff}=0.61$ indicates a small ($\sim 1.4\sigma$) deviation from the prediction of the Planck $\Lambda$CDM model, the low-redshift constraint is in good agreement with Planck $\Lambda$CDM. This paper is part of a set that analyses the final galaxy clustering dataset from BOSS. The measurements and likelihoods presented here are combined with others in~\citet{Alam2016} to produce the final cosmological constraints from BOSS.
\end{abstract}


\section{introduction}
\label{sec:intro}

Clustering in the matter density field carries an enormous amount of information about cosmological parameters. The growth of the matter clustering amplitude is directly sensitive to the sum of the neutrino masses~\citep[e.g.,][]{Lesgourgues:2006nd,Beutler:2014}, the dark energy equation of state and the nature of Gravity~\citep{Kaiser:1987qv,Peacock:2001gs,Guzzo:2008ac}. Galaxy redshift surveys sample the underlying matter density field with galaxies as tracer particles. A measurement of the matter clustering amplitude, $\sigma_8$, could be compared to the precise measurement of the matter clustering amplitude at the recombination redshift, measured in the Cosmic Microwave Background (CMB) providing a long lever-arm with which to test the growth of structure. The formation of galaxies is correlated with the underlying density field, but the galaxy formation processes are complicated and do not allow us to infer these correlations from first principles. I.e., while the clustering amplitude of the galaxy density field can be measured to percent level precision, it cannot straightforwardly be related to the clustering amplitude of the matter density field due to the uncertainties in the bias relation.

Despite these limitations, galaxy redshift surveys still allow constraints on the matter clustering amplitude due to redshift-space distortions (RSD). RSD are caused by the underlying peculiar velocity field along the line-of-sight that Doppler-shifts the features in the spectral energy distribution of a galaxy. If the redshift is used to estimate the distance to the galaxy using Hubble's law, peculiar velocity contributions to the redshift introduce errors in the physical coordinates along the line-of-sight. Since it is nearly impossible to estimate the line-of-sight peculiar velocities of millions of galaxies individually,  with a precision anywhere near the redshift uncertainty, rather than correcting the effect in the redshift measurements, we  account for the resulting statistical distortions in the observed clustering signal. 

Peculiar velocities trace the gravitational potential field, itself produced by the underlying matter density field. Therefore the distortion effect due to the peculiar velocity field is correlated with the density field. Since the peculiar velocity field affects only the line-of-sight positions without affecting the angular positions, the distortions generate anisotropy in the observed clustering. In the linear regime, redshift-space distortions lead to an angle-dependent increase in the power spectrum amplitude of $(1 + \beta\mu^2)^2$~\citep{Kaiser:1987qv}, where $\mu$ is the cosine of the angle to the line-of-sight and $\beta = f/b$ is the growth rate $f$ divided by the galaxy bias $b$. Using the angular dependence of the RSD signal we can constrain the parameter combination $f(z)\sigma_8(z)$. The  redshift-space distortion signal is now considered one of the most powerful observables in large-scale structure~\citep{Peacock:2001gs,Hawkins:2002sg,Tegmark:2006az,Guzzo:2008ac,Yamamoto:2008gr,Blake:2011rj,Beutler:2012px,Reid:2012sw,Samushia:2012iq,Chuang:2013hya,Nishimichi:2013aba}.

Another source of anisotropy in the galaxy clustering signal is known as the Alcock-Pazynski (AP) effect~\citep{Alcock:1979mp}. The AP effect is imprinted in the clustering measurement when converting from observable (redshifts and angles) to physical coordinates, which requires a fiducial cosmology. If that fiducial cosmology deviates from the true cosmology, it distorts the clustering signal differently along the line-of-sight and in angular scales. The two effects would be difficult to separate with a featureless power spectrum. By measuring the distortion in the distinct Baryon Acoustic Oscillations (BAO) feature present in the power spectrum, we can constrain the AP signal, thereby isolating the anisotropy in the clustering amplitude due to the RSD~\citep{Matsubara:1996nf,Ballinger:1996cd,Padmanabhan:2008ag,Okumura:2007br}. The AP test through the BAO signal constrains the geometry, i.e., $D_V(z)/r_s(z_d)$ and $F_{\rm AP}(z) = (1+z)D_A(z)H(z)/c$, where $D_V(z) = [(1+z)^2D_A^2(z)cz/H(z)]^{1/3}$ is the angle averaged distance depending on the angular diameter distance $D_A(z)$ and the Hubble parameter $H(z)$.

In this paper we use the final data release (\citealt{Alam:2015mbd}, DR12) of the Baryon Oscillation Spectroscopic Survey~\citep{Dawson:2012va}, the largest galaxy redshift dataset available to date to measure the anisotropy in the galaxy power spectrum. Our analysis framework follows our DR11 analysis~\citep{Beutler:2013yhm} with several modifications: (1) In addition to the monopole and quadrupole we now include the hexadecapole, (2) we modified the fitting range to $k = 0.01$ - $0.15\ihMpc$ for the monopole and quadrupole and $k = 0.01$ - $0.10\ihMpc$ for the hexadecapole, (3) we modified our method to include the effect due to the discrete \textit{k}-space grid when estimating the power spectrum, (4) we use a computationally more efficient way to include window function effects, (5) we use larger \textit{k}-bins to reduce the noise in the covariance matrix estimation, (6) we use a new set of mock catalogues called MultiDark-Patchy, which have been introduced in~\citet{Kitaura:2015uqa} and (7) we employ the FFT based power spectrum estimator suggested by~\citet{Bianchi:2015oia} and~\citet{Scoccimarro:2015bla} instead of the $\mathcal{O}(N^2)$ algorithm we used previously to speed up the computation.

Our companion paper,~\cite{Beutler:2016}, presents a BAO-only analysis, where we use the same power spectrum multipole measurements, but isolate the BAO signal by marginalising over the shape of the power spectrum (including RSD). While the BAO-only analysis does not capture the information on RSD, it will allow measurements of $D_V(z)/r_s(z_d)$ and $F_{\rm AP}(z) = (1+z)D_A(z)H(z)/c$ that do not depend on our understanding of redshift-space distortions. Without the need to model RSD in detail, the BAO-only analysis can use a wider range of wave numbers ($k_{\rm max} = 0.3 \ihMpc$), and improve the BAO signal by using the BAO reconstruction technique~\citep{Eisenstein:2006nk}. BAO measurements obtained using the monopole and quadrupole correlation function are presented in~\citet{Ross:2016}, while~\citet{Vargas-Magana2016} diagnoses the level of theoretical systematic uncertainty in the BOSS BAO measurements. Beside this paper there are three more measurements of the growth of structure~\citep{Grieb:2016, Sanchez:2016, Satpathy2016}. \citet{Alam2016} combines the results of these seven papers (including this work) into a single likelihood that can be used to test cosmological models.
 
This paper is organised as follows: Section~\ref{sec:data} presents the BOSS dataset used in this analysis. Section~\ref{sec:estimator} describes the power spectrum estimator, followed by the treatment of the window function in section~\ref{sec:win}. Section~\ref{sec:model} introduces our power spectrum model, which is based on renormalised perturbation theory. Section~\ref{sec:mocks} discusses the mock catalogues used to derive covariance matrices for our measurements. In Section~\ref{sec:sys} we use these mock catalogues as well as N-body simulations to test our power spectrum model. The analysis together with the fitting results is presented in section~\ref{sec:analysis}, while section~\ref{sec:dis} discusses the result and compares to previous studies. We conclude in section~\ref{sec:conclusion}.

The fiducial cosmological parameters, which we use to turn the observed angles and redshifts into co-moving coordinates and to generate our linear power spectrum models as an input for the power spectrum templates, follow a flat $\Lambda$CDM model with $\Omega_m=0.31$, $\Omega_bh^2=0.022$, $h=0.676$, $\sigma_8=0.8$, $n_s=0.96$, $\sum m_{\nu}=0.06\,$eV and $r_s^{\rm fid}(z_d) = 147.78\,$Mpc.

\section{The BOSS DR12 dataset}
\label{sec:data}

\begin{figure}
\begin{center}
\epsfig{file=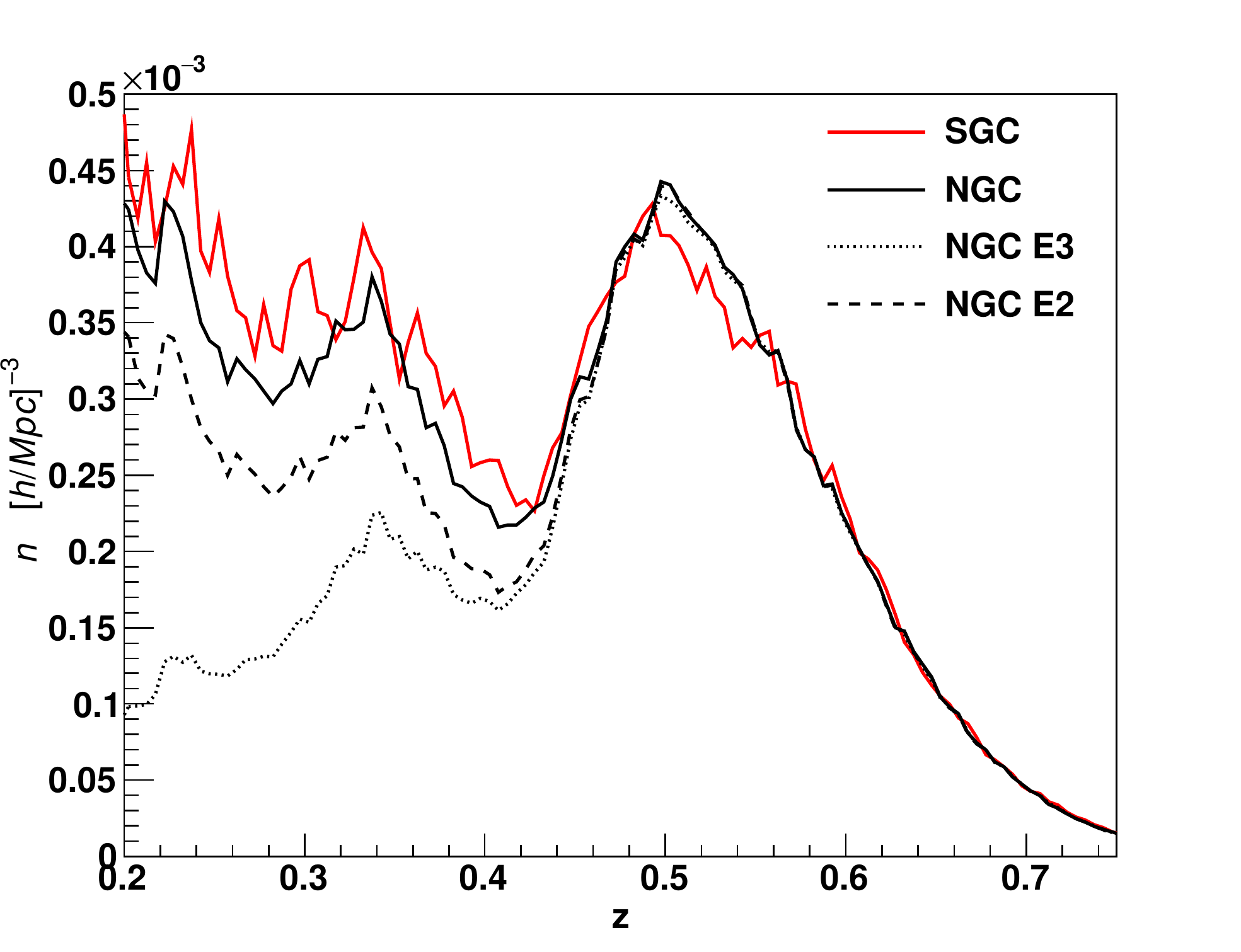,width=8cm}
\caption{Galaxy density distribution for the North Galactic Cap (NGC, red), South Galactic Cap (SGC, black) as well as the early (E) region 2 and 3, consisting of chunks 2-6 (for details see section~\ref{sec:data}).}
\label{fig:nbar}
\end{center}
\end{figure}

The Baryon Oscillation Spectroscopic Survey (BOSS) is part of SDSS-III~\citep{Eisenstein:2011sa,Dawson:2012va} and measured spectroscopic redshifts of $1\,198\,006$ million galaxies using the SDSS multi-fibre spectrographs~\citep{Bolton:2012hz,Smee:2012wd}. The galaxies are selected from multi-colour SDSS imaging~\citep{Fukugita:1996qt,Gunn:1998vh,Smith:2002pca,Gunn:2006tw,Doi:2010rf}.

\begin{table*}
\begin{center}
\caption{The number of galaxies $N_{\rm gal}$, the weighted (incompleteness weight) number of galaxies $N'_{\rm gal}$ and the effective volume for the three redshift bins used in this analysis. Eq.~\ref{eq:zeff} produces the effective redshifts of $z_{\rm eff} = 0.38, 0.51$ and $0.61$ for the three redshift bins, respectively.}
	\begin{tabular}{lllllll}
     		\hline
		 & \multicolumn{2}{c}{$0.2 < z < 0.5$} & \multicolumn{2}{c}{$0.4 < z < 0.6$} & \multicolumn{2}{c}{$0.5 < z < 0.75$}\\
		 & NGC & SGC & NGC & SGC & NGC & SGC\\
		\hline
		$N_{\rm gal}$ & $429\,182$ & $174\,820$ & $500\,875$ & $185\,500$ & $435\,742$ & $158\,262$ \\
		$N'_{\rm gal}$ & $445\,261$ & $182\,678$ & $534\,725$ & $197\,084$ & $467\,504$ & $169\,907$ \\
		$V_{\rm eff} [\text{Gpc}^3]$ & $2.7$ & $1.0$ & $3.1$ & $1.1$ & $3.0$ & $1.1$ \\
		\hline
		\hline
	  \end{tabular}
	  \label{tab:samples}
\end{center}
\end{table*}

The survey is optimised for the measurement of the BAO scale and hence covers a large cosmic volume with a density of $\overline{n} \approx 3\times10^{-4} \itMpc$, high enough to ensure that shot noise is not the dominant error contribution at the BAO scale. Most BOSS galaxies are red with a prominent $4000\,$\AA\; break in their spectral energy distribution; this feature allows a reliable redshift detection with a short exposer, providing a fast survey speed.

The BOSS-DR12 sample covers a redshift range of $0.2 < z < 0.75$ over $10\,252\deg^2$ divided in two patches on the sky. The North Galactic Cap (NGC) contains $864\,924$ galaxies and the South Galactic Cap (SGC) contains $333\,082$. Although the BOSS galaxy sample consists of two separate selection algorithms, so-called LOWZ and CMASS, we utilise the whole galaxy sample by combining these two samples. We refer to~\citet{Reid:2015gra} for concrete definitions of selection criteria for LOWZ and CMASS. The completeness of this sample in terms of stellar mass is studied in~\citet{Leauthaud:2015fva} and~\citet{Saito:2015eka} using the deeper galaxy sample of S82-MGC~\citep{Bundy:2014vpa}. 

The DR12 data release also includes a few regions on the sky (early regions, chunk 2-6), which have not been included in previous data releases. Galaxies in these regions were selected with different algorithms from those of subsequent data. We now create separate masks for chunk 2 (LOWZE2) and chunks 3-6 (LOWZE3) and combine these regions with the rest of the BOSS dataset. However, the density for chunk 2-6 is generally lower compared to the rest of the dataset, as shown in Figure~\ref{fig:nbar}. Since chunks 2-6 are located in the North Galactic Cap, the expected bias parameters are different for NGC and SGC. We will address this issue further in section~\ref{sec:parametrization}. Details about the selection differences between chunks 2-6 and the rest of the BOSS dataset as well as the mask creation can be found in~\citet{Reid:2015gra}.

We include three different incompleteness weights to account for shortcomings of the BOSS dataset (see~\citealt{Ross:2012qm} and~\citealt{Anderson2.0} for details): A redshift failure weight, $w_{\rm rf}$, a fibre collision weight, $w_{\rm fc}$ and a systematics weight, $w_{\rm sys}$, which is a combination of a stellar density weight and a seeing condition weight. Each galaxy is thus counted as 
\begin{equation}
w_c = (w_{\rm rf} + w_{\rm fc} - 1)w_{\rm sys}.
\end{equation} 
Accounting for redshift failure and fibre collisions, the weighted galaxy count is $1\,265\,350$ (see Table~\ref{tab:samples}).

We divide the BOSS dataset into three overlapping redshift bins defined by $0.2 < z < 0.5$, $0.4 < z < 0.6$ and $0.5 < z < 0.75$. The effective redshift for these samples can be calculated as
\begin{equation}
z_{\rm eff} = \frac{\sum^{N_{\rm gal}}_i w_{\rm FKP}(\vc{x}_i)w_c(\vc{x}_i)z_i}{\sum_i^{N_{\rm gal}}w_{\rm FKP}(\vc{x}_i)w_c(\vc{x}_i)},
\label{eq:zeff}
\end{equation}
where $w_{\rm FKP}=1/(1+n(z)P_0)$ is a signal-to-noise ratio weight suggested by~\citet{Feldman:1993ky} (we adopt $P_0 = 10\,000h^{-3}$Mpc$^3$). With the above definition of the effective redshift we find $z_{\rm eff} = 0.38, 0.51$, and $0.61$ for the three redshift bins.

\section{The power spectrum estimator}
\label{sec:estimator}

We employ the Fast Fourier Transform (FFT)-based anisotropic power spectrum estimator suggested by~\citet{Bianchi:2015oia} and~\citet{Scoccimarro:2015bla}. This estimator follows the ideas of~\citet{Feldman:1993ky}, but also allows one to estimate the higher order multipoles by decomposing the power spectrum estimate into its spatial vector components and performing a series of FFTs for each component. This approach accounts for the different line-of-sights for different galaxy pairs within the local plane parallel approximation\footnote{We define the local plane parallel approximation to be the assumption that the position vectors of a given galaxy pair can be treated as parallel such that $\vc{x}_h=\frac{\vc{x}_1+\vc{x}_2}{2}\approx \vc{x}_1 \approx \vc{x}_2$ , while the lines of sight vary  for different pairs. The global plane parallel approximation assumes that the line-of-sight is fixed for all galaxy pairs. See~\citet{Beutler:2013yhm} section 3.1 for more details.}. By using FFTs rather than summing over all galaxy pairs~\citep{Yamamoto:2005dz,Beutler:2013yhm}, this estimator allows a computational complexity of $\mathcal{O}(N\log N)$, which is much faster than a naive pair counting analysis (here $N$ is the number of grid cells used to bin the data).

The power spectrum multipoles can be calculated as~\citep{Feldman:1993ky,Yamamoto:2005dz,Bianchi:2015oia,Scoccimarro:2015bla}

\begin{align}
P_{0}(\vc{k}) &= \frac{1}{2A}\bigg[F_{0}(\vc{k})F_{0}^*(\vc{k})  - S\bigg],\\
P_{2}(\vc{k}) &= \frac{5}{4A}F_{0}(\vc{k})\bigg[3F_2^*(\vc{k}) - F_0^*(\vc{k})\bigg],\\
P_{4}(\vc{k}) &= \frac{9}{16A}F_{0}(\vc{k})\bigg[35F_4^*(\vc{k}) - 30F_2^*(\vc{k}) + 3F_0^*(\vc{k})\bigg],
\end{align}
with 
\begin{align}
F_{0}(\vc{k}) &= A_0(\vc{k}),\\
\begin{split}
F_{2}(\vc{k}) &= \frac{1}{k^2}\bigg[k_x^2 B_{xx} + k_y^2 B_{yy} + k_z^2B_{zz}\\
&+2\bigg(k_zk_y B_{xy} + k_xk_zB_{xz} + k_yk_zB_{yz}\bigg)\bigg],
\end{split}\\
\begin{split}
F_{4}(\vc{k}) &= \frac{1}{k^4}\bigg[k_x^4 C_{xxx} + k_y^4 C_{yyy} + k_z^4C_{zzz} \\
&+4\bigg(k^3_xk_y C_{xxy} + k^3_xk_zC_{xxz} + k^3_yk_xC_{yyx}\\
&+k^3_yk_z C_{yyz} + k^3_zk_xC_{zzx} + k^3_zk_yC_{zzy}\bigg) \\
&+6\bigg(k^2_xk^2_y C_{xyy} + k^2_xk^2_zC_{xzz} + k^2_yk^2_zC_{yzz}\bigg) \\
&+12k_xk_yk_z\bigg(k_x C_{xyz} + k_yC_{yxz} + k_zC_{zxy}\bigg) \bigg]
\end{split}
\end{align}
and
\begin{align}
A_0(\vc{k}) &= \int d\vc{r} F(\vc{r})e^{i\vc{k}\cdot \vc{r}},\\
B_{xy}(\vc{k}) &= \int d\vc{r} \frac{r_xr_y}{r^2}F(\vc{r})e^{i\vc{k}\cdot \vc{r}},\\
C_{xyz}(\vc{k}) &= \int d\vc{r} \frac{r^2_xr_yr_z}{r^4}F(\vc{r})e^{i\vc{k}\cdot \vc{r}}.
\end{align}
$F(\vc{r})$ is the over-density field calculated from the data and random galaxies as 
\begin{equation}
F(\vc{r}) = G(\vc{r}) - \alpha'R(\vc{r}),
\end{equation}
where $\vc{r}$ is defined on a 3D Cartesian grid, in which we bin all data and random galaxies. The function $G(\vc{r})$ gives the number of weighted galaxies at the location $\vc{r}$, while $R(\vc{r})$ is the equivalent function for the random galaxies. The normalisation of the random field is given by $\alpha' = N'_{\rm ran}/N'_{\rm gal}$; $N'_{\rm ran}$ and  $N'_{\rm gal}$ are the number of weighted random and data galaxies, respectively. All the integrals above can be solved with Fast Fourier Transforms. The normalisation is given by
\begin{equation}
A = \alpha'\sum_i^{N_{\rm ran}}n'_g(\vc{x}_i)w^2_{\rm FKP}(\vc{x}_i),
\label{eq:norm}
\end{equation}
where $n'_g$ is the weighted galaxy number density. The shot noise term is only relevant for the monopole and is given by
\begin{align}
S &= \sum^{N_{\rm gal}}_i \bigg[f_cw_{c}(\vc{x}_i)w_{\rm sys}(\vc{x}_i)w_{\text{\tiny{FKP}}}^2(\vc{x}_i)\\
&+ (1-f_c)w_c^2(\vc{x}_i)w^2_{\rm FKP}(\vc{x}_i)\bigg]\\
&+ \alpha'^2\sum^{N_{\rm ran}}_i w_{\text{\tiny{FKP}}}^2(\vc{x}_i),
\end{align}
where $f_c$ is the probability of a fibre collided galaxy being associated with its nearest neighbour, which we set to $0.5$ based on the study by~\citet{Guo:2011ai}. \citet{Guo:2011ai}, however, studied the fibre collision only for the CMASS sample and their fibre collision correction assumes a uniform tiling algorithm. Although this definition of the shot noise deviates from the one used in~\citet{Beutler:2013yhm}, the difference does not actually impact our analysis since we marginalise over any residual shot noise component (see section~\ref{sec:model}).

The final power spectrum is calculated as the average over spherical \textit{k}-space shells
\begin{equation}
P_{\ell}(k) = \langle P_{\ell}(\vc{k})\rangle = \frac{1}{N_{\rm modes}}\sum_{k-\frac{\Delta k}{2} < |\vc{k}| < k+\frac{\Delta k}{2}}P_{\ell}(\vc{k}),
\label{eq:averaging}
\end{equation}
where $N_{\rm modes}$ is the number of $\vc{k}$ modes in that shell. In our analysis we use $\Delta k = 0.01\ihMpc$. We employ a Triangular Shaped Cloud method to assign galaxies to the 3D Cartesian grid and correct for the aliasing effect following~\citet{Jing:2004fq}. The grid configuration implies a Nyquist frequency of $k_{\rm Ny}=0.6\ihMpc$, four times as large as the largest scale used in our analysis ($k_{\rm max} = 0.15\hMpc$), and the expected error on the power spectrum monopole at $k = 0.15\hMpc$ due to aliasing is $< 0.001\%$~\citep{Sefusatti:2015aex}.

\section{The survey window function}
\label{sec:win}

\begin{figure*}
\begin{center}
\epsfig{file=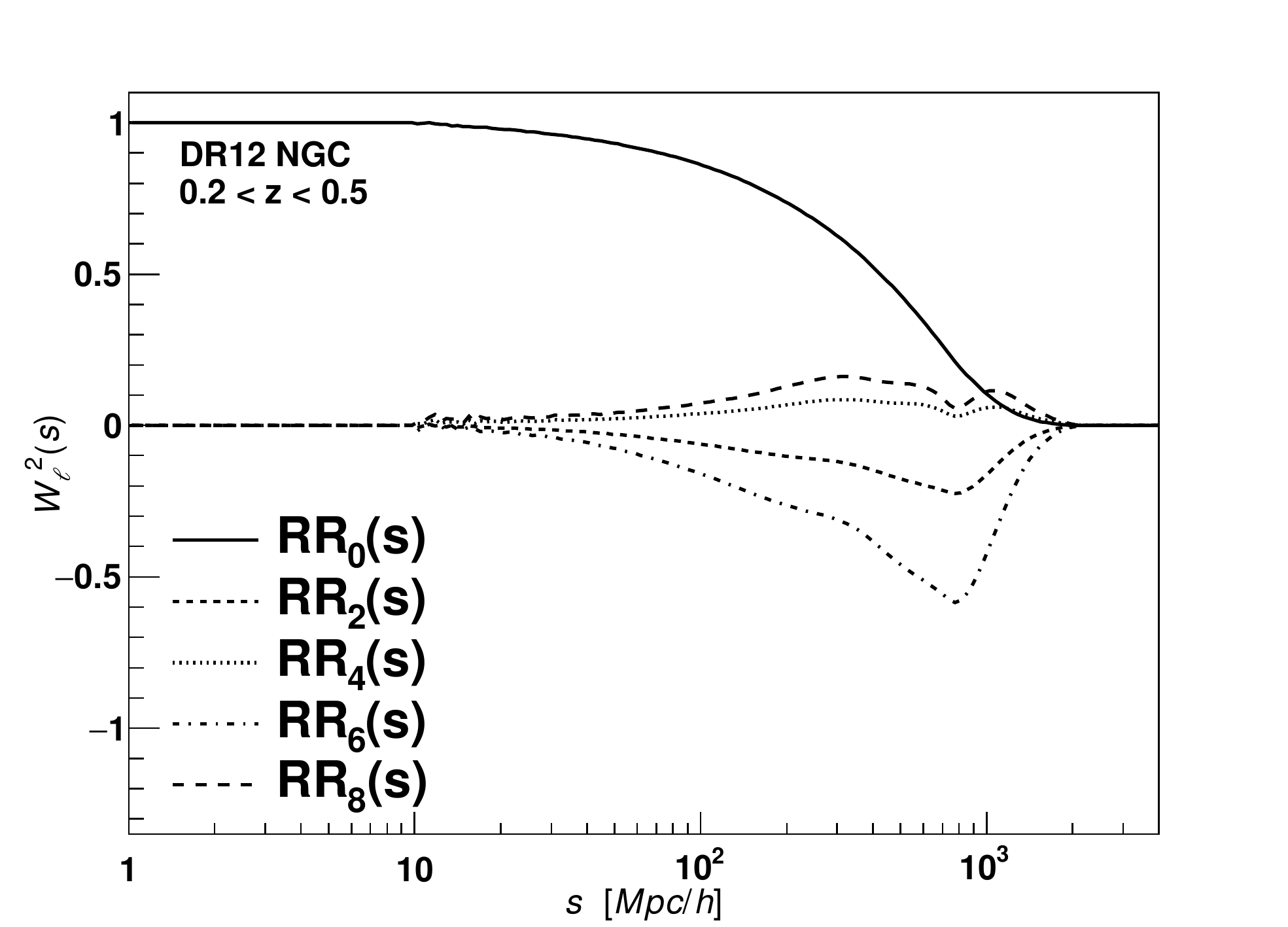,width=5.8cm}
\epsfig{file=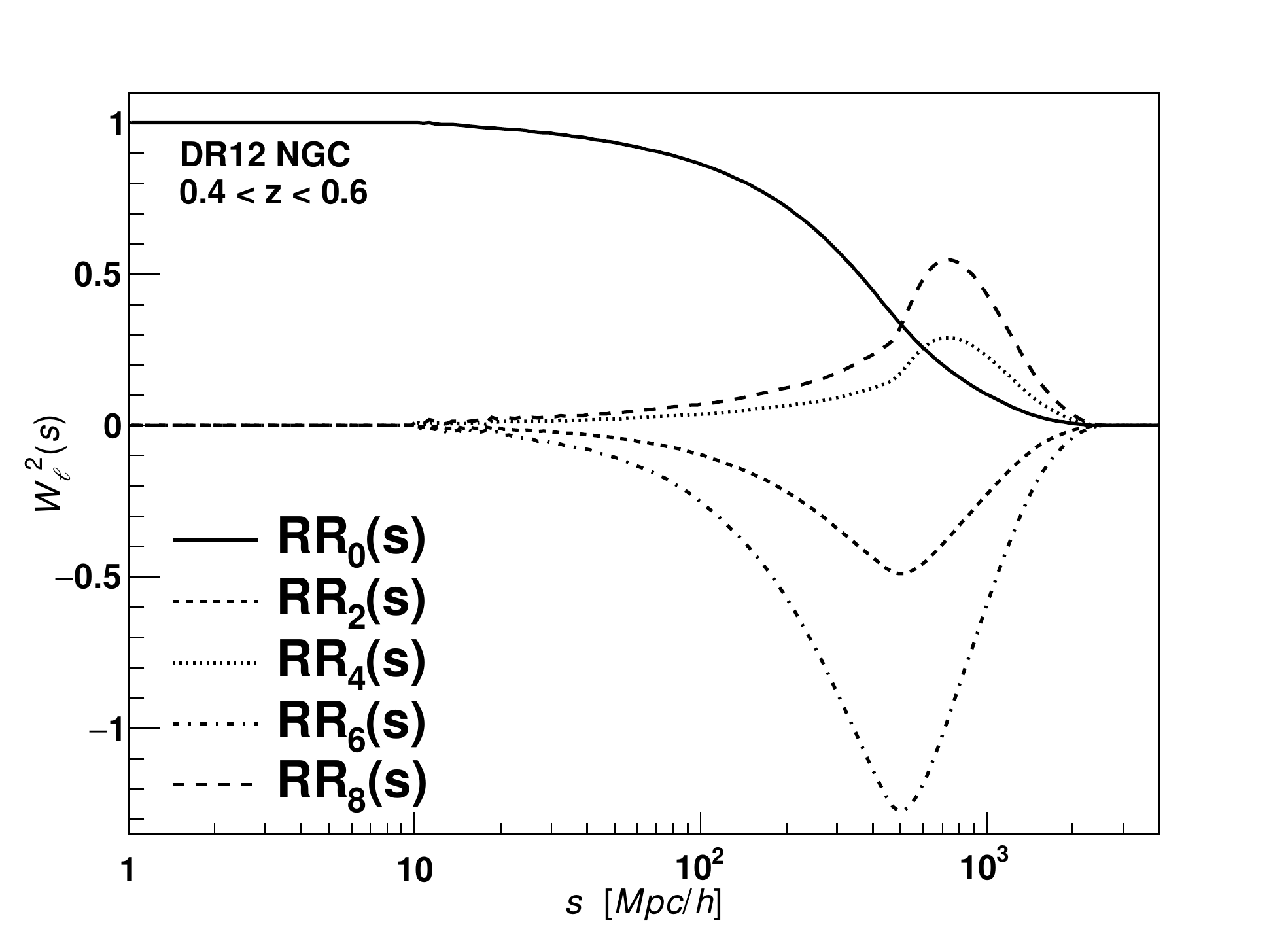,width=5.8cm}
\epsfig{file=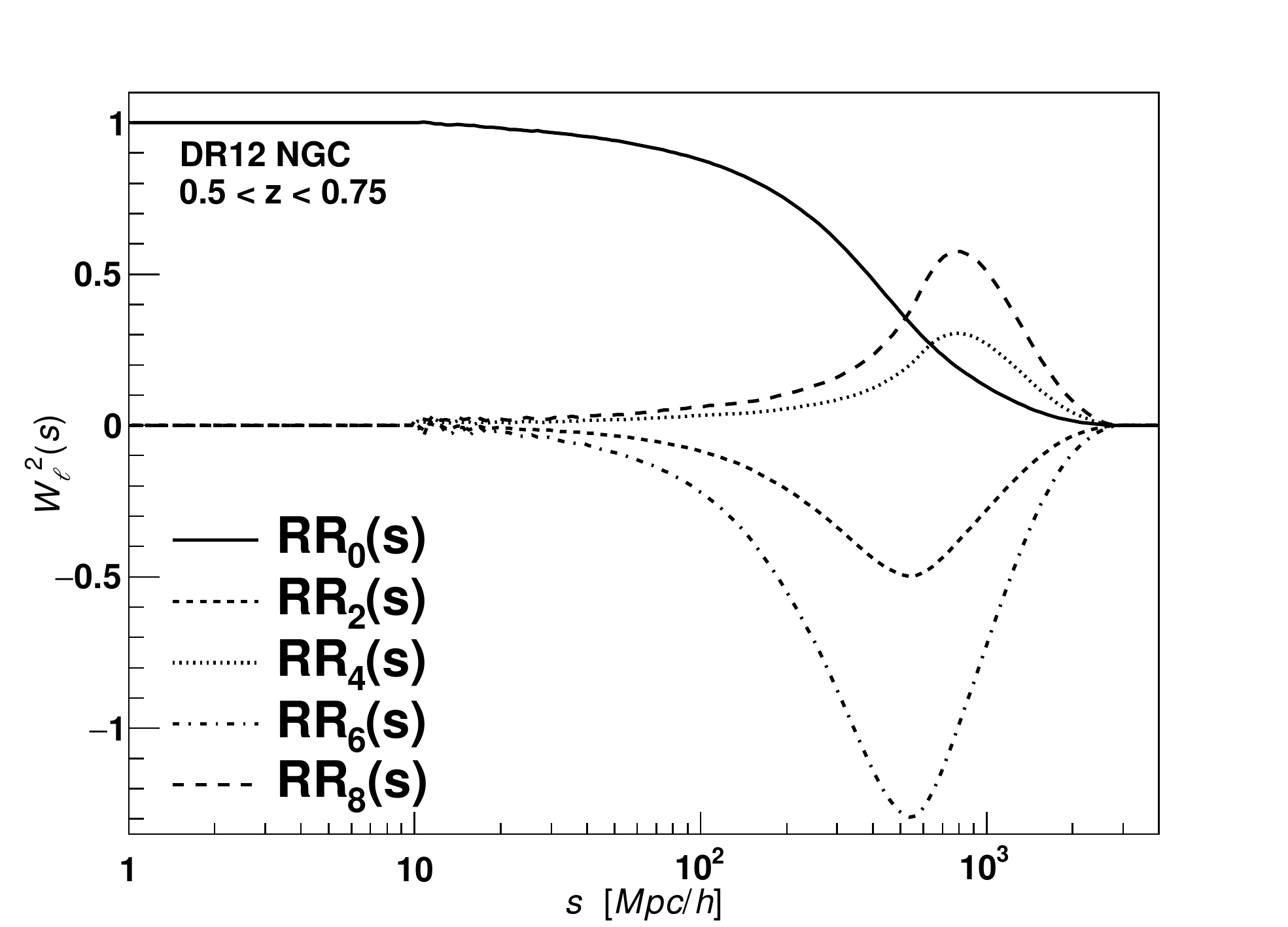,width=5.8cm}\\
\epsfig{file=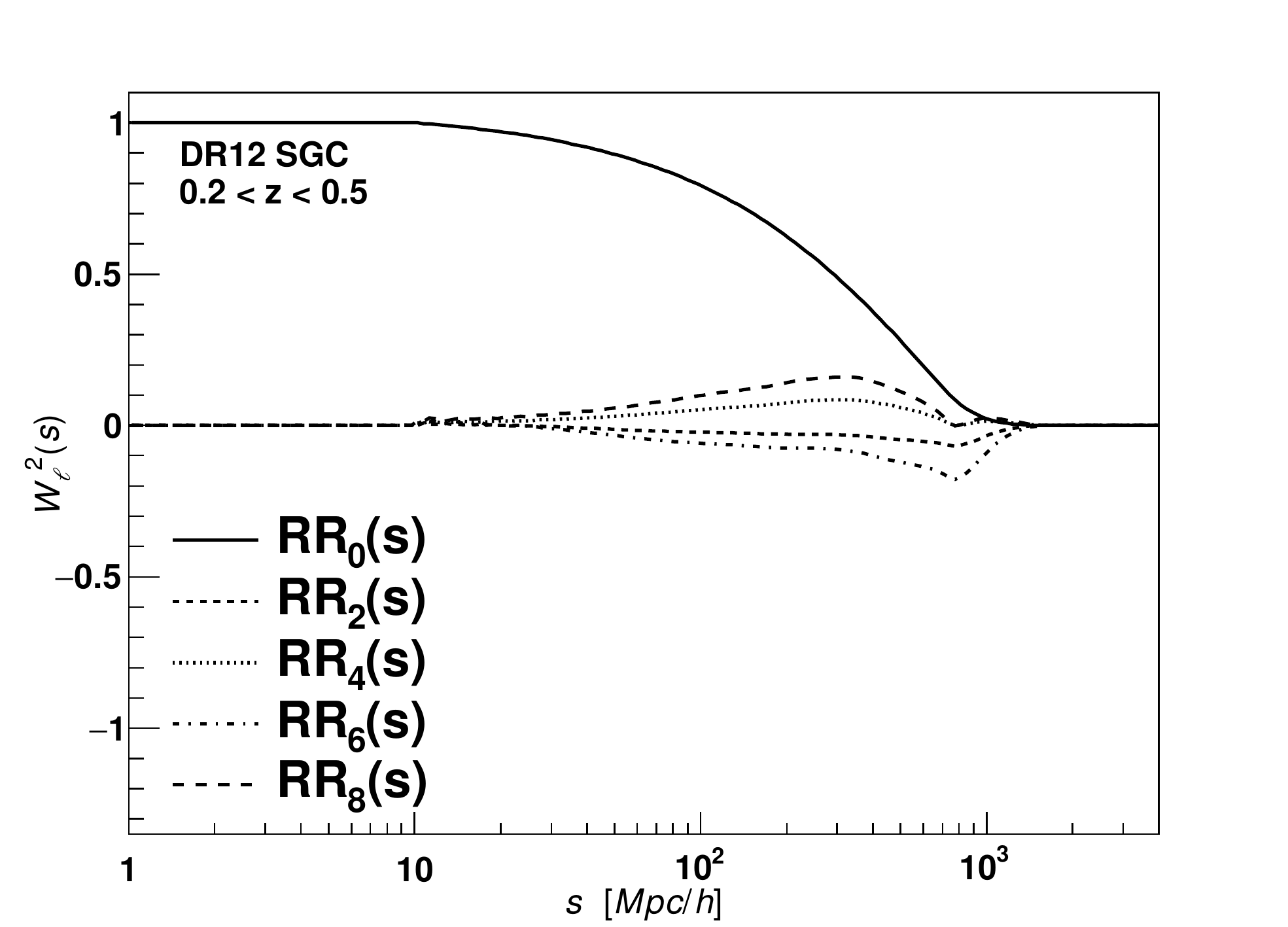,width=5.8cm}
\epsfig{file=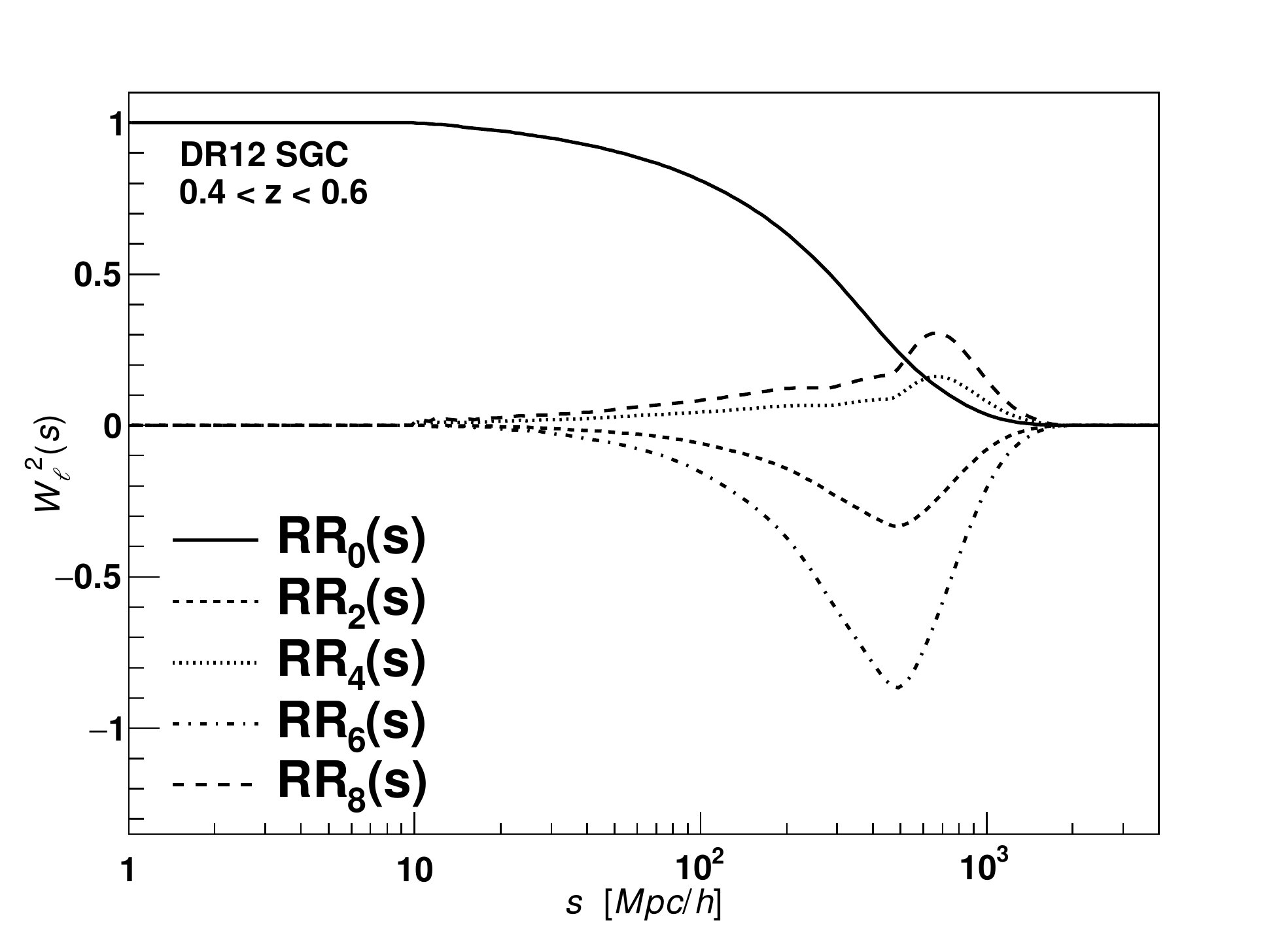,width=5.8cm}
\epsfig{file=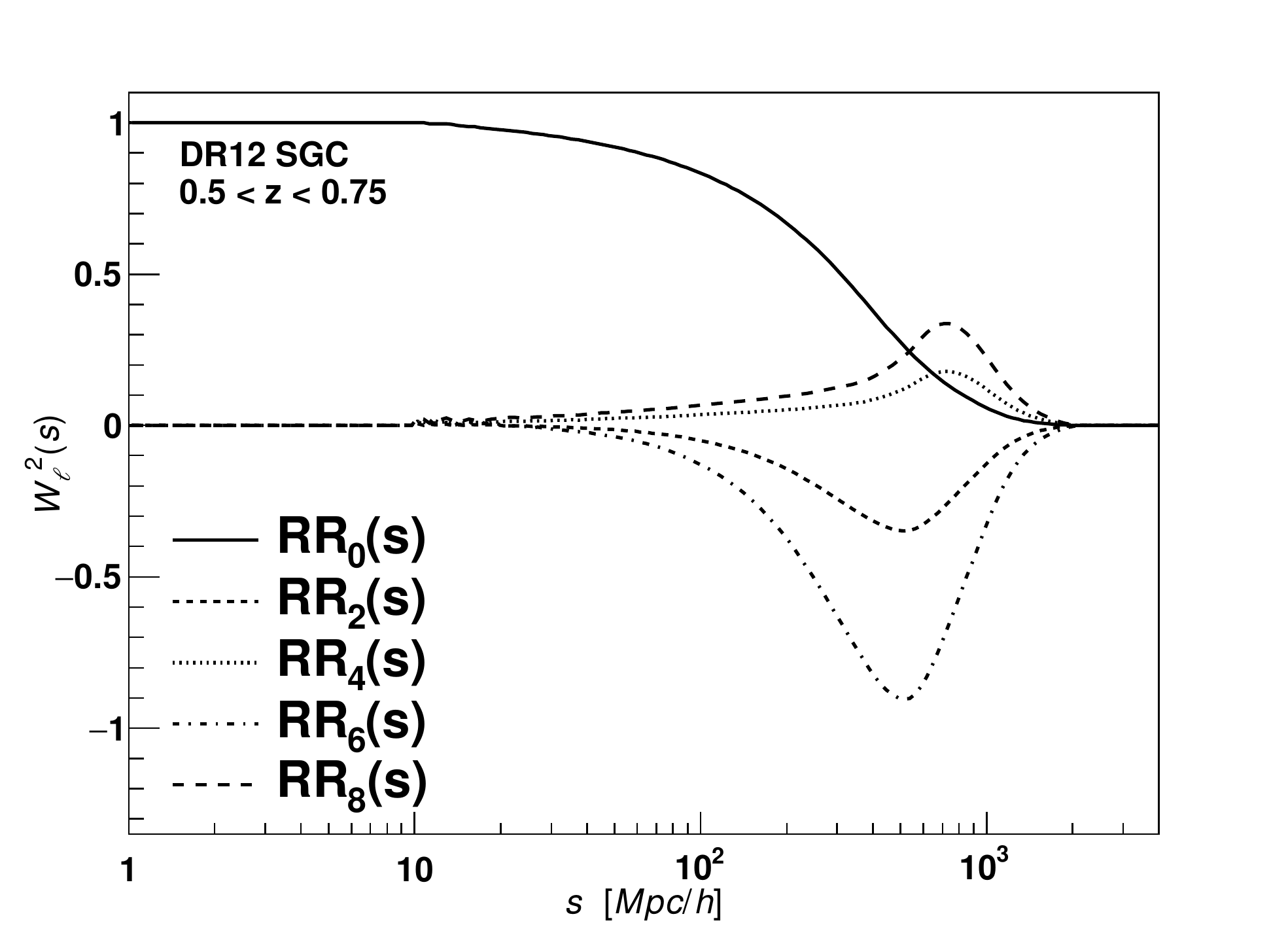,width=5.8cm}
\caption{Window function multipoles for BOSS DR12 as given in eq.~\ref{eq:Well} and used for the convolved correlation functions in eq.~\ref{eq:conv1} - \ref{eq:conv3}. The top panels display the window functions for the North Galactic Cap (NGC) in the three redshift bins used in this analysis; the bottom panels show the window functions for the South Galactic Cap (SGC).}
\label{fig:win_NGCSGC}
\end{center}
\end{figure*}

\begin{figure}
\begin{center}
\epsfig{file=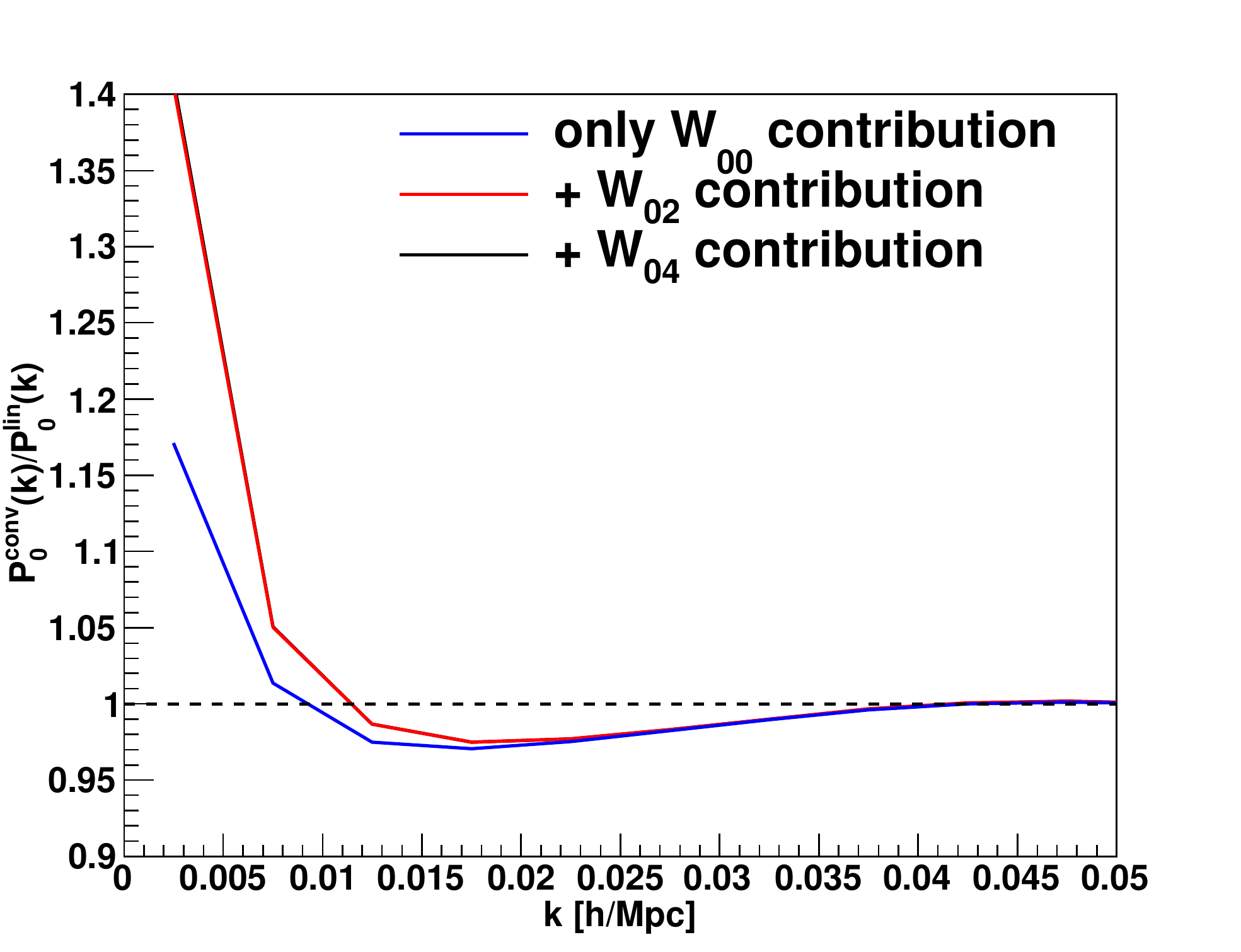,width=8cm}
\caption{The leakage of higher order multipoles to the observed monopole power spectrum due to the window function effect. The blue line shows the contribution from the monopole to the observed monopole; the red line indicates the contributions from the monopole and quadrupole and the black line represents the contributions from the monopole, quadrupole and hexadecapole. The total window function effect (black) is of the order of $2\%$ for $k \lesssim 0.04\ihMpc$ and significantly increases for $k \lesssim 0.01\ihMpc$. The quadrupole contribution (from the difference between the blue and red lines) becomes significant at $k \lesssim 0.015\ihMpc$. The hexadecapole contribution (from the difference between the red and the black lines) is negligible on all scales.}
\label{fig:win_mono}
\end{center}
\end{figure}

\begin{figure}
\begin{center}
\epsfig{file=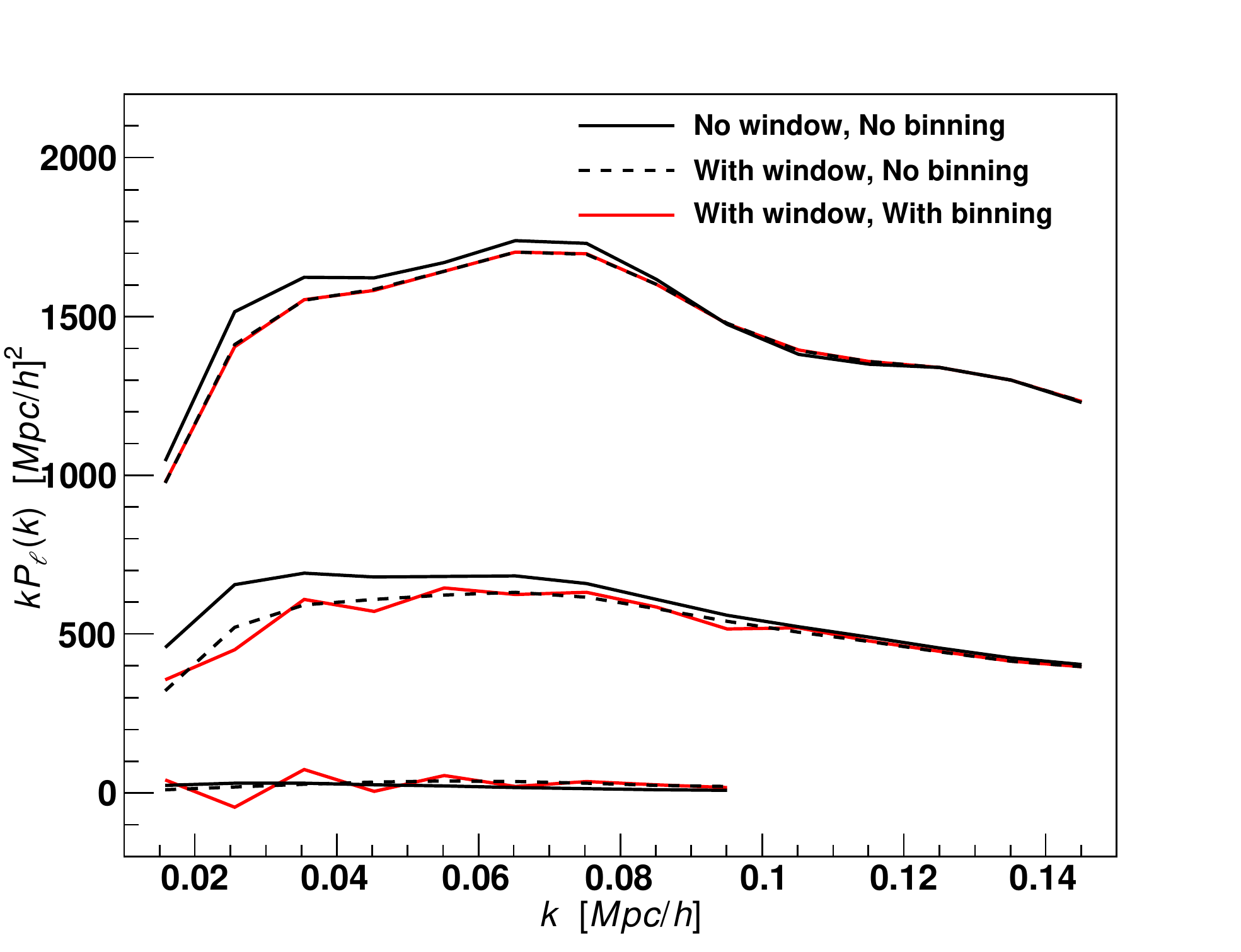,width=8cm}
\caption{Illustration of the discreteness and the window function effects on the power spectrum multipoles for the low redshift bin in the South Galactic Cap (SGC). The monopole (top), the quadrupole (middle) and hexadecapole (bottom), are displayed in the range used in this analysis. The solid black lines show the input linear power spectrum multipoles using a linear bias of $b_1 = 2$ and a growth rate of $f = 0.7$, while the black dashed lines are the same power spectra convolved with the window function. The main effect of the window function is a damping at small $k$. The red line also includes the discreteness effect using eq.~\ref{eq:binning}. The discreteness effect is caused by the finite \textit{k}-space grid, used to estimate the power spectrum multipoles (see section~\ref{sec:binning}).}
\label{fig:binning}
\end{center}
\end{figure}

\begin{figure}
\begin{center}
\epsfig{file=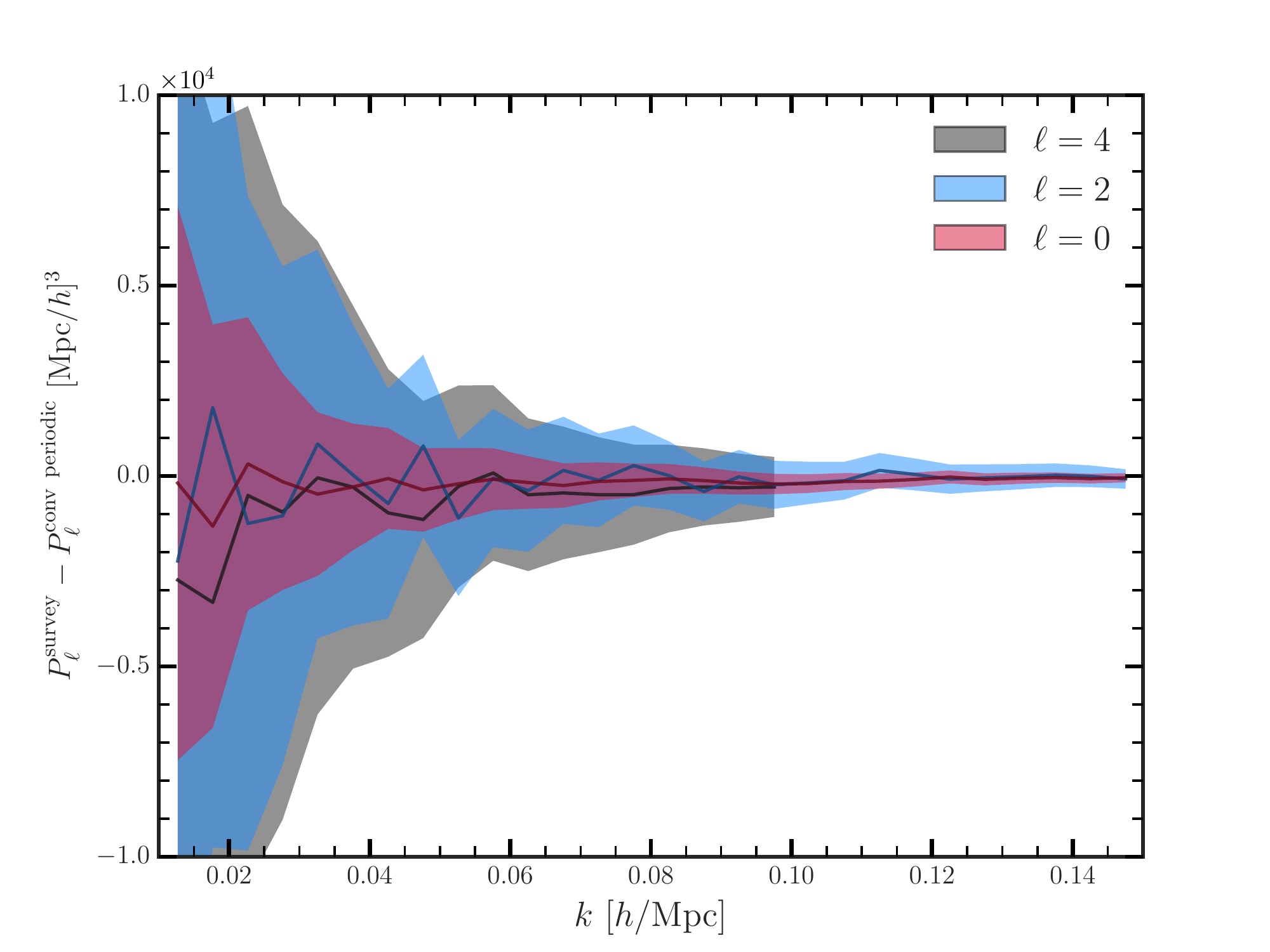,width=8cm}
\caption{Here we show the difference between the mean power spectrum multipoles of a set of CMASS-like mock catalogues and the mean power spectrum multipoles of the corresponding periodic boxes convolved with the window function. A value of zero indicates that our window function convolution method does correctly model the effects introduced by the survey geometry. The colour bands indicate the uncertainties as given by the diagonal terms of the NGC covariance matrix for the second redshift bin.}
\label{fig:win_test}
\end{center}
\end{figure}

In this paper, we include the window function effect in the power spectrum model rather than attempting to remove the effect from the data. Unlike~\citet{Beutler:2013yhm}, our window function treatment follows the method suggested by~\citet{Wilson:2015lup}. However,~\citet{Wilson:2015lup} employs the global plane parallel approximation in their derivation by setting $\mu = \hat{k}\cdot\hat{\eta}$, where $\eta$ defines a fixed global line-of-sight vector. The global plane parallel approximation might be an acceptable approximation for small angle surveys\footnote{\citet{Wilson:2015lup} apply their formalism to the VIPERS survey~\citep{Garilli:2013eoa}.}, but is generally not valid for wide area surveys such as BOSS. In appendix~\ref{app:window} we will re-derive this window function treatment within the local plane parallel approximation, which justifies its use in our analysis.

The three main steps to include the effect of the window function in our power spectrum model are:
\begin{enumerate}
\item For each model power spectrum, which we intend to compare to the data, we first calculate the model power spectrum multipoles and Fourier transform them to determine the correlation function multipoles $\xi_{\ell}^{\rm model}(s)$.
\item We calculate the ``convolved'' correlation function $\hat{\xi}^{\rm model}_{\ell}(s)$ by multiplying the correlation function with the window function. 
\item We Fourier transform the convolved correlation function multipoles back into Fourier space to get the convolved power spectrum multipoles, $\hat{P}_{\ell}(k)$.
\end{enumerate}

The convolved power spectrum multipoles are given by 
\begin{equation}
\hat{P}_{\ell}(k) = 4\pi (-i)^\ell \int ds\,s^2\hat{\xi}_{\ell}(s) j_{\ell}(sk).
\label{eq:PWell}
\end{equation}
For our analysis we need to calculate the convolved monopole, quadrupole, and hexadecapole power spectra. The convolved correlation function multipoles in eq.~\ref{eq:PWell}, relevant for our analysis, are
\begin{align}
	\hat{\xi}_{0}(s) &= \xi_{0}W_{0}^2  + \frac{1}{5}\xi_2W^2_2 + \frac{1}{9}\xi_4W^2_4 +...
	\label{eq:conv1}\\
	\begin{split}
		\hat{\xi}_{2}(s) &= \xi_{0} W_{2}^2  + \xi_2\left[W^2_0 + \frac{2}{7}W^2_2 + \frac{2}{7}W^2_4\right]\\
		&\;\;\;\;\;\;\;\;\;\;\;\;\;\,+\xi_4\left[\frac{2}{7}W^2_2 + \frac{100}{693}W^2_4 + \frac{25}{143}W^2_6\right]\\
		&\;\;\;\;\;\;\;\;\;\;\;\;\;\,+...
		\label{eq:conv2}
	\end{split}\\
	\begin{split}
		\hat{\xi}_{4}(s) &= \xi_{0} W_{4}^2  + \xi_2\left[\frac{18}{35}W^2_2 + \frac{20}{77}W^2_4 + \frac{45}{143}W^2_6\right]\\
		&\;\;\;\;\;\;\;\;\;\;\;\;\;\,+\xi_4\bigg[W^2_0 + \frac{20}{77}W^2_2 + \frac{162}{1001}W^2_4\\
		&\;\;\;\;\;\;\;\;\;\;\;\;\;\,+ \frac{20}{143}W^2_6 + \frac{490}{2431}W^2_8 \bigg]\\
		&\;\;\;\;\;\;\;\;\;\;\;\;\;\,+...
		\label{eq:conv3}
	\end{split}
\end{align}
We truncate after the hexadecapole contribution of the correlation function but use all window function multipoles up to $\ell = 8$.

The different window function multipoles included in our analysis can be derived from the random pair distribution as 
\begin{equation}
W_{\ell}^2(s) \propto  \sum_{\vec{x}_1} \sum_{\vec{x}_2}   RR(s,\mu_s)\mathcal{L}_{\ell}(\mu_s).
\label{eq:Well}
\end{equation}
The first five non-zero window functions used in our analysis are shown in Figure~\ref{fig:win_NGCSGC}. The shape of these functions can be understood by investigating eq.~\ref{eq:Well}. The monopole is a spherically averaged function, and on small scales, where survey edge effects don't matter, the window will be equal to $1$ (given the choice of our normalisation). Similarly the quadrupole, which is an integral over a function oscillating around zero, will average to zero on small scales, while as soon as it reaches scales as large as the survey, it will no longer average to zero.

Figure~\ref{fig:binning} shows the unconvolved and convolved power spectrum multipoles for the SGC in the lowest redshift bin. Given that the SGC in the smallest redshift bin has the smallest volume, we expect window function effects to be largest in this case.

When one reaches the scale of the survey, the window function becomes crucial for the correct interpretation of the data. We find a significant leakage of power from the quadrupole to the monopole on large scales due to the window function. The effect is shown in Figure~\ref{fig:win_mono}, which presents the ratio of the convolved power spectrum monopole relative to the unconvolved (linear) power spectrum monopole. The window function effect is of the order of $2\%$ for $k \lesssim 0.04\ihMpc$ and significantly increases for $k \lesssim 0.01\ihMpc$. The quadrupole contribution to the monopole becomes significant at $k \lesssim 0.015\ihMpc$. The hexadecapole contribution (see the difference between the red and the black lines in Figure~\ref{fig:win_mono}) is negligible on all scales. While these effects are of minor importance for our analysis given that we have a minimum $k$ value of $0.01\ihMpc$, they can be quite important for studies of non-Gaussianity, where the observable is at very small $k$.

We also tested our window function treatment by comparing the power spectrum multipoles for $84$ CMASS-like mock catalogues with the convolved multipoles of the corresponding periodic boxes (see~Figure~\ref{fig:win_test}). The agreement is far better than the measurement uncertainties and confirms that our convolution method is capturing the geometric effects correctly.

\begin{figure*}
\begin{center}
\epsfig{file=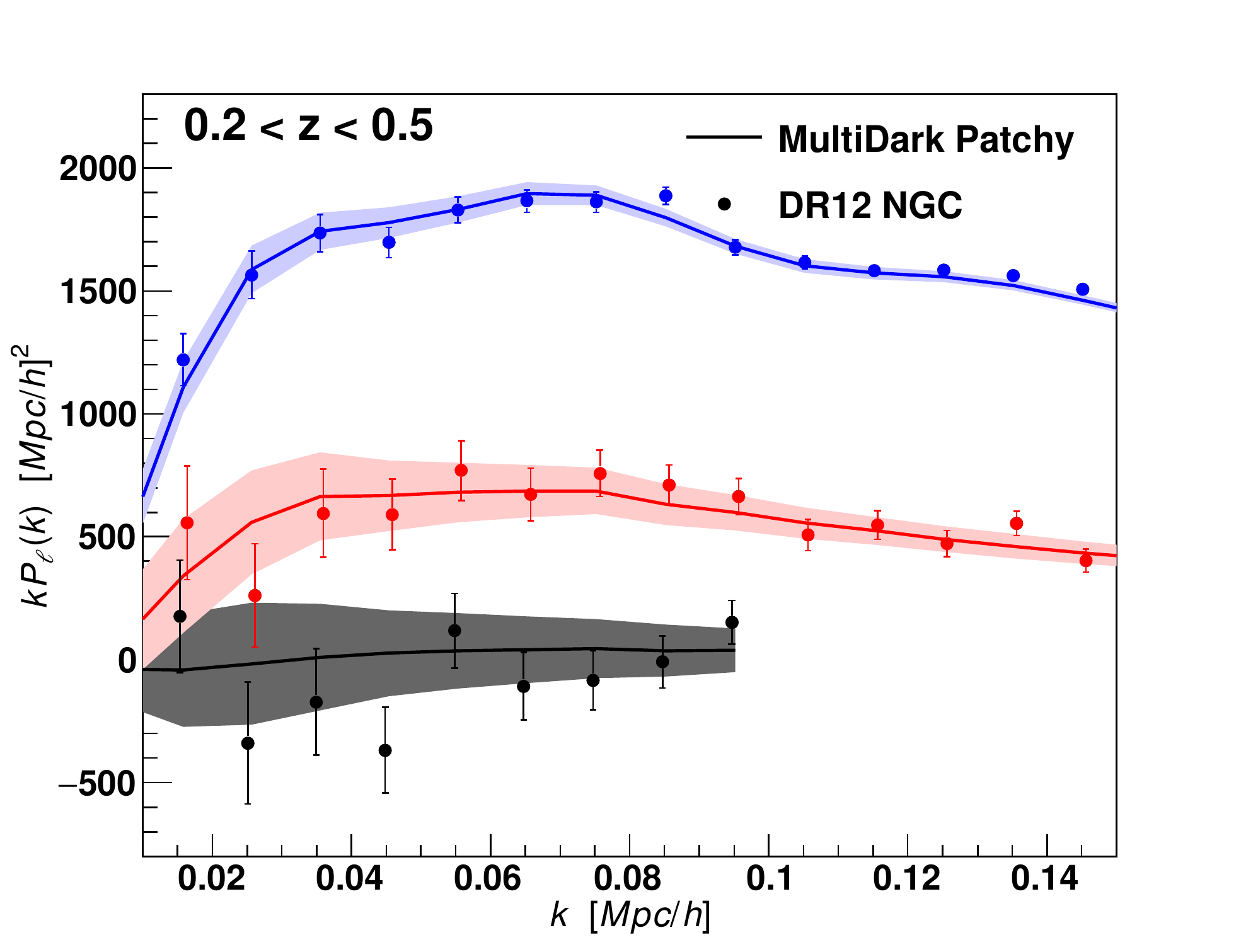,width=5.8cm}
\epsfig{file=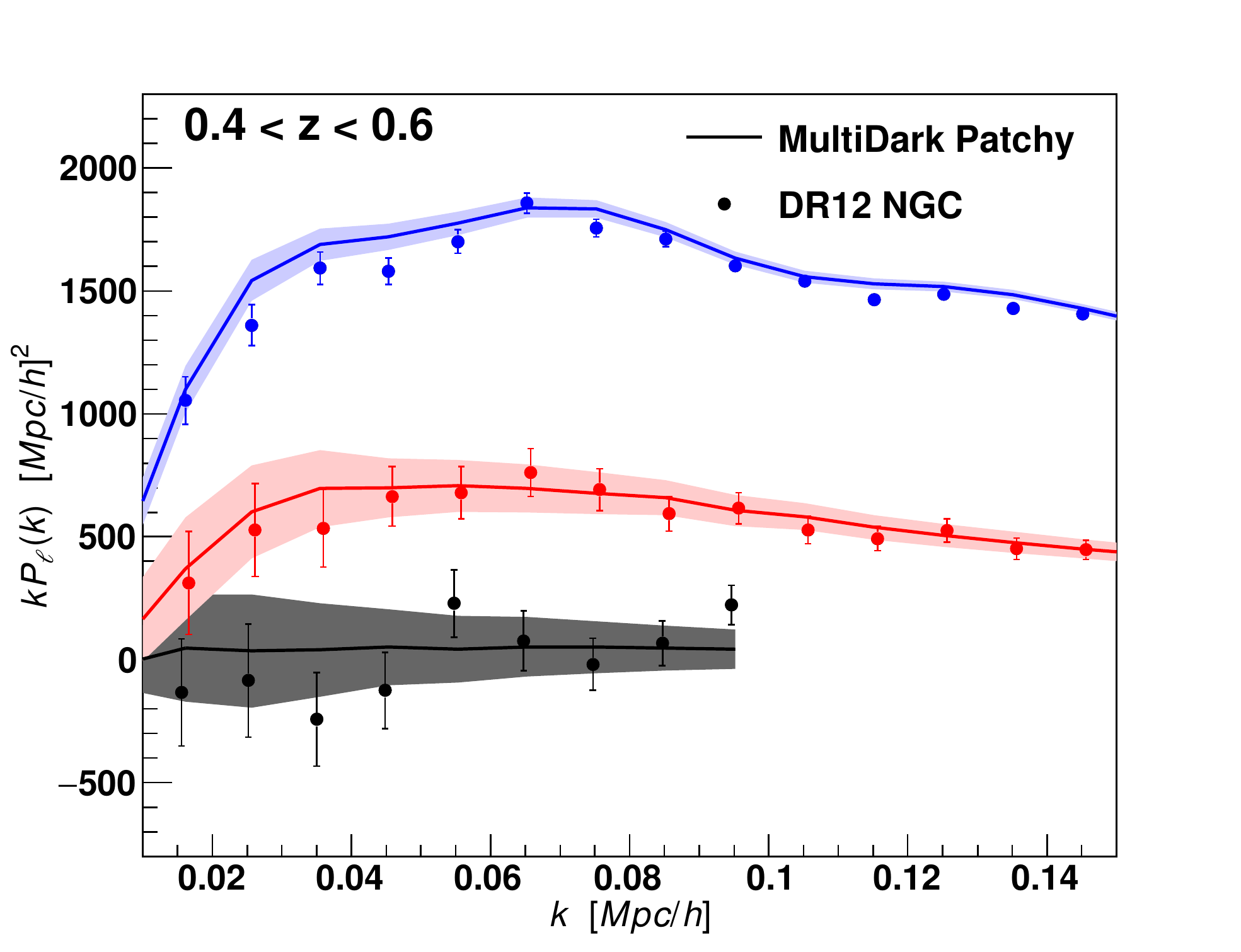,width=5.8cm}
\epsfig{file=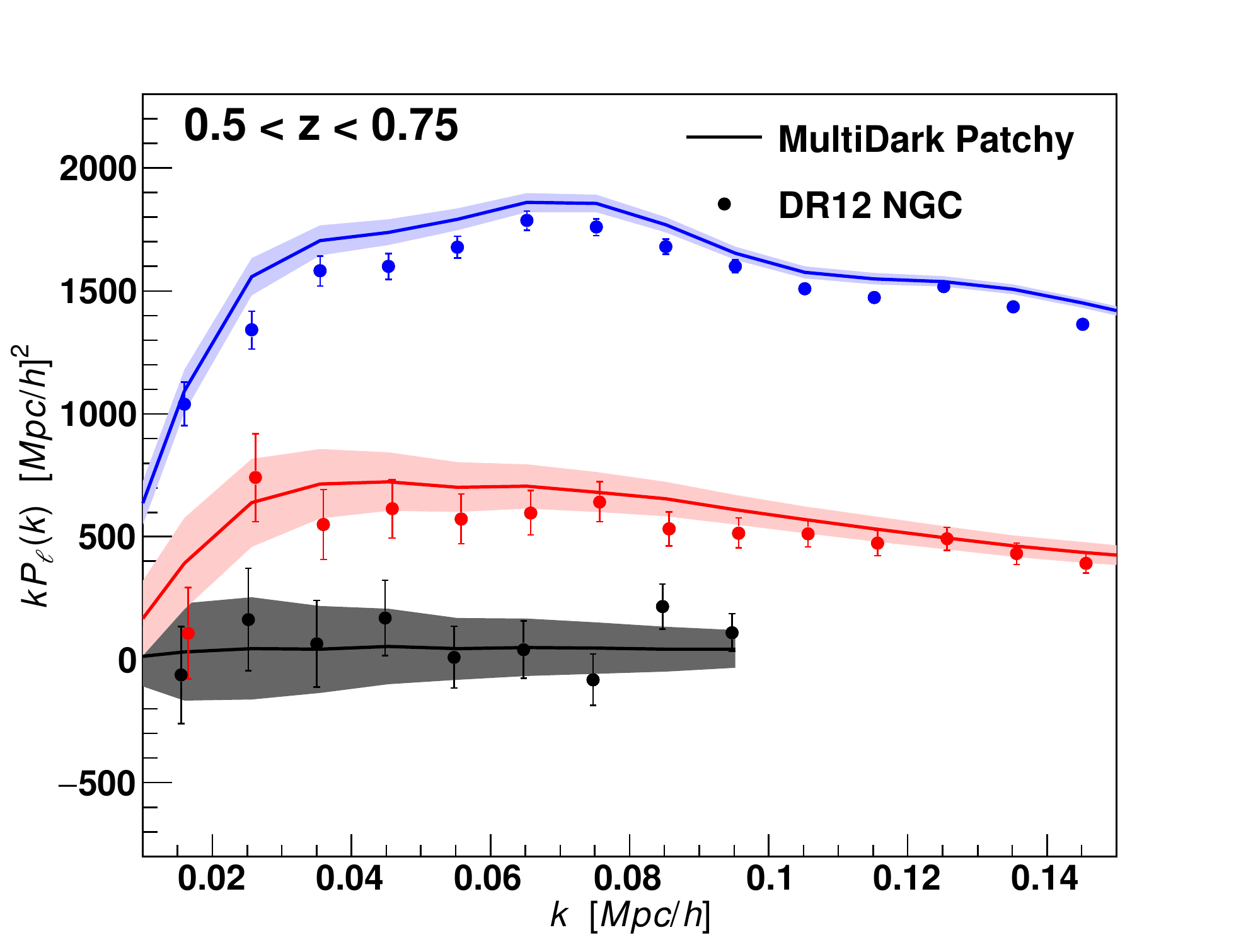,width=5.8cm}\\
\epsfig{file=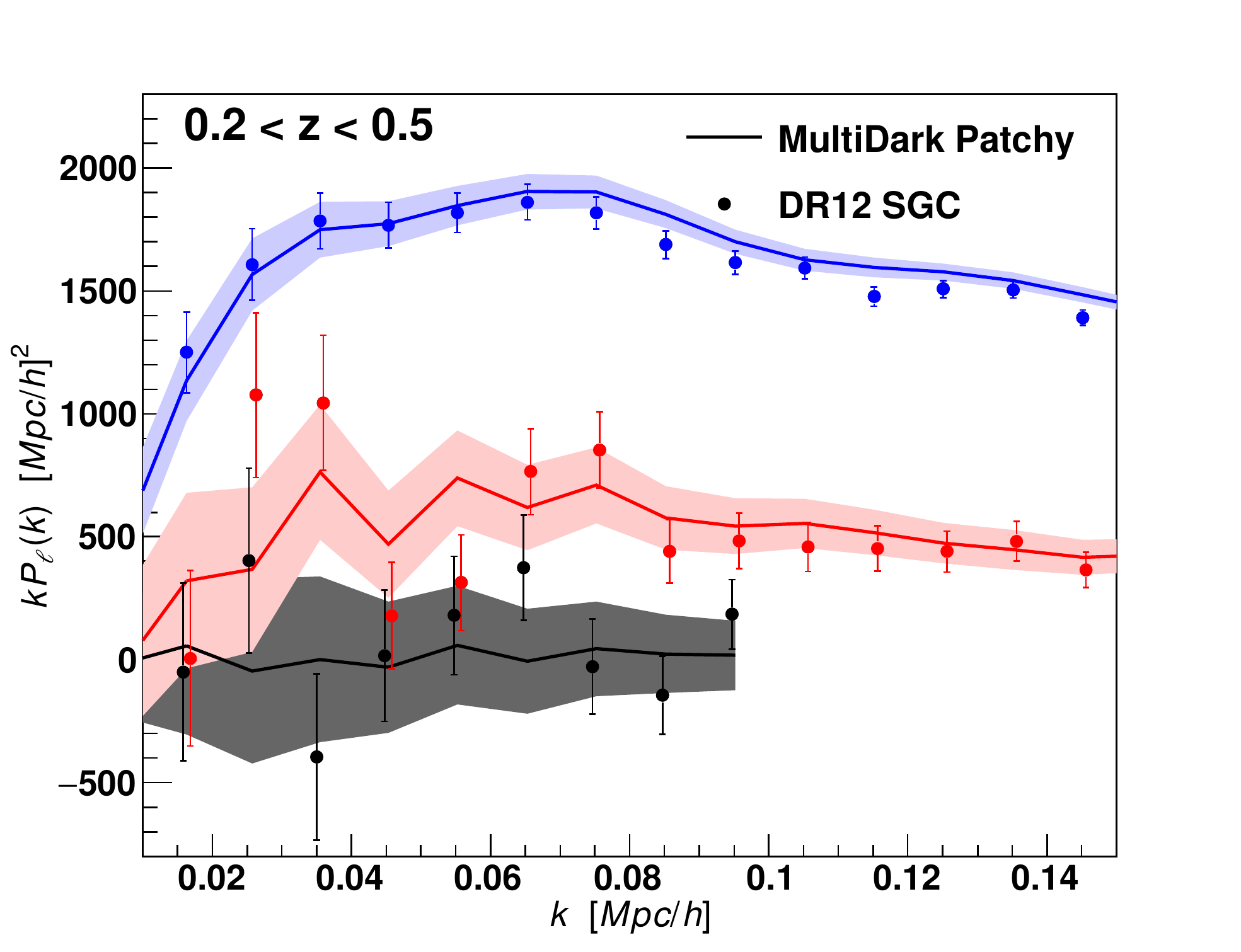,width=5.8cm}
\epsfig{file=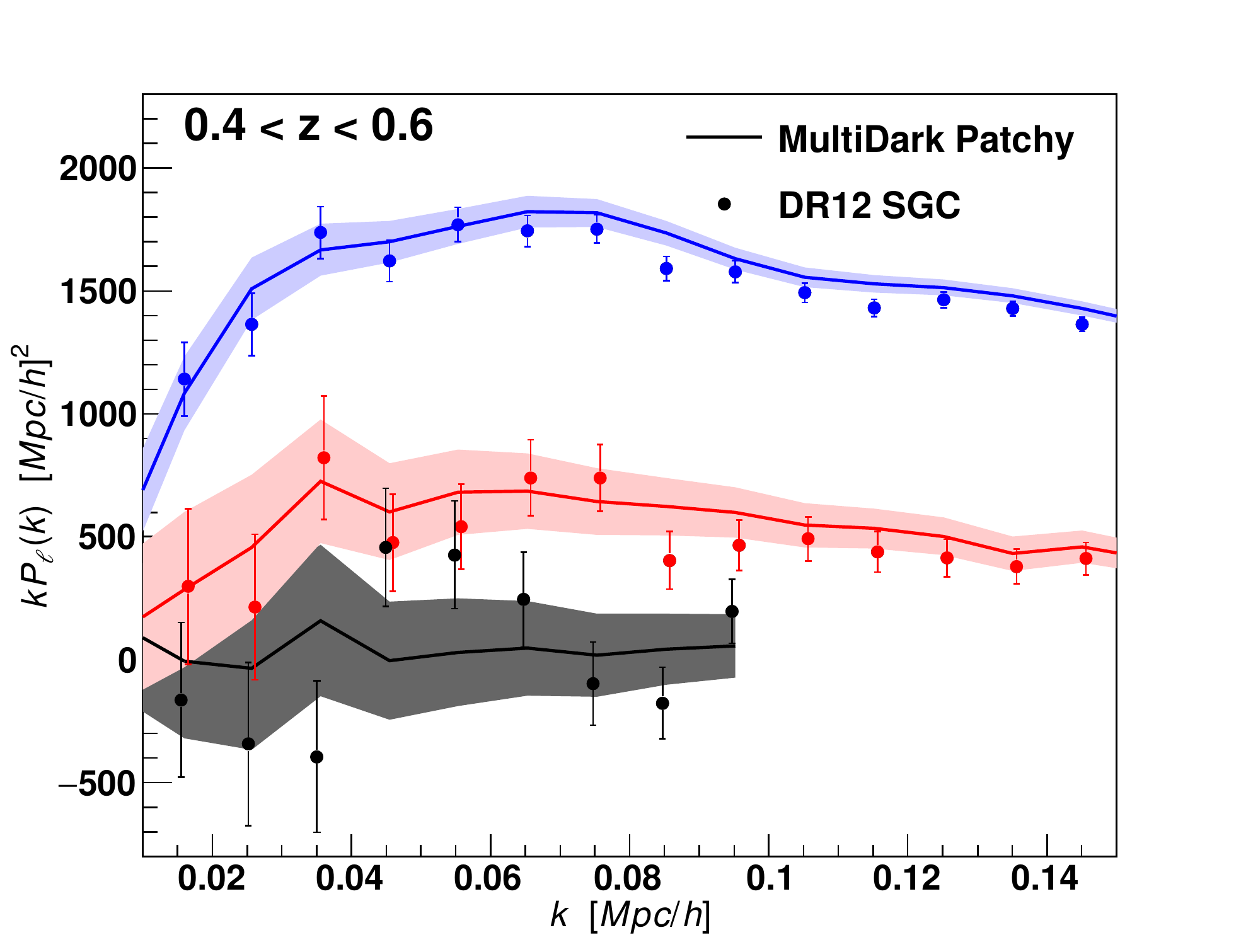,width=5.8cm}
\epsfig{file=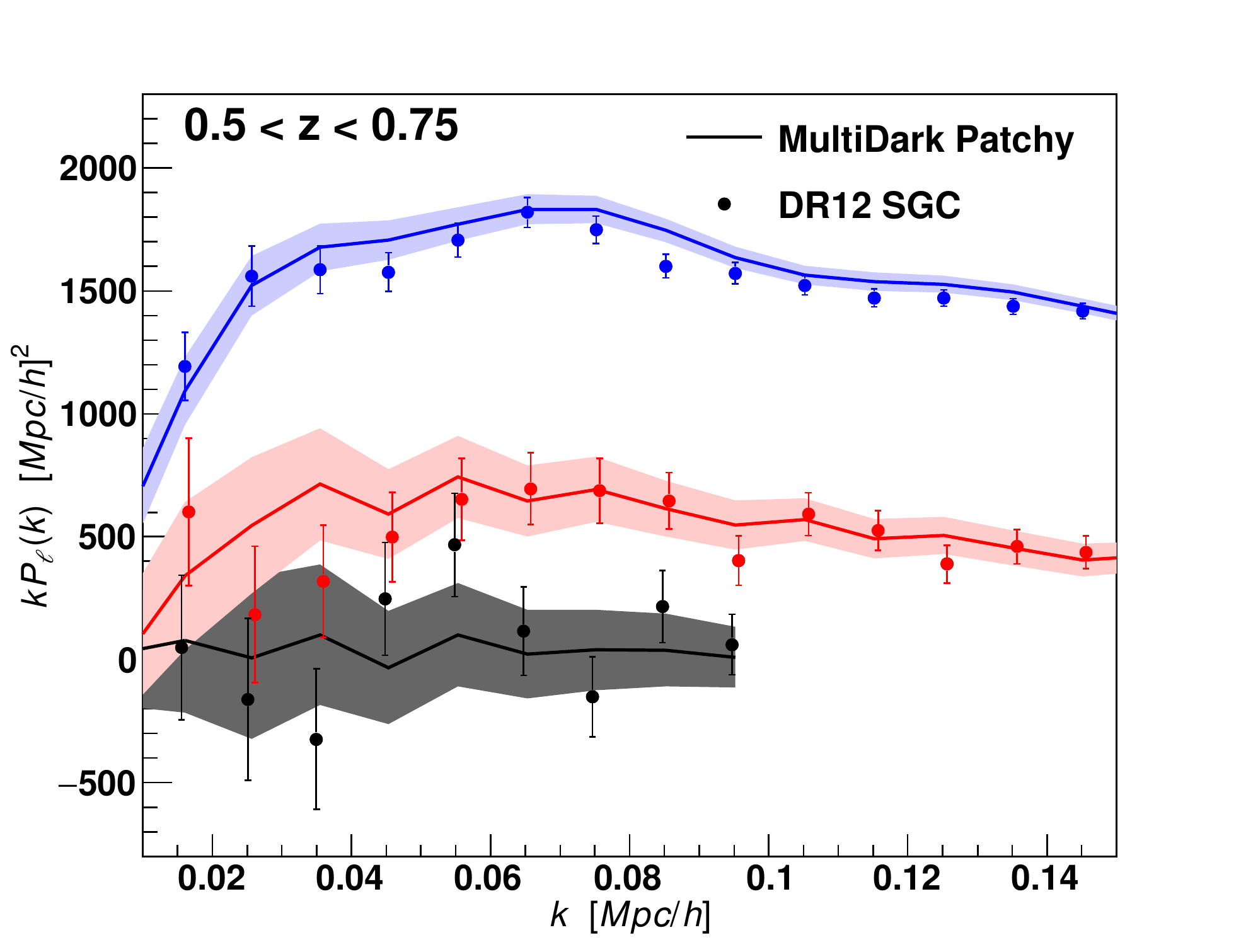,width=5.8cm}
\caption{Comparison of the BOSS DR12 power spectrum multipoles (coloured data points) and the mean of the MultiDark-Patchy mock catalogues (coloured solid lines) with the same selection function as the data. The top panels show the power spectrum multipoles for the three redshift bins in the North Galactic Cap (NGC) and the bottom panels are the same measurements for South Galactic Cap (SGC). The different multipoles are colour coded, where blue represents the monopole, red represents the quadrupole and black shows the hexadecapole. The shaded area is the variance between all mock catalogues and is identical to the extent of the error bars on the data points. For SGC (bottom panels), the mock catalogues show some correlated fluctuations at small $k$, which is most prominent in the higher order multipoles. This feature is a discreteness effect, due to the finite number of modes at large scales. This effect is present in the data as well, and we discuss how to account for this effect in our power spectrum model in section~\ref{sec:binning}.}
\label{fig:ps_NGCSGC}
\end{center}
\end{figure*}

\section{The power spectrum model}
\label{sec:model}

Our model for the anisotropic galaxy power spectrum is based on the work of~\citet{Taruya:2010mx} (TNS) and is the same as used in our DR11 analysis~\citep{Beutler:2013yhm}. We summarise this model below but refer the reader to~\citet{Beutler:2013yhm} for more details. 

The anisotropic power spectrum is given by
\begin{equation}
\begin{split}
P_{\rm g}(k,\mu) &= \exp\left\{-(fk\mu\sigma_v)^2\right\}\big[P_{{\rm g},\delta\delta}(k)\\
&\;\;\;\; + 2f\mu^2P_{{\rm g},\delta\theta}(k) + f^2\mu^4P_{\theta\theta}(k)\\
&\;\;\;\; + b_1^3A(k,\mu,\beta) + b_1^4B(k,\mu,\beta)\big], 
\label{eq:taruya}
\end{split}
\end{equation}
where $\mu$ denotes the cosine of the angle between the wavenumber vector and the line-of-sight direction. The overall exponential factor represents the suppression due to the ``Finger of God'' effect, and we treat $\sigma_{v}$ as a free parameter.

The first three terms in the square brackets in eq.~\ref{eq:taruya} describe an extension to the simple Kaiser model. The density ($P_{\delta\delta}$), velocity divergence ($P_{\theta\theta}$) and their cross-power spectra ($P_{\delta\theta}$) are identical in linear theory,  while in the quasi non-linear regime, the density power spectrum increases and velocities are randomised on small scales which damps the velocity power spectrum~\citep{Scoccimarro:2004tg}. In addition to this fact, we need to relate the density and velocity fields for (dark) matter to those of galaxies. Here we assume no velocity bias, i.e., $\theta_{\rm g}=\theta$, but include every possible galaxy bias term at the next-to-leading order using symmetry arguments~\citep{McDonald:2009dh}: 
\begin{align}
\begin{split}
P_{{\rm g},\delta\delta}(k) &= b_1^2P_{\delta\delta}(k) + 2b_2b_1 P_{b2,\delta}(k) 
+ 2b_{s2}b_1P_{bs2,\delta}(k)\\
& + 2b_{\rm 3nl}b_1\sigma_3^2(k)P^{\rm lin}_{\rm m}(k) + b^2_2P_{b22}(k)\\
& + 2b_2b_{s2}P_{b2s2}(k) + b^2_{s2}P_{bs22}(k) + N,
\label{eq:40}
\end{split}\\
\begin{split}
P_{{\rm g},\delta\theta}(k) &= b_1P_{\delta\theta}(k) + b_2P_{b2,\theta}(k) 
+ b_{s2}P_{bs2,\theta}(k)\\
& + b_{\rm 3nl}\sigma_3^2(k)P^{\rm lin}_{\rm m}(k),
\label{eq:41}
\end{split}
\end{align}
where $P^{\rm lin}_{\rm m}$ is the linear matter power spectrum. Here we introduce five galaxy bias parameters: the re-normalised linear bias, $b_{1}$, $2$nd-order local bias, $b_2$, $2$nd-order non-local bias, $b_{s2}$, $3$rd-order non-local bias, $b_{\rm 3nl}$, and the constant stochasticity term, $N$. We evaluate the non-linear matter power spectra, $P_{\delta\delta}$, $P_{\delta\theta}$, $P_{\theta\theta}$, with the RegPT scheme at $2$-loop order~\citep{Taruya:2012ut}. The other bias terms are given by 
\begin{align}
\begin{split}
P_{b2,\delta}(k) &= \int \frac{d^3q}{(2\pi)^3}P^{\rm lin}_{\rm m}({q})P^{\rm lin}_{\rm m}(|\vc{k}-\vc{q}|)\\
&\;\;\;\;\;\times F^{(2)}_{\rm S}(\vc{q},\vc{k}-\vc{q}),
\end{split}\\
\begin{split}
P_{b2,\theta}(k) &= \int \frac{d^3q}{(2\pi)^3}P^{\rm lin}_{\rm m}({q)}P^{\rm lin}_{\rm m}(|\vc{k}-\vc{q}|)\\
&\;\;\;\;\;\times G^{(2)}_{\rm S}(\vc{q},\vc{k}-\vc{q}),
\end{split}\\
\begin{split}
P_{bs2,\delta}(k) &= \int \frac{d^3q}{(2\pi)^3}P^{\rm lin}_{\rm m}({q})P^{\rm lin}_{\rm m}(|\vc{k}-\vc{q}|)\\
&\;\;\;\;\;\times F^{(2)}_{\rm S}(\vc{q},\vc{k}-\vc{q})S^{(2)}(\vc{q},\vc{k}-\vc{q}),
\end{split}\\
\begin{split}
P_{bs2,\theta}(k) &= \int \frac{d^3q}{(2\pi)^3}P^{\rm lin}_{\rm m}({q})P^{\rm lin}_{\rm m}(|\vc{k}-\vc{q}|)\\
&\;\;\;\;\;\times G^{(2)}_{\rm S}(\vc{q},\vc{k}-\vc{q})S^{(2)}(\vc{q},\vc{k}-\vc{q}),
\end{split}\\
\begin{split}
P_{b22}(k) &= \frac{1}{2}\int \frac{d^3q}{(2\pi)^3}P^{\rm lin}_{\rm m}({q})\Big[P^{\rm lin}_{\rm m}(|\vc{k}-\vc{q}|)\\
&\;\;\;\;\; - P^{\rm lin}_{\rm m}({q})\Big],
\end{split}\\
\begin{split}
P_{b2s2}(k) &= -\frac{1}{2}\int \frac{d^3q}{(2\pi)^3}P^{\rm lin}_{\rm m}({q})\Big[\frac{2}{3}P^{\rm lin}_{\rm m}(\vc{q})\\
&\;\;\;\;\; - P^{\rm lin}_{\rm m}(|\vc{k}-\vc{q}|)S^{(2)}(\vc{q},\vc{k}-\vc{q})\Big],
\end{split}\\
\begin{split}
P_{bs22}(k) &= -\frac{1}{2}\int \frac{d^3q}{(2\pi)^3}P^{\rm lin}_{\rm m}({q})\Big[\frac{4}{9}P^{\rm lin}_{\rm m}(\vc{q})\\
&\;\;\;\;\; - P^{\rm lin}_{\rm m}(|\vc{k}-\vc{q}|)S^{(2)}(\vc{q},\vc{k}-\vc{q})^2\Big],
\end{split}
\label{eq:nlterms}
\end{align}
where the symmetrised $2$nd-order PT kernels, $F^{(2)}_{S}$, $G^{(2)}_{S}$, 
and $S^{(2)}$ are given by
\begin{align}
F^{(2)}_{\rm S}(\vc{q}_1,\vc{q}_2) &= \frac{5}{7} + \frac{\vc{q}_1\cdot \vc{q}_2}{2q_1q_2}
\left(\frac{q_1}{q_2} + \frac{q_2}{q_1}\right) + \frac{2}{7}\left(\frac{\vc{q}_1\cdot \vc{q}_2}{q_1q_2}\right)^2,\\
G^{(2)}_{\rm S}(\vc{q}_1,\vc{q}_2) &= \frac{3}{7} + \frac{\vc{q}_1\cdot \vc{q}_2}{2q_1q_2}
\left(\frac{q_1}{q_2} + \frac{q_2}{q_1}\right) + \frac{4}{7}\left(\frac{\vc{q}_1\cdot \vc{q}_2}{q_1q_2}\right)^2,\\
S^{(2)}(\vc{q}_1,\vc{q}_2) & = \left(\frac{\vc{q}_1\cdot \vc{q}_2}{q_1q_2}\right)^2 - \frac{1}{3}.
\end{align}
If we additionally define 
\begin{equation}
D^{(2)}(\vc{q}_1,\vc{q}_2) = \frac{2}{7}\left[S^{(2)}(\vc{q}_1,\vc{q}_2) - \frac{2}{3}\right],
\end{equation}
we can express $\sigma^2_3(k)$ of eq.~\ref{eq:41} as
\begin{equation}
                \sigma_{3}^{2}(k) = \frac{105}{16}\int \frac{d^3 q}{(2\pi)^{3}}P^{\rm lin}_{\rm m}({q})
                \left[  D^{(2)}(-\vc{q},\vc{k})S^{(2)}(\vc{q}, \vc{k}-\vc{q}) + \frac{8}{63}  \right].
\end{equation}
In the case of the local Lagrangian bias picture, we can predict the amplitude of the non-local bias as~\citep{Chan:2012jj, Baldauf:2012hs, Saito2013}
\begin{align}
b_{s2} &= -\frac{4}{7}(b_1 - 1),\\
b_{\rm 3nl} &= \frac{32}{315}(b_1 - 1), 
\end{align}
which are in good agreement with the values measured in simulations. In this work, we adopt these relations for simplicity, while we take $b_{1}$, $b_{2}$ and $N$ as independent parameters to vary. Since we measure the amplitude of the biased clustering, the actual free parameters used are $b_{1}\sigma_8(z)$, $b_{2}\sigma_8(z)$ and $N$ at each redshift bin, as discussed in \S~\ref{sec:parametrization}.

Our RSD model is based on the local distant observer approximation, i.e., without accounting for the wide angle effect. The wide angle effect has been shown to be negligible compared to the sample variance for surveys such as BOSS~\citep{Samushia:2011cs, Beutler:2011hx, Beutler:2012px, Yoo:2013zga}.

Recently, potential improvements for the model discussed above have been proposed. For the nonlinear RSD model,~\citet{Zheng:2016zxc} try to improve the TNS model by further examining our FoG suppression term and directly comparing the correction terms between perturbation theory and simulations. For the nonlinear galaxy bias,~\citet{Lazeyras:2015lgp} study the separate universe simulations which enable to directly measure and assess the nonlinear local bias of dark matter halos (see also~\citealt{Li:2015jsz}). They also discuss the importance of the $k^{2}$ bias term which we ignore just for simplicity (see also~\citealt{McDonald:2009dh,Biagetti:2013hfa,Schmidt:2015gwz} etc.). Also, the developments in terms of the distribution function approach (e.g.,~\citealt{Okumura:2015fga}) and the effective field theory approach (e.g,~\citealt{Lewandowski:2015ziq}) are ongoing and can be complementary to our model. 

\subsection{Correction for the irregular $\mu$ distribution}
\label{sec:binning}

Because the survey volume is not infinite, the measured power spectra are estimated on a finite and discrete \textit{k}-space grid. Performing FFTs in a Cartesian lattice makes the angular distribution of the Fourier modes irregular and causes increasing deviation from the isotropic distribution at smaller $k$. As a result, fluctuation-like deviations appear in the measured power spectrum multipoles that are not caught by the window function, as shown in the bottom panel (SGC) of Figure \ref{fig:ps_NGCSGC}. The effect is larger for the quadrupole than the monopole since the quadrupole is more sensitive to an anisotropy. Our DR11 analysis corrected the measured data for this effect, while here we include this effect in our power spectrum model.  When integrating the model power spectrum $P(k,\mu)\mathcal{L}(\mu)$ in eq.~\ref{eq:binning}  over $\mu $, we weight each $\mu$ bin by the normalised number of modes $N(k,\mu)$ counted on the \textit{k}-space grid used to estimate the power spectrum.
\begin{equation}
P_{\ell}(k) = \int_{-1}^{1} d\mu\;P(k,\mu)\frac{N_{\rm modes}(k,\mu)}{N_{\rm bin}(k)}\mathcal{L}_{\ell}(\mu)
\label{eq:binning}
\end{equation}
with the normalisation for each $k$ given by
\begin{equation}
N_{\rm bin}(k) = \int^{1}_{-1} d\mu\; N_{\rm modes}(k,\mu).
\end{equation}
This $P_{\ell}(k)$ is used to calculate $\xi_{\ell}$ in eq.~\ref{eq:conv1} - \ref{eq:conv3}. Figure~\ref{fig:binning} shows the effect of irregular $\mu$ distribution in the three power spectrum multipoles. While the effect is most pronounced in the higher order multipoles, it never exceeds the measurement uncertainties and hence is not a dominant effect.

The inclusion of a $\mu$-dependent function in eq.~\ref{eq:binning} is inconsistent with our derivation of the window function convolution in eq.~\ref{eq:win1}. A completely consistent approach would include the effect of irregular $\mu$ distribution after the window function convolution, or would properly include this function in eq~\ref{eq:win1}. We tested the impact of this assumption by including the discreteness effect after the convolution (using multipole expansion) and found that this does not change our results. 

\subsection{The Alcock-Paczynski effect}
\label{sec:alcock}

When transforming our observables, such as celestial position and redshift, into physical coordinates, we assume specific relations between the redshift and the line-of-sight distance (i.e., the Hubble parameter) and between the angular separation and the distance perpendicular to the line-of-sight (i.e., the angular diameter distance) given by the fiducial cosmological model. Therefore, if we assume a fiducial cosmology that is different from the true cosmology, it will produce geometric warping and artificially introduce an anisotropy in an otherwise isotropic feature in the galaxy clustering, independently from the effect of redshift space distortions. This behaviour is known as the Alcock-Paczynski (AP) effect~\citep{Alcock:1979mp} and can be used to measure cosmological parameters~\citep{Matsubara:1996nf,Ballinger:1996cd}. The anisotropy due to the AP effect is often difficult to separate from the RSD effect for a featureless power spectrum given the uncertainties in the models for redshift-space distortions~\citep{Seo:2003,Shoji:2009}. The presence of the BAO feature in the power spectrum, however, helps to break this degeneracy.

To account for the Alcock-Paczynski effect due to the different geometric scaling along and perpendicular to the line-of-sight directions between the true and fiducial cosmology, we introduce the scaling factors 
\begin{align}
\alpha_{\parallel} &= \frac{H^{\rm fid}(z)r^{\rm fid}_s(z_d)}{H(z)r_s(z_d)},\\
\alpha_{\perp} &= \frac{D_A(z)r^{\rm fid}_s(z_d)}{D^{\rm fid}_A(z)r_s(z_d)},
\end{align}
where $H^{\rm fid}(z)$ and $D^{\rm fid}_A(z)$ are the fiducial values for the Hubble parameter and angular diameter distance at the effective redshifts of the dataset, and $r^{\rm fid}_s(z_d)$ is the fiducial value of the sound horizon scale at the drag epoch assumed in the power spectrum template. By using the sound horizon scale as the reference scale for the AP test, we are assuming that the main feature that contributes to the AP test is the BAO. The true wave-numbers $k_{\parallel}'$ and $k_{\perp}'$ are then related to the observed wave-numbers by $k_{\parallel}' = k_{\parallel}/\alpha_{\parallel}$ and $k_{\perp}' = k_{\perp}/\alpha_{\perp}$. Transferring this information into scalings for the absolute wavenumber $k = \sqrt{k^2_{\parallel} + k^2_{\perp}}$ and the cosine of the angle to the line-of-sight $\mu$, we can relate the true ($k'$, $\mu'$) and observed values ($k$, $\mu$) by~\citep{Ballinger:1996cd}
\begin{align}
k' &= \frac{k}{\alpha_{\perp}}\left[1 + \mu^2\left(\frac{1}{F^2} - 1\right)\right]^{1/2},
\label{eq:scaling1}\\
\mu' &= \frac{\mu}{F_{}}\left[1 + \mu^2\left(\frac{1}{F^2} - 1\right)\right]^{-1/2}
\label{eq:scaling2}
\end{align}
with $F = \alpha_{\parallel}/\alpha_{\perp}$. The multipole power spectrum including the Alcock-Paczynski effect can then be written as
\begin{align}
P_{\rm \ell}(k) &= \left(\frac{r_s^{\rm fid}}{r_s}\right)^3 \frac{(2\ell + 1)}{2\alpha^2_{\perp}\alpha_{\parallel}}\int^1_{-1}d\mu\; P_{\rm g}\left[k'(k, \mu), \mu'(\mu)\right]\mathcal{L}_{\ell}(\mu),
\label{eq:multi}
\end{align}
where we use the model of section~\ref{sec:model} for $P_{\rm g}\left[k'(k, \mu),\mu'(\mu)\right]$. The factor $\left(\frac{r_s^{\rm fid}}{r_s}\right)^3\frac{1}{2\alpha^2_{\perp}\alpha_{\parallel}}$ accounts for the difference in the cosmic volume in different cosmologies. The ratio of sound horizon scales is needed to compensate for the sound horizon scale included in the definitions of the $\alpha$ values. To treat this $r_s$ properly, we could apply the Planck measurement~\citep{Ade:2015xua} on $r_s$ as a prior during the parameter fitting. Since the Planck uncertainty on $r_s$ is only at the level of $\sim 0.2\%$, fixing $r_s = 147.41\hMpc$ has a negligible effect on our measurements of $\alpha_{\parallel}$ and $ \alpha_{\perp} $.

The AP effect (from the anisotropic warping of the BAO) constrains the parameter combination $F_{\rm AP}(z) = (1+z)D_A(z)H(z)/c$, while the radial dilation of the BAO feature constrains the combination $D_V(z)/r_s(z_d) \propto \left[D^2_A(z)/H(z)\right]^{1/3}$. Together these two signals allow one to break the degeneracy between $D_A(z)$ and $H(z)$. 

\subsection{Model parameterization}
\label{sec:parametrization}

Based on the discussion of our model in section~\ref{sec:model} we have four nuisance parameters, $b_1\sigma_8$, $b_2\sigma_8$, $\sigma_v$ and $N$, which we fit to our measurements together with the three cosmological parameters $f\sigma_8$, $\alpha_\parallel$ and $\alpha_{\perp}$. The two $\alpha$ parameters carry the BAO and AP information; we can re-phrase these parameters to 
\begin{align}
F_{\rm AP} &= (1+z_{\rm eff}) D_A(z_{\rm eff})H(z_{\rm eff})/c \\
&= \frac{\alpha_{\perp}}{\alpha_{\parallel}}(1+z_{\rm eff})D_A^{\rm fid}(z_{\rm eff})H^{\rm fid}(z_{\rm eff})/c
\label{eq:FAP}
\end{align}
and
\begin{equation}
D_V(z_{\rm eff})\frac{r_s^{\rm fid}(z_d)}{r_s(z_d)} = \left(\alpha_{\perp}^2\alpha_{\parallel}[(1 + z_{\rm eff})D_A^{\rm fid}(z_{\rm eff})]^2\frac{cz_{\rm eff}}{H^{\rm fid}(z_{\rm eff})}\right)^{\frac{1}{3}}.
\label{eq:DV}
\end{equation}
At low redshift the BOSS galaxies follow a slightly different selection in the SGC and NGC (see section~\ref{sec:data},~\citealt{Reid:2015gra} and~\citealt{Ross:2016}). These differences lead to different power spectrum amplitudes in the SGC and NGC. To account for this issue we marginalise over the four nuisance parameters independently for NGC and SGC, while we use the same cosmological parameters. Tests on mock catalogues demonstrated that using separate nuisance parameters for the NGC and SGC does not degrade our cosmological constraints. We therefore have a total of $11$ parameters for each redshift bin in our analysis: $b_1^{\rm NGC}\sigma_8(z)$, $b_1^{\rm SGC}\sigma_8(z)$, $b_2^{\rm NGC}\sigma_8(z)$, $b_2^{\rm SGC}\sigma_8(z)$, $\sigma_v^{\rm NGC}$, $\sigma_v^{\rm SGC}$, $N^{\rm NGC}$, $N^{\rm SGC}$, $f(z)\sigma_8(z)$, $\alpha_\parallel$ and $\alpha_{\perp}$.

\section{Mock catalogues}
\label{sec:mocks}

To derive a covariance matrix for the power spectrum multipoles we use the MultiDark-Patchy mock catalogues~\citep{Kitaura:2015uqa}. These mock catalogues have been calibrated to a $N$-body based reference sample using approximate gravity solvers and analytical-statistical biasing models. The reference catalogue is extracted from one of the BigMultiDark simulations~\citep{Klypin:2014kpa}, which used $3\,840^3$ particles on a volume of ($2.5\hMpc$)$^3$ assuming a $\Lambda$CDM cosmology with $\Omega_M = 0.307115$, $\Omega_b = 0.048206$, $\sigma_8 = 0.8288$, $n_s = 0.9611$, and a Hubble constant of $H_0 = 67.77\,\text{km}s^{-1}\text{Mpc}^{-1}$. 

Halo abundance matching is used to reproduce the observed BOSS two and three point clustering measurements~\citep{Rodriguez-Torres:2015vqa}. This technique is applied at different redshift bins to reproduce the BOSS DR12 redshift range. These mock catalogues are combined into light cones, also accounting for selection effects and masking. In total we have $2045$ mock catalogues available for the NGC and $2048$ mock catalogues for the SGC.

The mean power spectrum multipoles for MultiDark-Patchy mock catalogues are shown in Figure~\ref{fig:ps_NGCSGC} (lines with shaded area) for the NGC (top panels) and SGC (bottom panels) together with the BOSS measurements (coloured points with error bars). The mock catalogues closely reproduce the data power spectrum multipoles for the entire range of wave-numbers relevant for this analysis. 

The SGC mock catalogues show some correlated fluctuations in the power spectra at small $k$, which are more prominent in the quadrupole and hexadecapole. This behaviour is a discreteness effect due to the finite number of Fourier modes, which is more severe at small $k$. This effect is present in the data as well; we discuss how to account for this effect in our power spectrum model in section~\ref{sec:binning}.

\subsection{The covariance matrix}
\label{sec:cov}

\begin{figure*}
\begin{center}
\epsfig{file=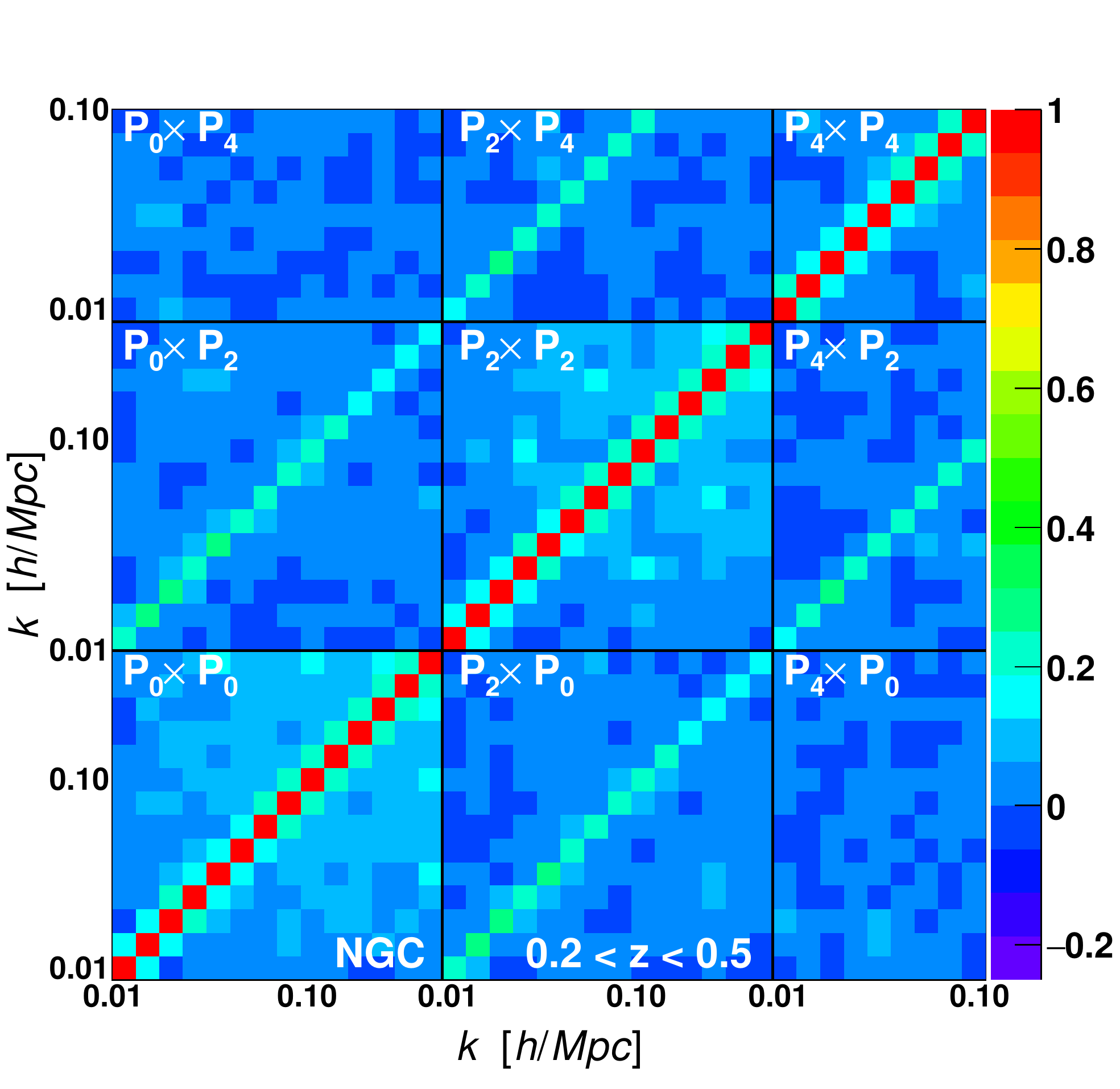,width=5.8cm}
\epsfig{file=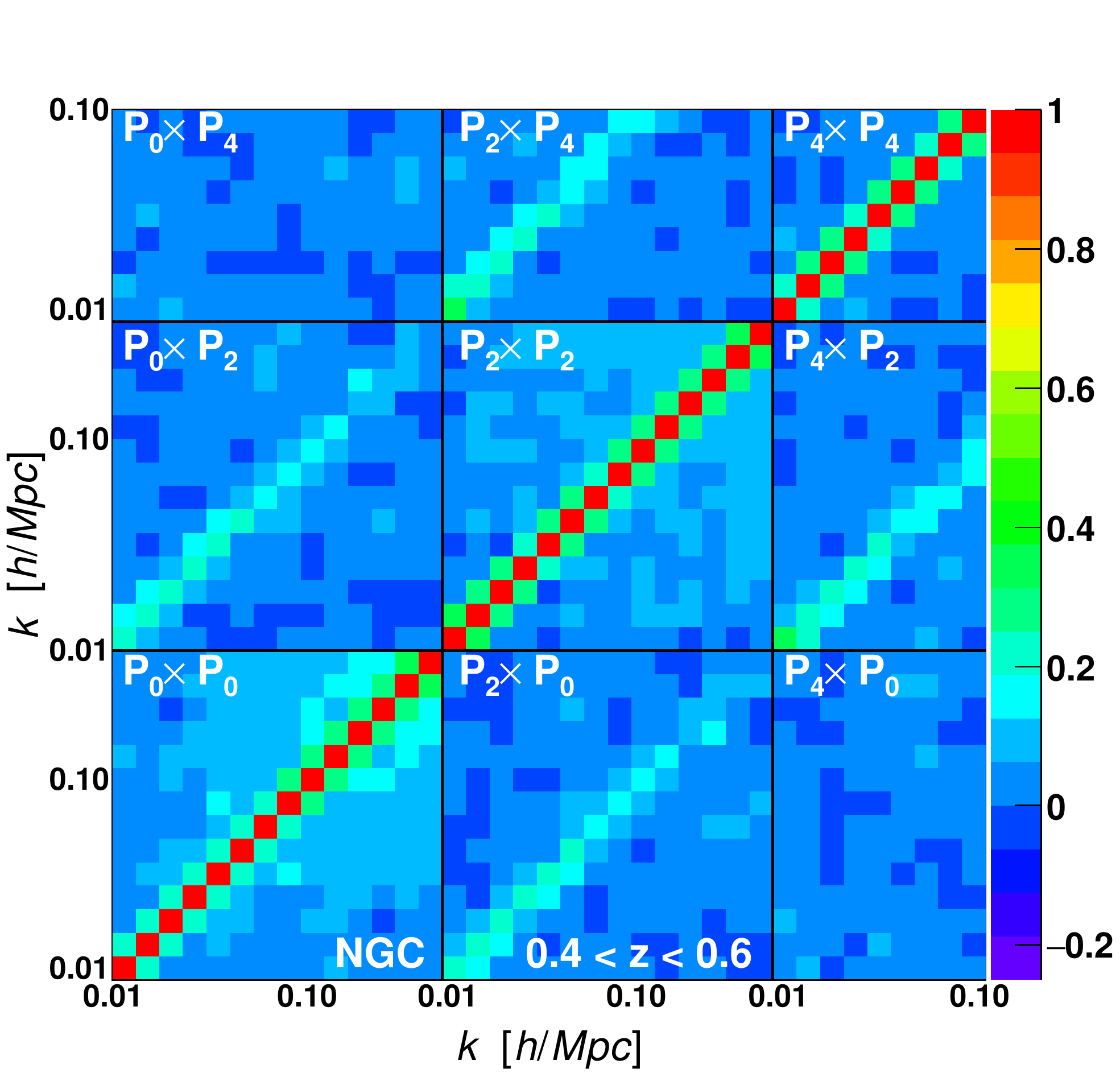,width=5.8cm}
\epsfig{file=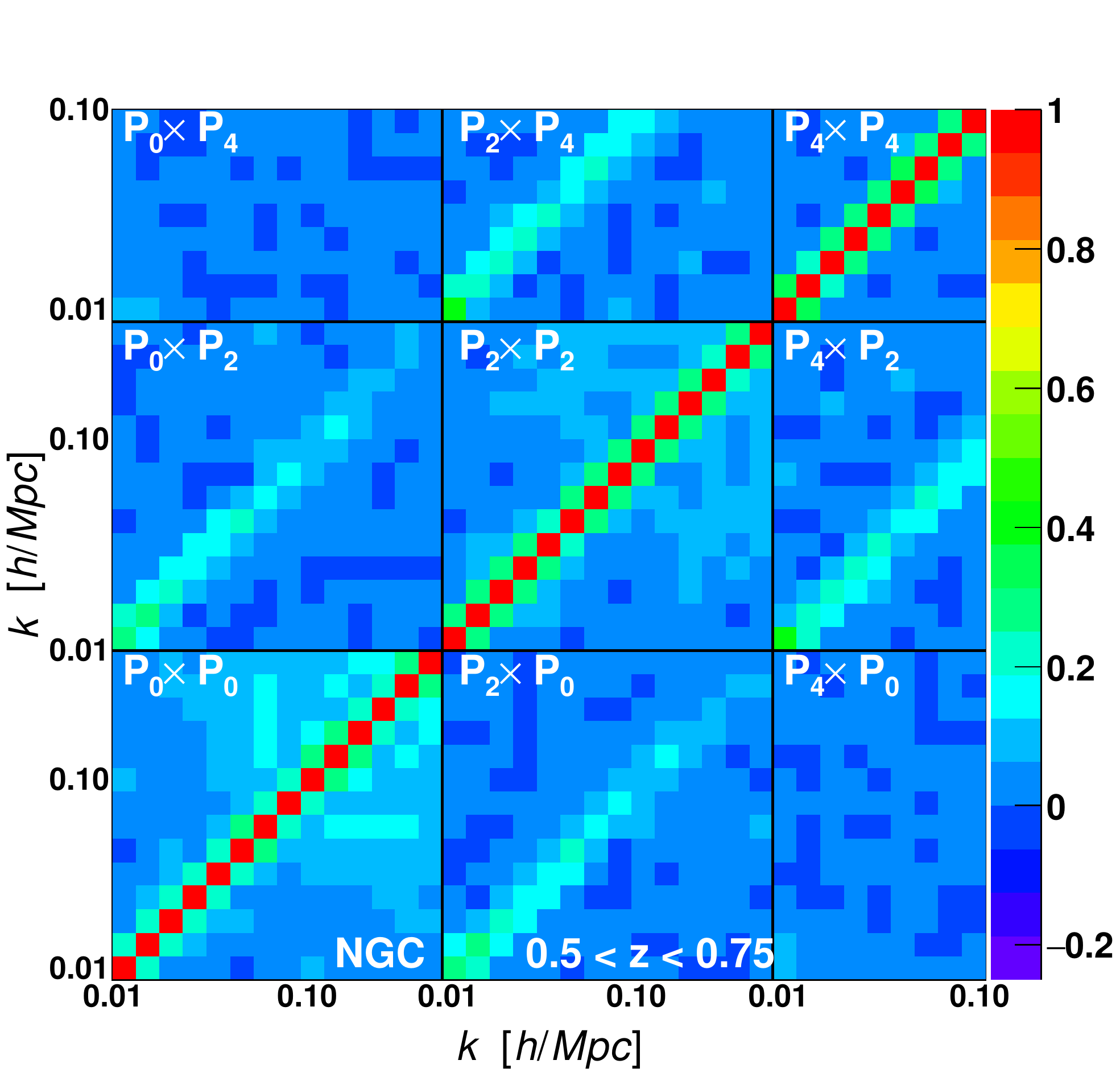,width=5.8cm}\\
\epsfig{file=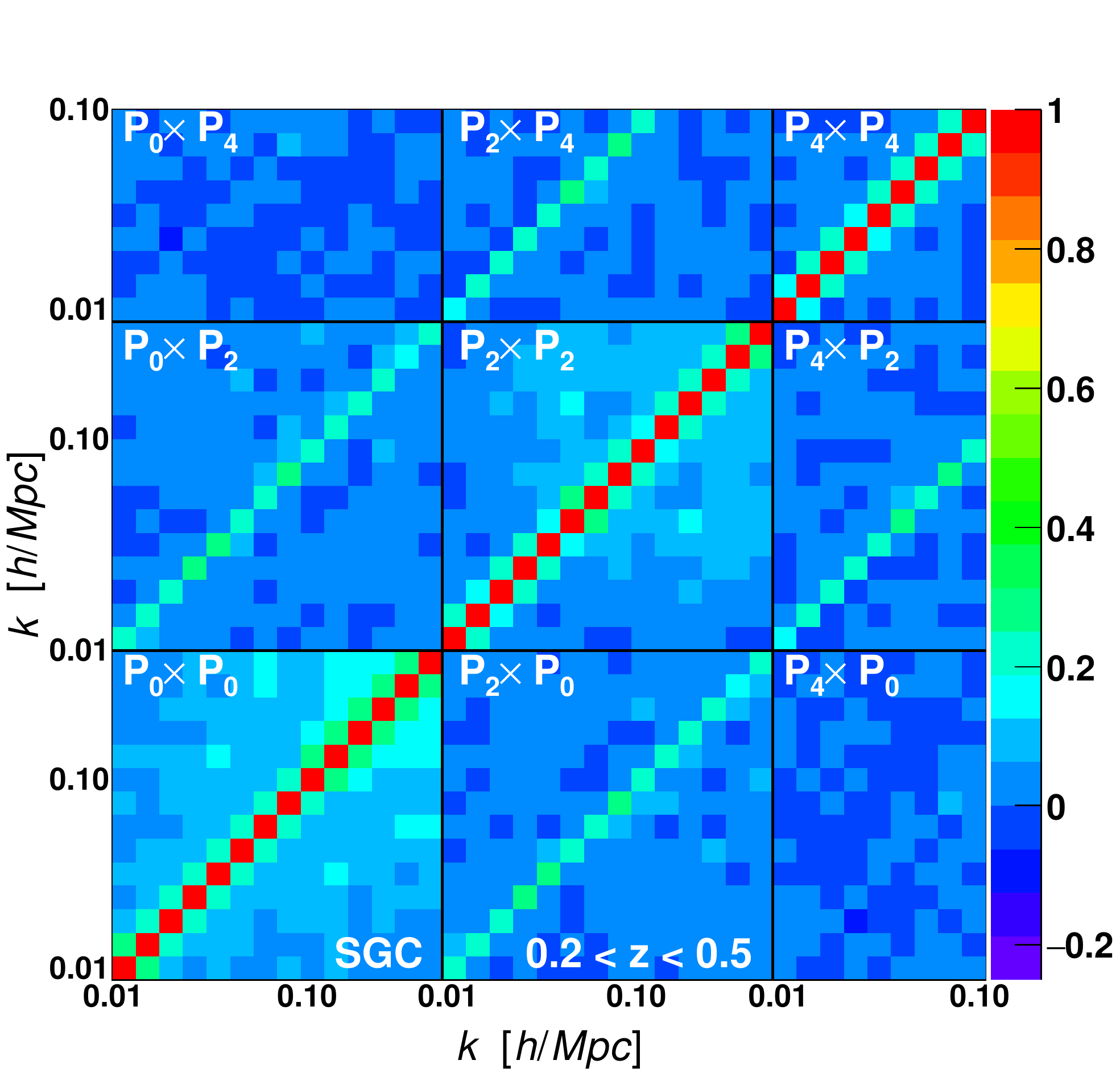,width=5.8cm}
\epsfig{file=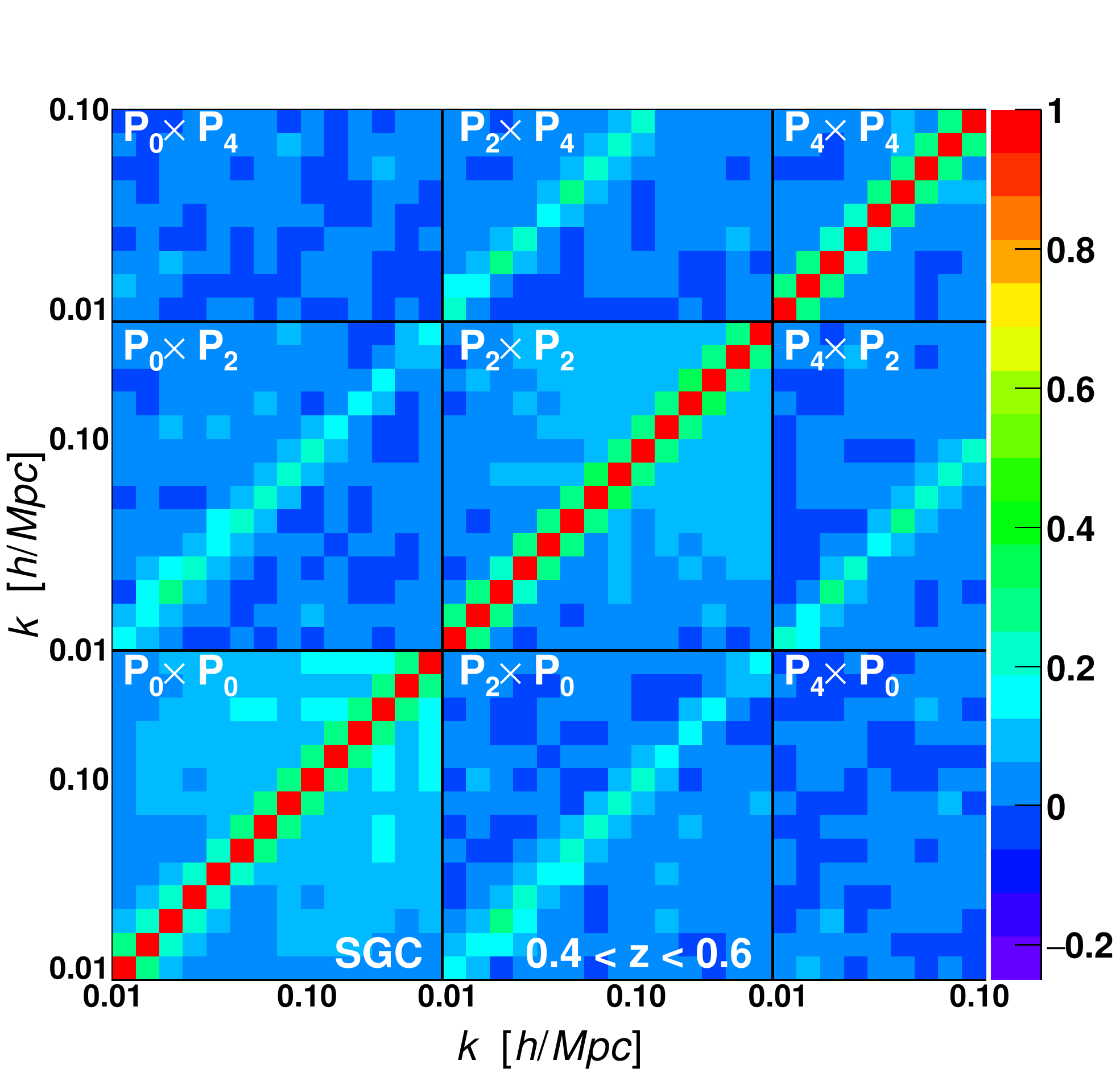,width=5.8cm}
\epsfig{file=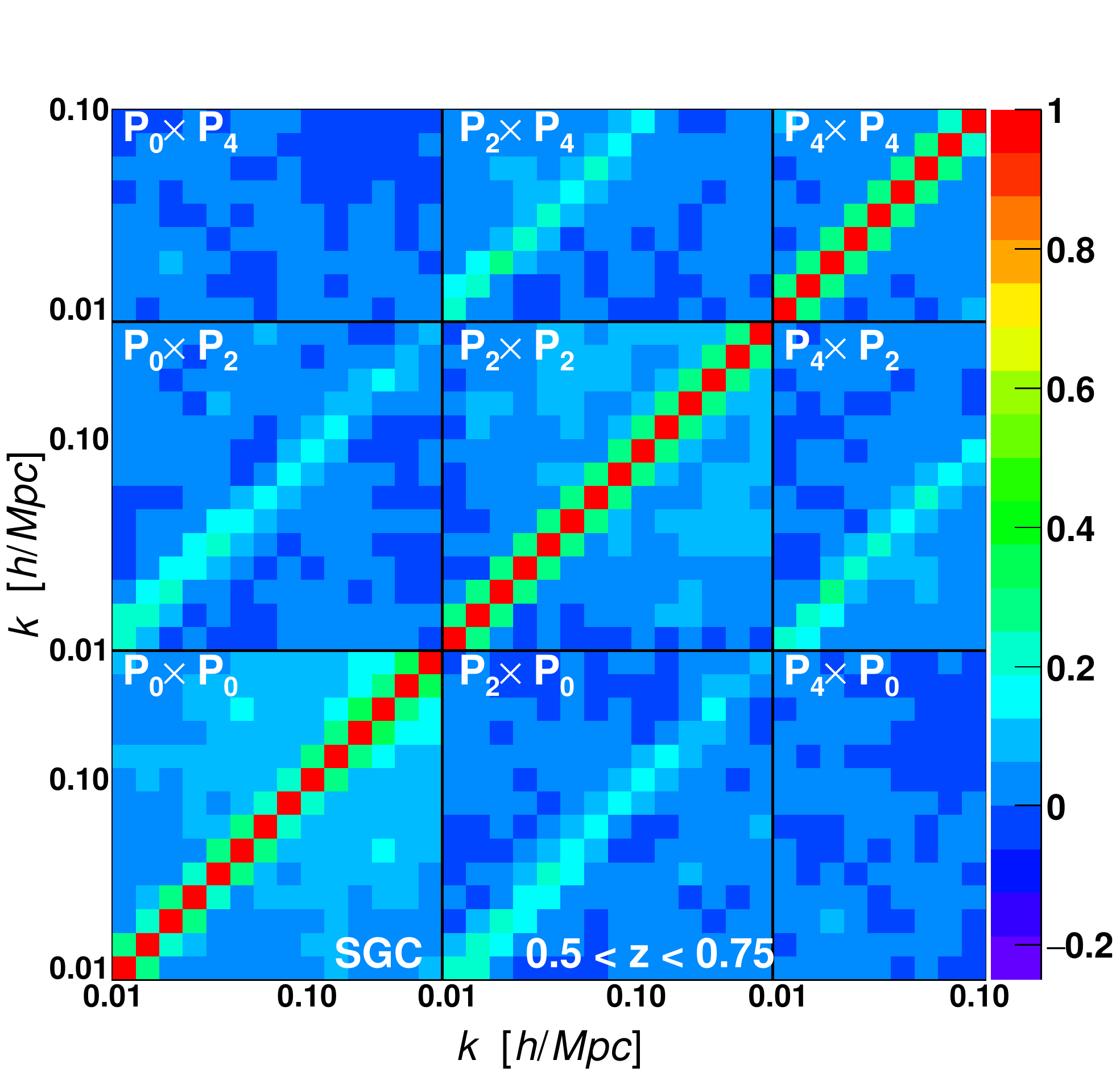,width=5.8cm}
\caption{Covariance matrices including the monopole, quadrupole and hexadecapole of the North Galactic Cap (NGC, top) and South Galactic Cap (SGC, bottom) for the three redshift bins used in this analysis. We include all bins between $k = 0.01$ - $0.15\hMpc$ for the monopole and quadrupole and all bins between $k = 0.01$ - $0.10\hMpc$ for the hexadecapole. The colour indicates the level of correlation, where red represents $100\%$ correlation and blue-magenta means low level of anti-correlation.}
\label{fig:cov}
\end{center}
\end{figure*}

We can derive a covariance matrix from the set of mock catalogues described in the last section as
\begin{equation}
\begin{split}
C_{xy} = \frac{1}{N_s - 1} \sum^{N_s}_{n=1}&\left[P_{\ell,n}(k_i) - \overline{P}_{\ell}(k_i)\right]\times\\
&\left[P_{\ell',n}(k_j) - \overline{P}_{\ell'}(k_j)\right]
\end{split}
\label{eq:cov}
\end{equation}
with $N_s$ being the number of mock catalogues. Our covariance matrix contains the monopole, quadrupole and hexadecapole uncertainties as well as their covariances. The elements of the matrices are given by $(x, y) = (\frac{n_b\ell}{2} + i, \frac{n_b\ell'}{2} + j)$, where $n_b$ is the number of bins in each multipole power spectrum. Our fitting range is $k = 0.01$ - $0.15h\,$Mpc$^{-1}$ for the monopole and quadrupole ($n_b =14$), and $k = 0.01$ - $0.10h\,$Mpc$^{-1}$ for the hexadecapole ($n_b = 9$), hence the dimensions of the covariance matrices are $37\times 37$. The mean of the power spectra is defined as
\begin{equation}
\overline{P}_{\ell}(k_i) = \frac{1}{N_s}\sum^{N_s}_{n=1}P_{\ell, n}(k_i).
\end{equation}
Since the mock catalogues have the same selection function as the data, they automatically incorporate the window function and integral constraint effects present in the data.

Figure~\ref{fig:cov} presents the correlation matrices for  the three redshift bins of BOSS NGC (top panels) and SGC (bottom panels), where the correlation coefficient is defined as
\begin{equation}
r_{xy} = \frac{C_{xy}}{\sqrt{C_{xx}C_{yy}}}.
\end{equation}
Each panel shows a matrix with three horizontal and vertical division lines. The first column displays the correlation between $k$ bins in the monopole with itself (bottom), with the quadruple (middle) and with the hexadecapole (top). The second column is the correlations for the quadrupole and the third column presents the correlations for the hexadecapole. There is a correlation between the monopole and quadrupole, as well as a correlation between the quadrupole and hexadecapole, while the correlation between the monopole and hexadecapole is quite weak.

Since the estimated covariance matrix $C$ is inferred from mock catalogues, its inverse, $C^{-1}$, provides a biased estimate of the true inverse covariance matrix, due to the skewed nature of the inverse Wishart distribution~\citep{Hartlap:2006kj}. To correct for this bias we rescale the inverse covariance matrix as
\begin{equation}
C^{-1}_{ij,\rm Hartlap} = \frac{N_s - n_b - 2}{N_s - 1}C^{-1}_{ij},
\label{eq:hartlap}
\end{equation}
where $n_b$ is the number of power spectrum bins. With these covariance matrices we can perform a standard $\chi^2$ minimisation to find the best fitting parameters. In our analysis we have $N_s = 2048\;(2045)$ and $n_b = 37$, which yield a Hartlap factor of $\sim 0.98$, representing an increase in the variance of about $1\%$.

\subsection{Fitting preparation}

Using the covariance matrix derived in section~\ref{sec:cov} we perform a $\chi^2$ minimisation to find the best fitting parameters. In addition to the scaling of the covariance matrix of eq.~\ref{eq:hartlap}, we have to propagate the error in the covariance matrix to the error on the estimated parameters. This test is accomplished by scaling the variance for each parameter by~\citep{Percival:2013}
\begin{equation}
M_1 = \sqrt{\frac{1 + B(n_b - n_p)}{1 + A + B(n_p + 1)}},
\end{equation}
where $n_p$ is the number of parameters and
\begin{align}
A &= \frac{2}{(N_s - n_b - 1)(N_s - n_b - 4)},\\
B &= \frac{N_s - n_b - 2}{(N_s - n_b - 1)(N_s - n_b - 4)}.
\end{align}
Taking the quantities which apply in our case ($N_s = 2048\;(2045)$, $n_b=76$, $n_p =11$) results in a modest correction of $M_1\approx 1.01$.

When dealing with the variance or standard deviation of a distribution of finite mock results which also has been fitted with a covariance matrix derived from the same mock results, the standard deviation from these mocks must be corrected as
\begin{equation}
M_2 = M_1\sqrt{\frac{N_s-1}{N_s-n_b-2}}.
\end{equation}  
When the error is estimated from the likelihood distribution, the resulting standard deviation is multiplied by $M_1$ alone since the second factor, i.e., Hartlap factor~\citep{Hartlap:2006kj}, is already included in eq~\ref{eq:hartlap}.

We calculate the power spectrum in bins of $\Delta k = 0.01\ihMpc$. Tests on mock datasets have shown that such a binning choice is small enough that it does not dilute any of the cosmologically relevant information, while sufficiently large that it keeps the~\citet{Hartlap:2006kj} correction factor small.

To derive the likelihood distribution for the different parameters given the measurements we use a Monte Carlo Markov Chain approach based on a modified version of the python \textit{emcee} package~\citep{ForemanMackey:2012ig}. We test the convergence of four chains run in parallel using the~\citet{Gelman:1992zz} convergence criterion.

\section{Testing the model with mock catalogues}
\label{sec:sys}

To confirm that our model is accurate enough to extract the true cosmological parameters within the measurement precision, we refer the reader to our DR11 analysis~\citep{Beutler:2013yhm}, where we already performed many tests of our analysis pipeline. Here we discuss two further investigations: tests on the MD-PATCHY mock catalogues, and the participation on a challenge exercise, conducted on a set of high fidelity mocks. For the purposes of the results presented in this paper, the latter can be considered a blind test. 

\subsection{Test using the Blind Mock Challenge results}

\begin{figure}
\begin{center}
\epsfig{file=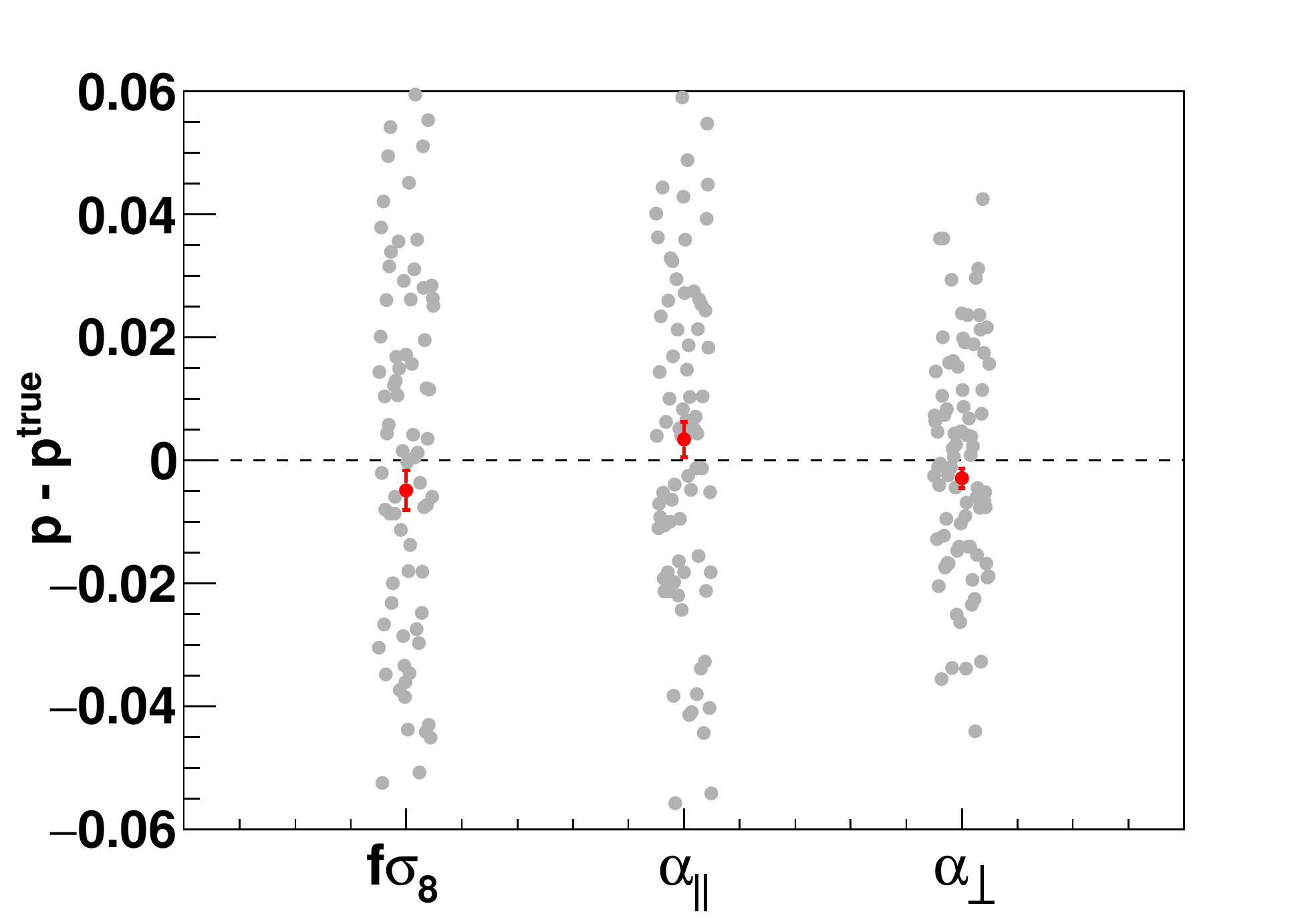,width=8.8cm}
\caption{Results of the the blind mock challenge for the power spectrum model used in this analysis when applied to $84$ CMASS like mock catalogues derived from N-body simulations. The grey data points are the results for all $84$ mock catalogues for the three parameters of interest. The red data points indicate the mean and error on the mean. The y-axis shows the (absolute) deviation from the true underlying cosmology.}
\label{fig:challenge4}
\end{center}
\end{figure}

We participated in a mock challenge within the BOSS galaxy clustering working group~\citep{Tinker:2015}. This activity was divided into two parts, where 1) we had to reproduce the correct cosmological parameters for several simulation boxes with different cosmologies and halo occupation distributions setup, and 2) we were required to reproduce the correct cosmological parameters for a set of $84$ N-body based mock catalogues with the same selection function as the CMASS dataset. The results for the second part of the mock challenge are displayed in Figure~\ref{fig:challenge4}. Assuming that the $84$ mock catalogues are uncorrelated\footnote{In reality, there is a small level of correlation between the different mocks, thus our systematic bias is slightly overestimated.}, we can use them to test for potential biases of the model up to the level of $\sqrt{84} = 9.16$ times smaller than the measurement uncertainties. Based on the offsets of our measurements from the true cosmology shown in Figure~\ref{fig:challenge4}, we can reproduce the correct cosmological parameters with a bias of  $\Delta f\sigma_8 = 0.00485$, $\Delta\alpha_{\perp} = 0.002924$ and $\Delta\alpha_{\parallel} = 0.0034$. These potential biases are $\lesssim 10\%$ of our measurement uncertainties. 

For comparisons of our model to other RSD studies and a more detailed discussion of the blind mock challenge, we refer the reader to~\citet{Tinker:2015}.

\subsection{Tests on the MultiDark-Patchy mock catalogues}

\begin{figure*}
\begin{center}
\epsfig{file=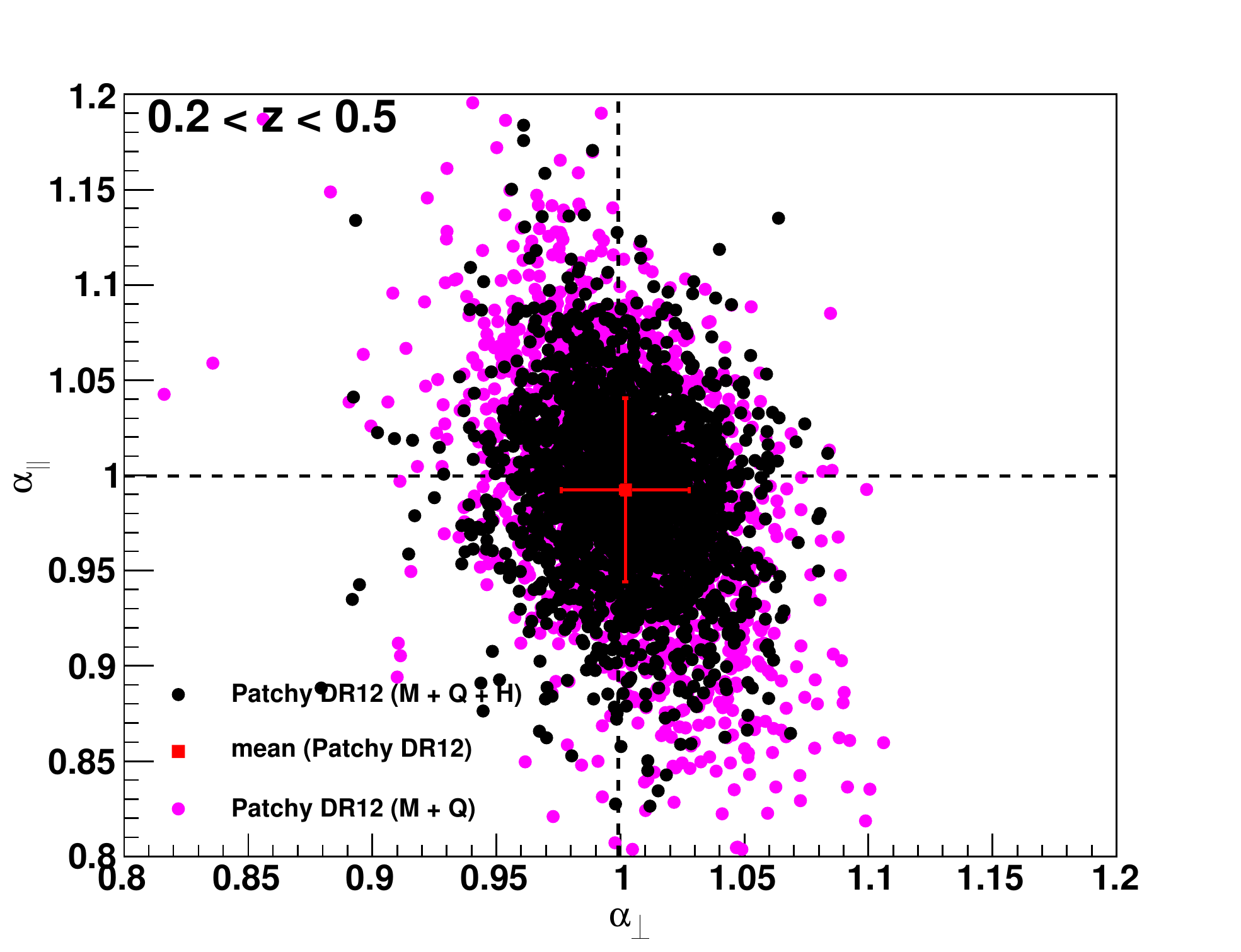,width=5.8cm}
\epsfig{file=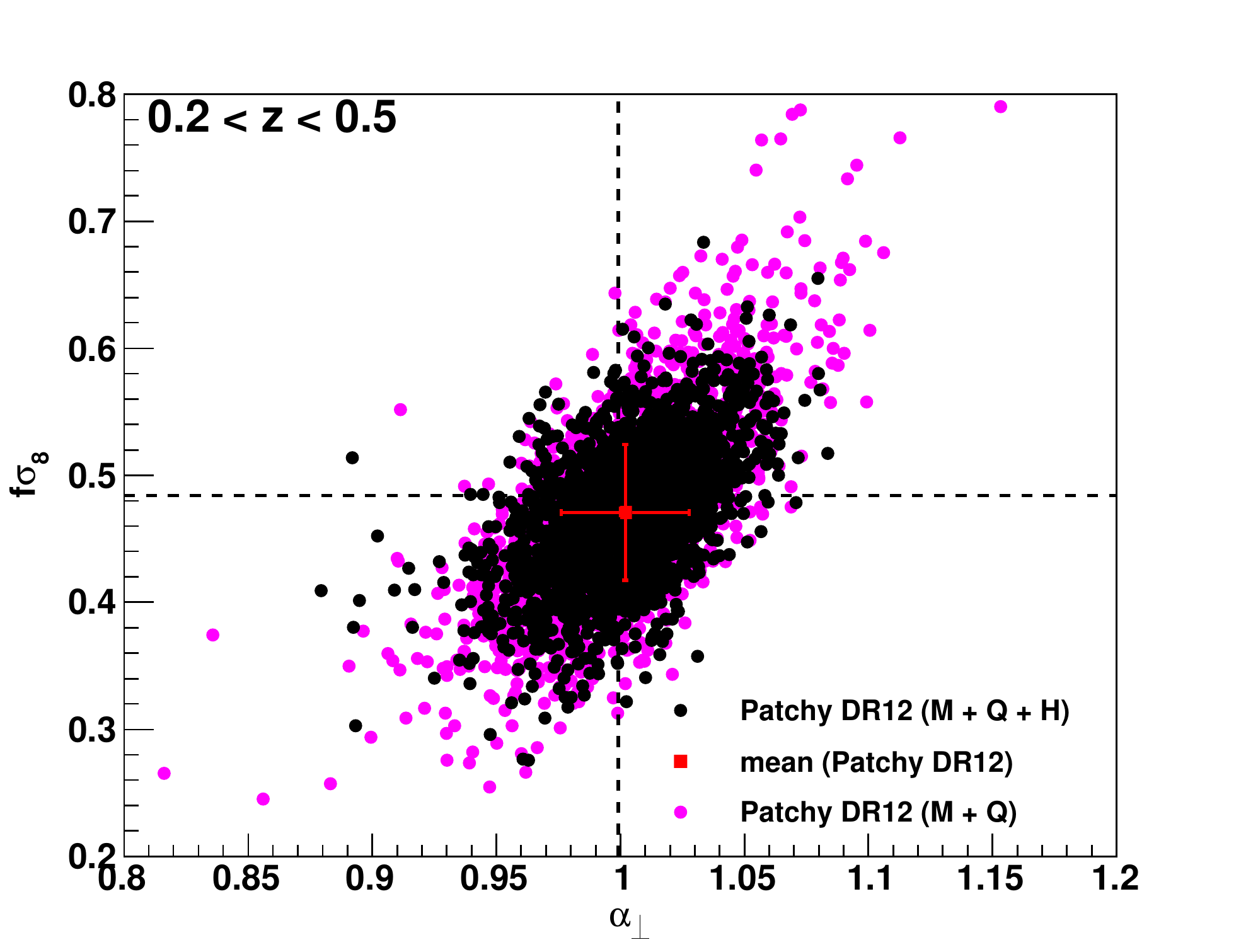,width=5.8cm}
\epsfig{file=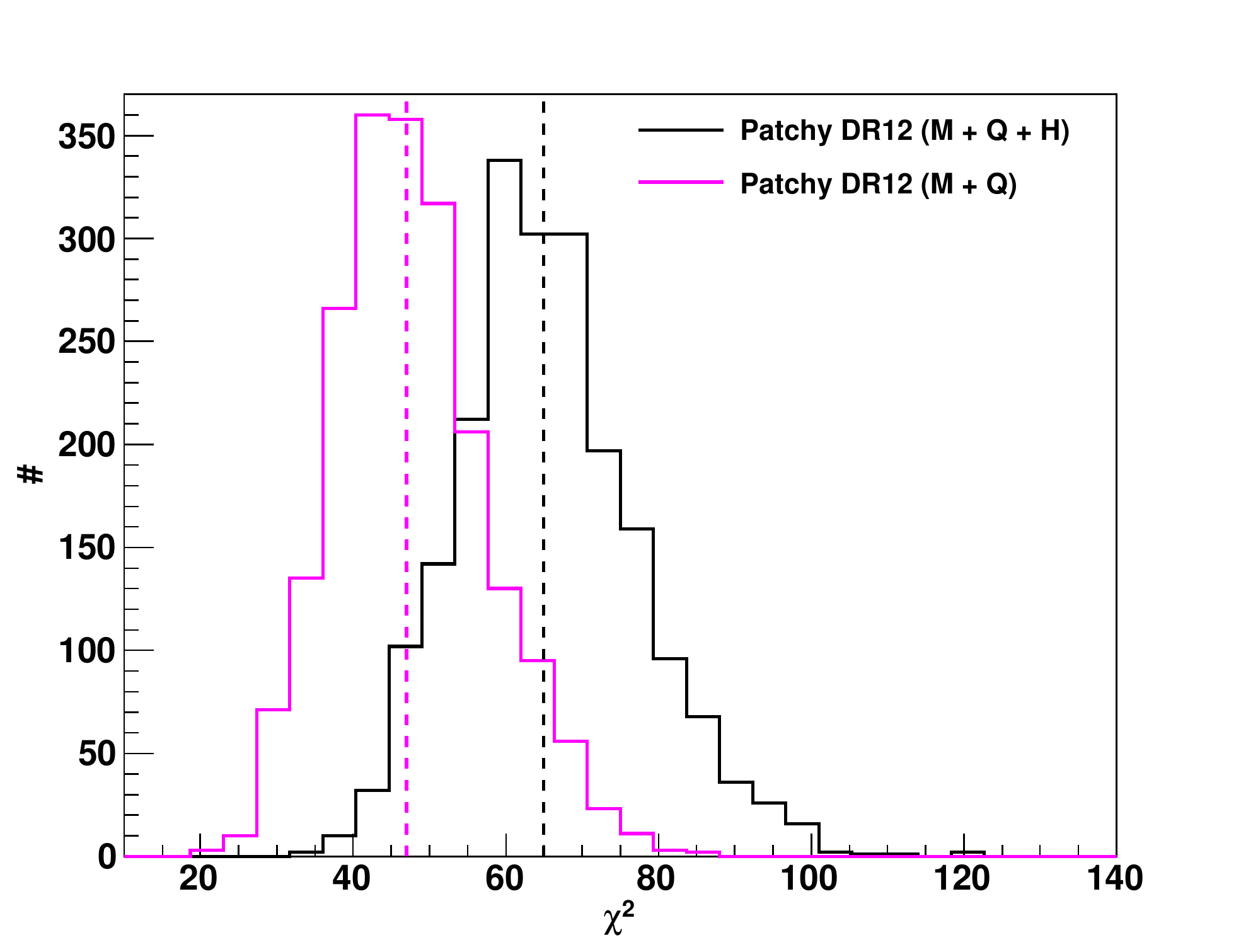,width=5.8cm}\\
\epsfig{file=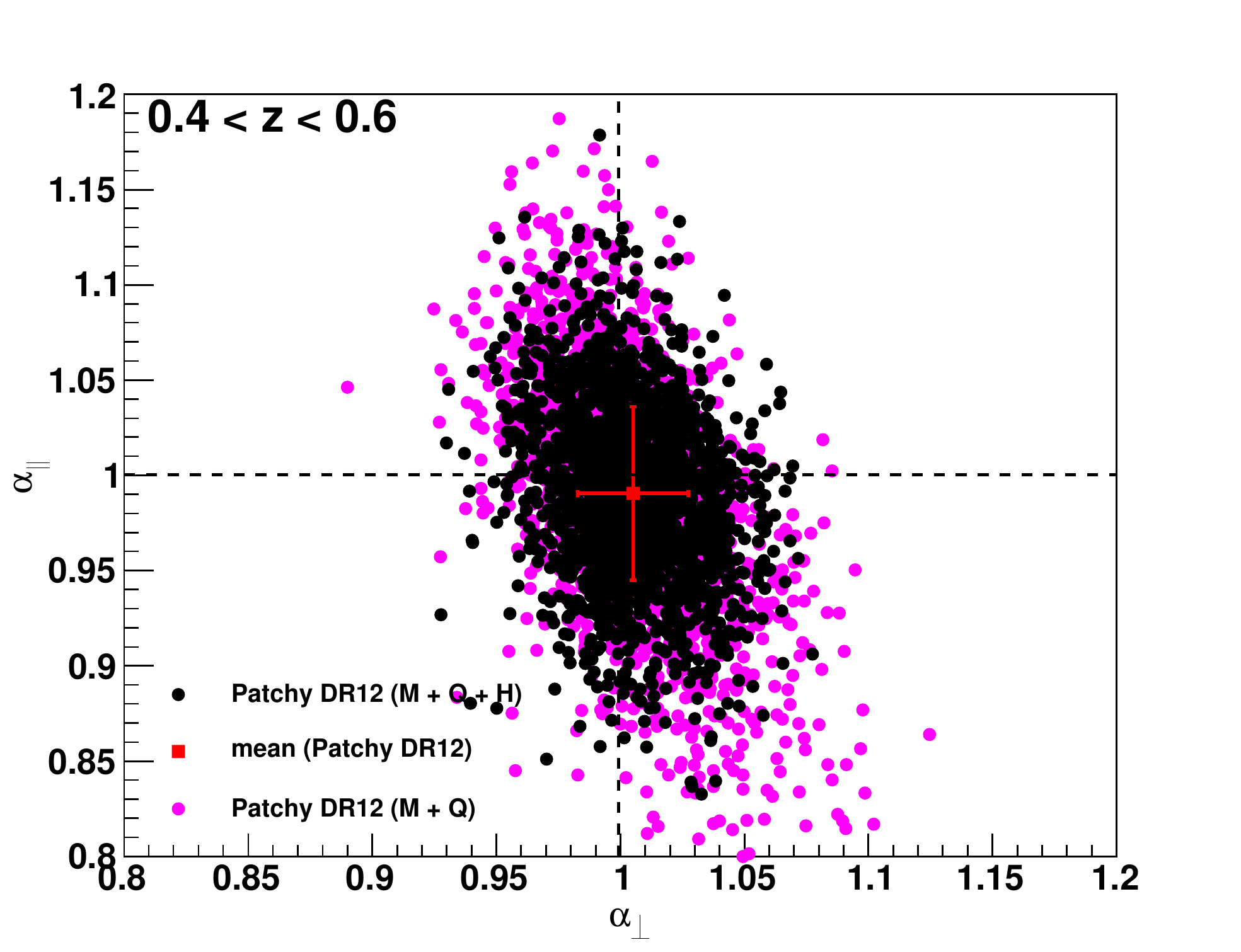,width=5.8cm}
\epsfig{file=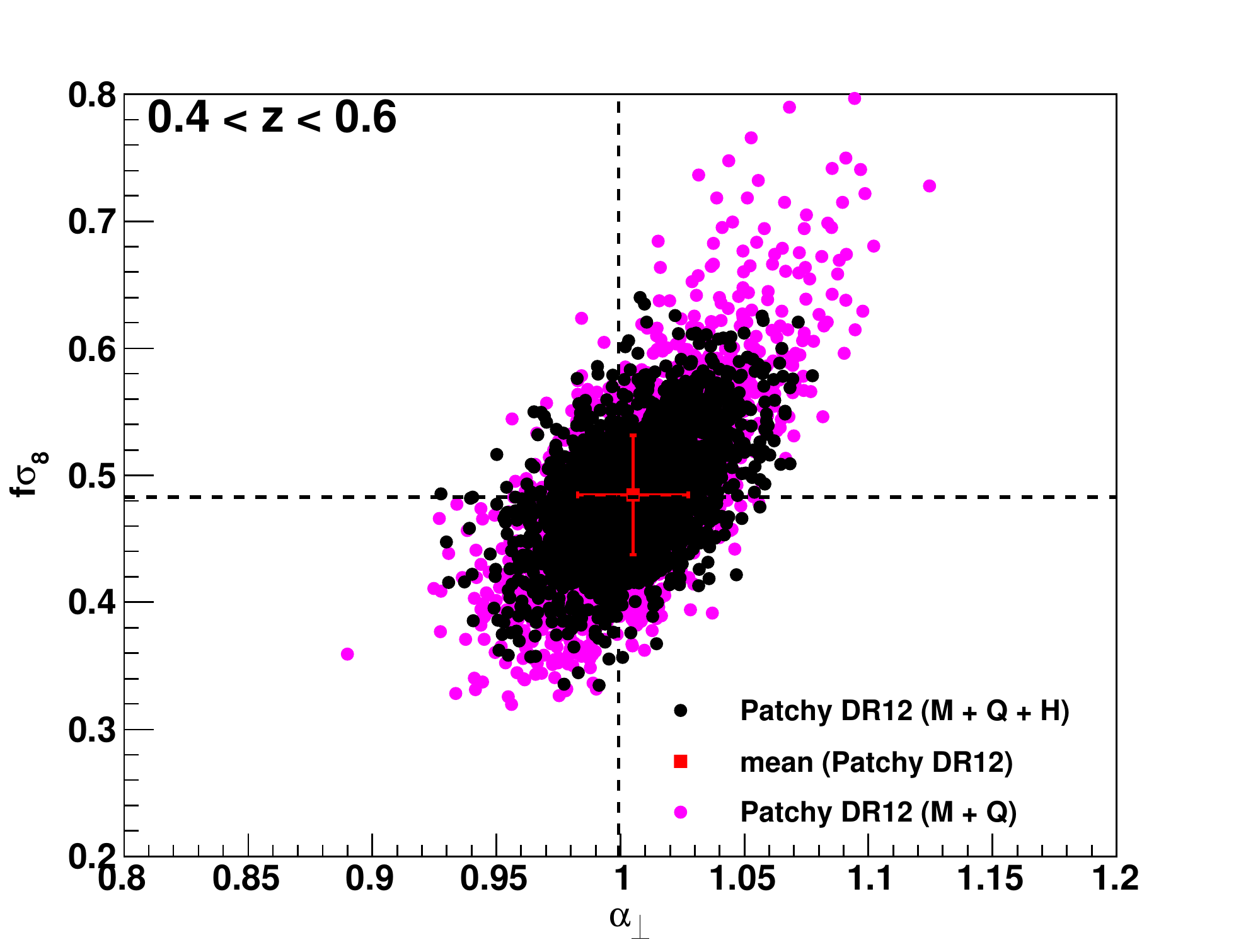,width=5.8cm}
\epsfig{file=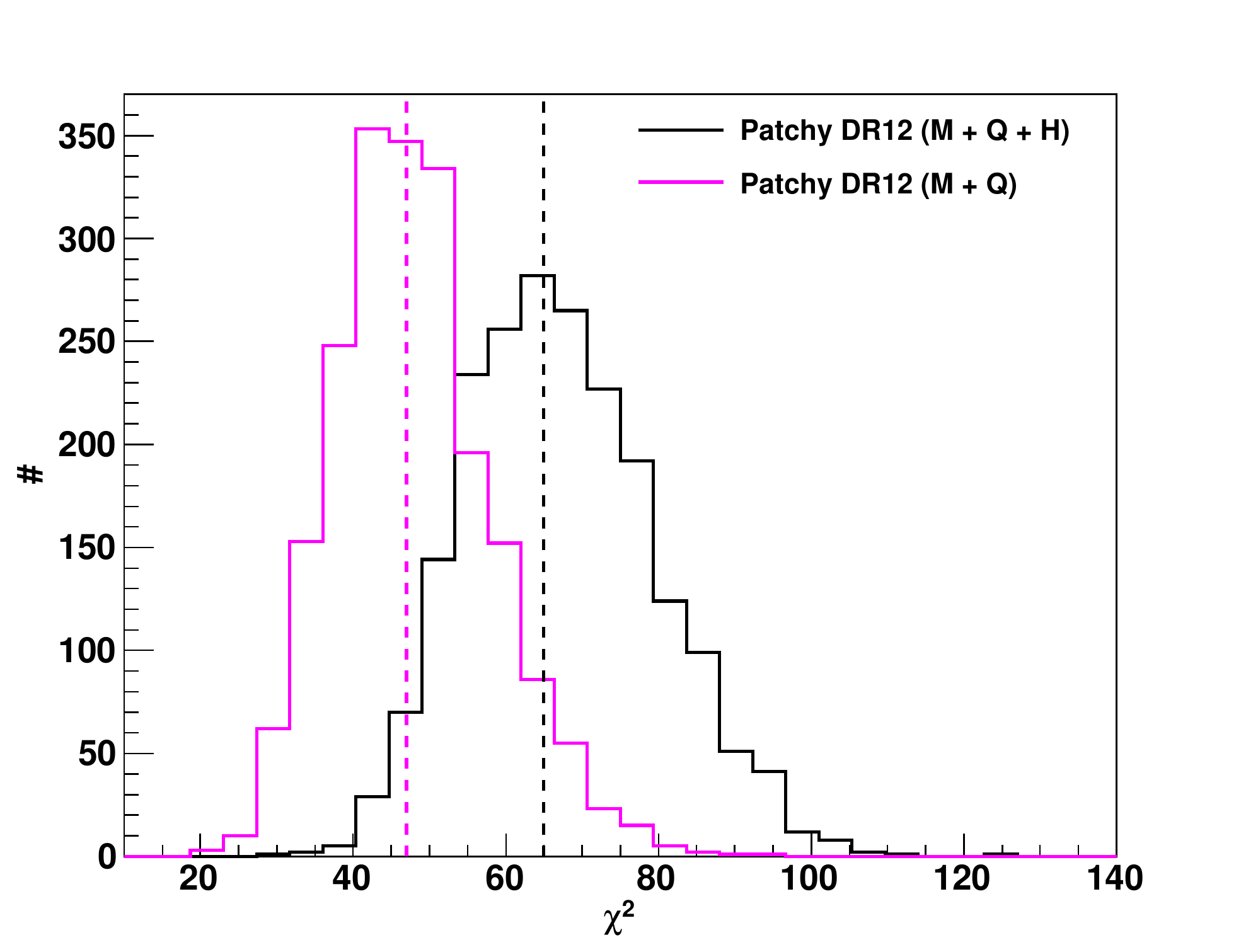,width=5.8cm}\\
\epsfig{file=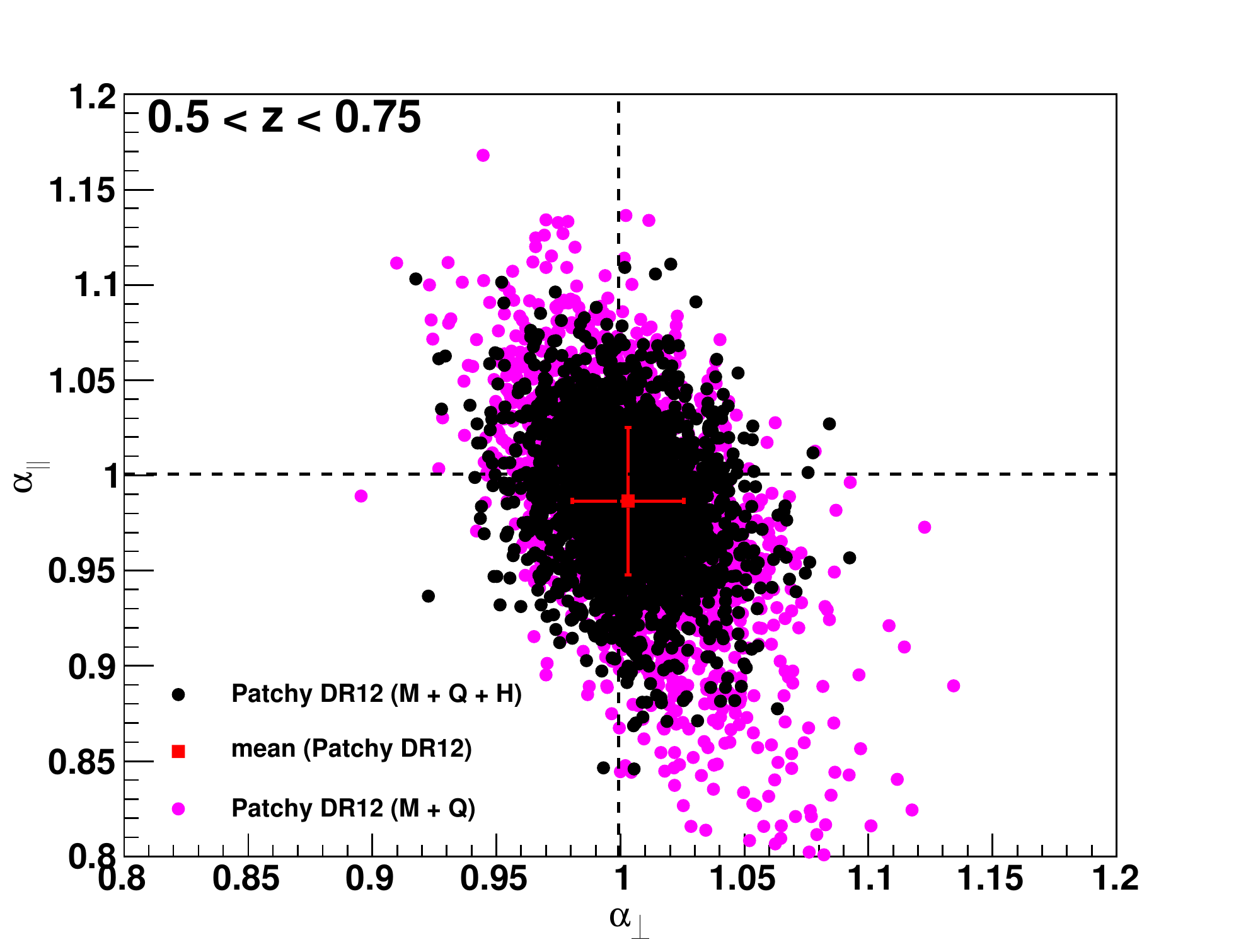,width=5.8cm}
\epsfig{file=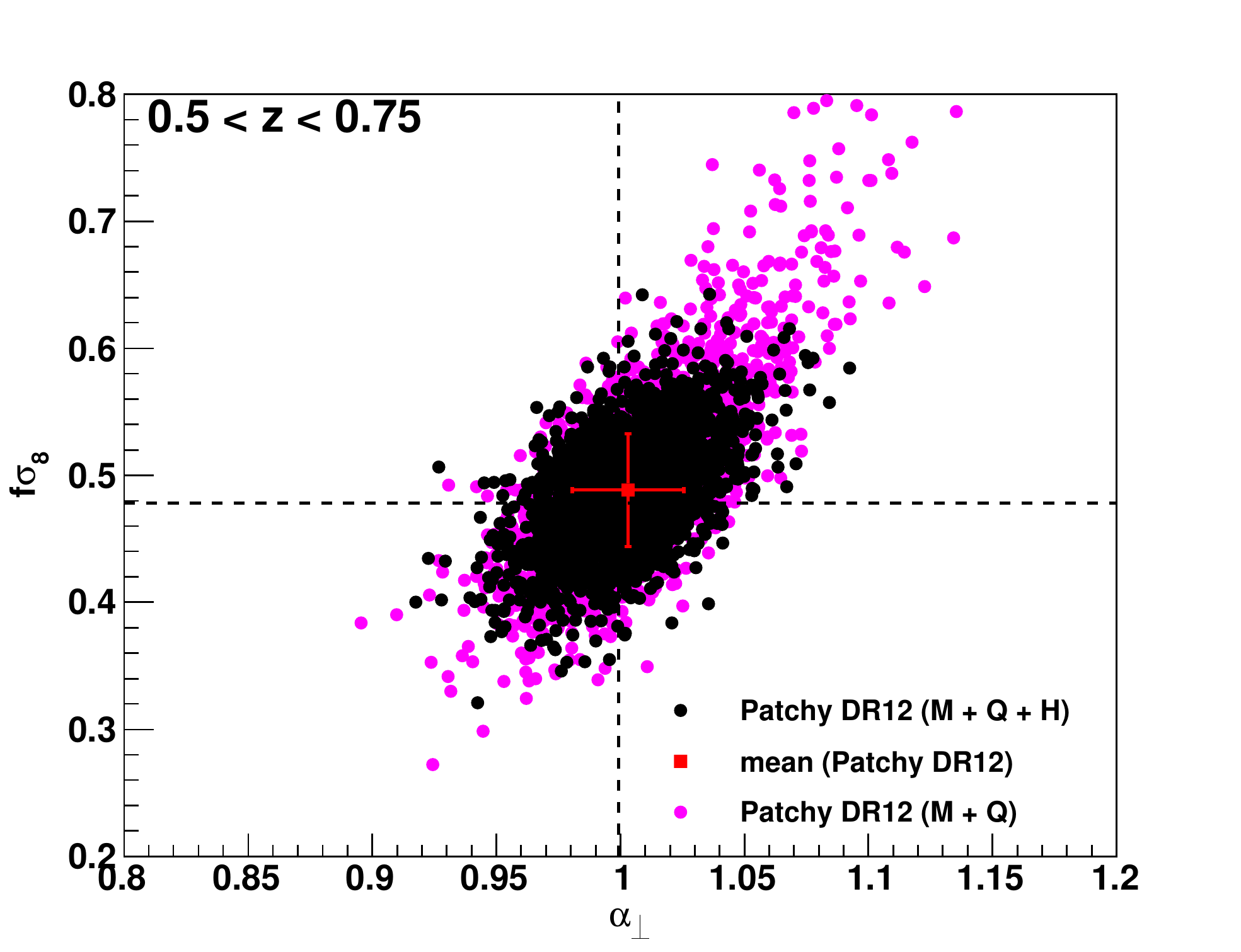,width=5.8cm}
\epsfig{file=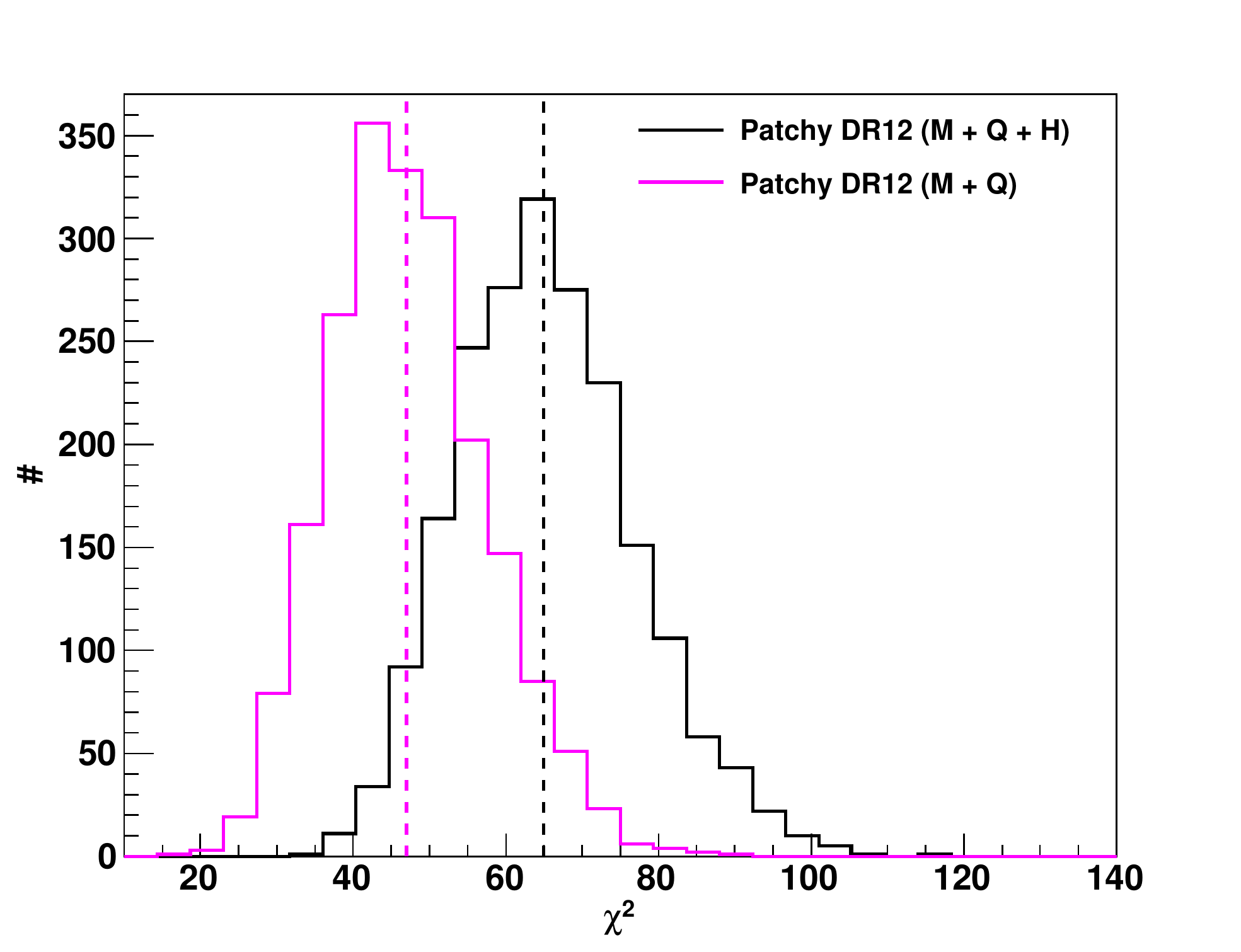,width=5.8cm}\\
\caption{The maximum likelihood results for all MultiDark-Patchy mock catalogues. The black points show the results when fitting the monopole, quadrupole and hexadecapole with the fitting range $k = 0.01$ - $0.15\ihMpc$ for the monopole and quadrupole and $k = 0.01$ - $0.10\ihMpc$ for the hexadecapole. The magenta data points are the result of fits to only the monopole and quadrupole between $k = 0.01$ - $0.15\ihMpc$. The red cross indicates the mean of the black points together with their variance. The histograms on the right show the corresponding $\chi^2$ distributions, where the dashed lines indicate the degrees of freedom.}
\label{fig:patchy_scatter}
\end{center}
\end{figure*}

\begin{table*}
\begin{center}
\caption{The results for the fits to the MultiDark-Patchy mock catalogues for the three redshift bins used in this analysis. For each bin we show the result for the fit to the monopole, quadrupole and hexadecapole (M+Q+H) as well as the fit excluding the hexadecapole (M+Q). The fitting range is $k = 0.01$ - $0.15\ihMpc$ for the monopole and quadrupole and $0.01$ - $0.10\ihMpc$ for the hexadecapole. The model to analyse the mock data is based on the BOSS fiducial cosmological model (see end of section~\ref{sec:intro}) i.e., the $\alpha$-values do not have to agree with unity. The expectation values for each redshift bin are given in the column labeled ``true''. The uncertainties represent the variance between all mock catalogues (not the error on the mean).}
	\begin{tabular}{llllllllll}
     		\hline
		 \multicolumn{10}{c}{Test on mock catalogues} \\
		 & \multicolumn{3}{c}{$0.2 < z < 0.5$} & \multicolumn{3}{c}{$0.4 < z < 0.6$} & \multicolumn{3}{c}{$0.5 < z < 0.75$}\\
		 & M+Q & M+Q+H & true& M+Q & M+Q+H & true & M+Q & M+Q+H & true\\
		\hline
		$f\sigma_{8}$ & $0.470\pm0.073$ & $0.471\pm0.053$ & $0.484$ & $0.485\pm0.068$ & $0.484\pm0.047$ & $0.483$ & $0.496\pm0.068$ & $0.488\pm0.044$ & $0.478$\\
		$\alpha_{\parallel}$ & $0.993\pm0.065$ & $0.992\pm0.048$ & $1.000$ &  $0.990\pm0.065$ & $0.990\pm0.046$ & $1.000$ & $0.982\pm0.060$ & $0.987\pm0.039$ & $1.001$ \\
		$\alpha_{\perp}$ & $1.001\pm0.032$ & $1.002\pm0.025$ & $0.999$ & $1.005\pm0.027$ & $1.005\pm0.022$ & $0.999$ & $1.006\pm0.029$ & $1.003\pm0.022$ & $0.999$\\
		\hline
		\hline
	  \end{tabular}
	  \label{tab:patchy}
\end{center}
\end{table*}

We applied our analysis pipeline to the MultiDark-Patchy mock catalogues; the results are shown in Figure~\ref{fig:patchy_scatter} and Table~\ref{tab:patchy}. Figure~\ref{fig:patchy_scatter} presents the maximum likelihood results for all MultiDark-Patchy mock catalogues including the hexadecapole (black data points) and excluding the hexadecapole (magenta data points). The red cross indicates the mean and variance between the black data points, while the black dashed lines show the fiducial parameters of the simulation. From the results, it is clear that by including the hexadecapole, we improve the scatter by $\sim 30\%$.

We can reproduce the fiducial parameters to a similar level as for the mock challenge discussed earlier. However, the MultiDark-Patchy mock catalogues are not real N-body simulations but use approximate methods to allow a large number of mocks to be produced. Therefore we adopt the blind mock challenge results to determine the potential systematic biases.

\section{BOSS DR12 Data analysis}
\label{sec:analysis}

\begin{figure*}
\begin{center}
\epsfig{file=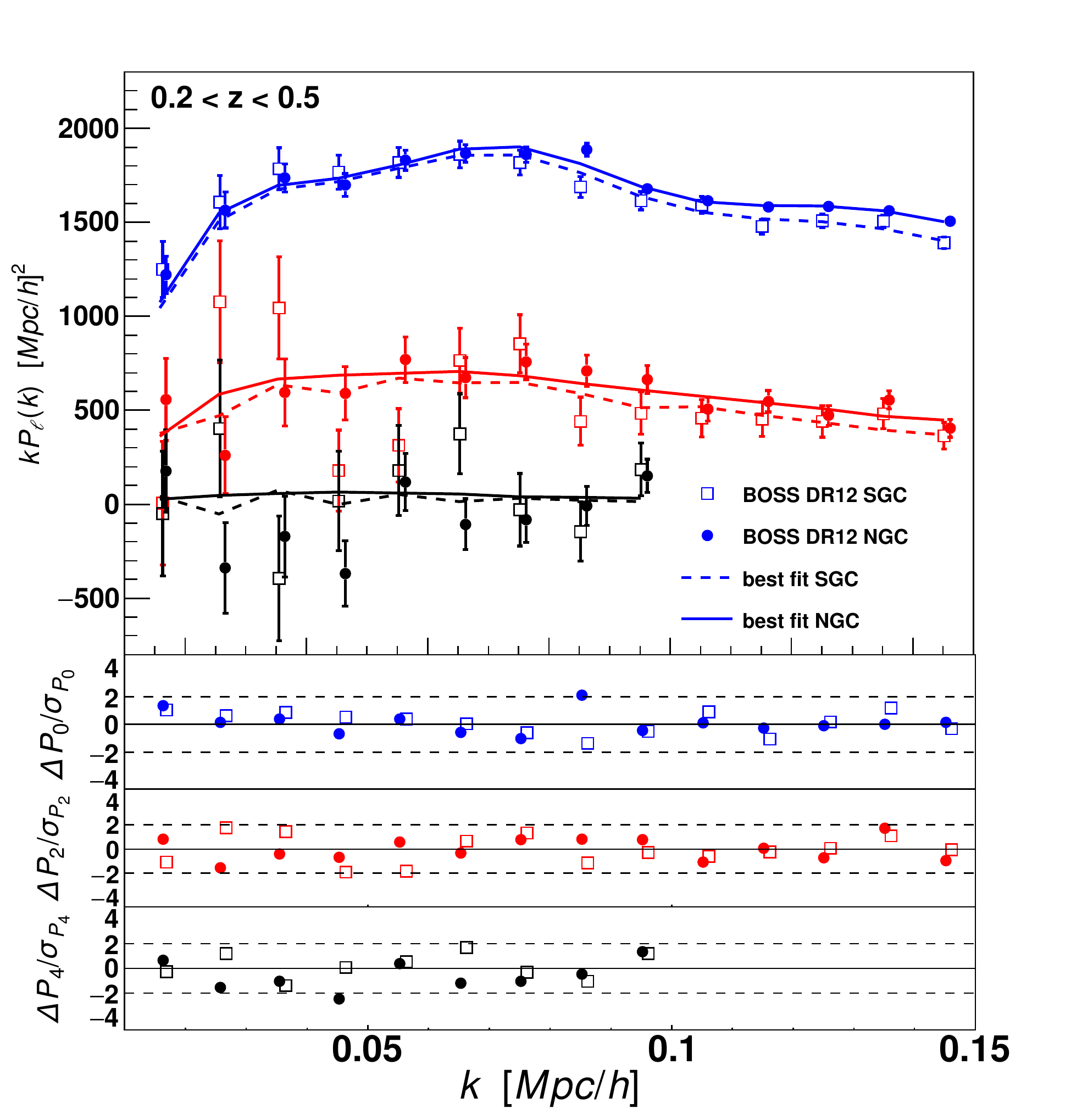,width=5.8cm}
\epsfig{file=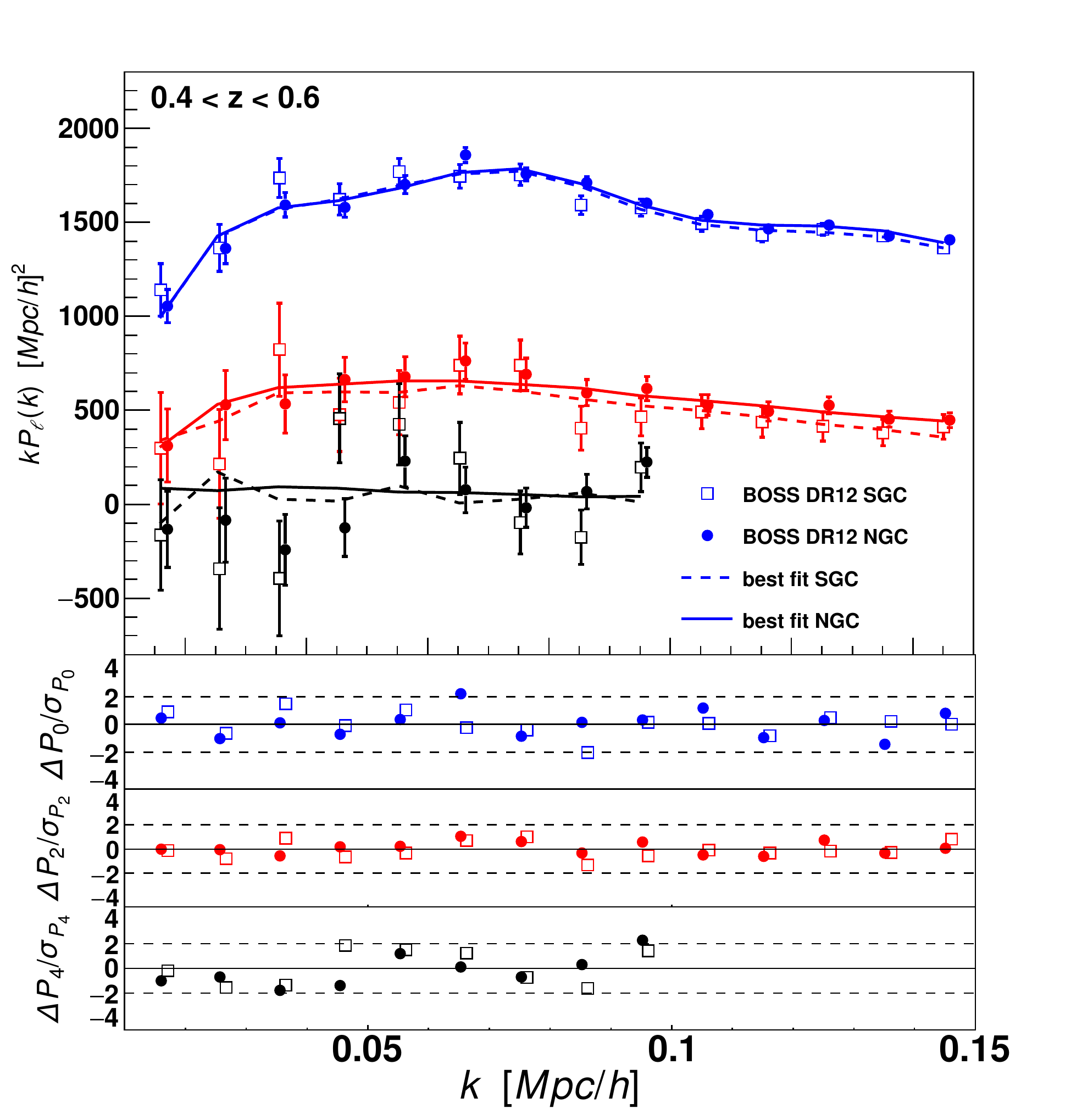,width=5.8cm}
\epsfig{file=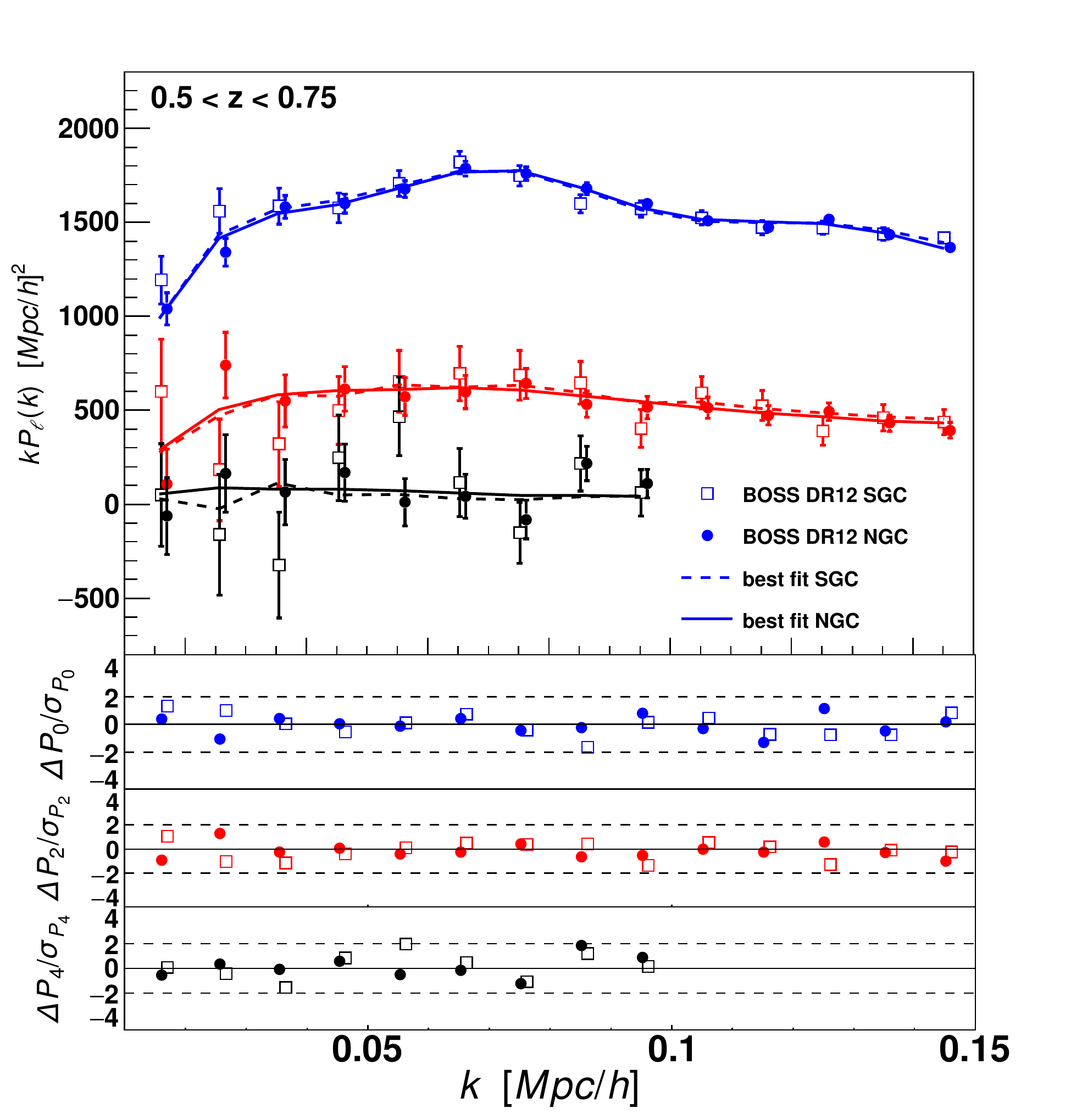,width=5.8cm}
\caption{The best fit power spectrum monopole (blue), quadrupole (red), and hexadecapole (black) models (lines) compared to the BOSS DR12 measurements (data points) in the three redshift bins used in this analysis. The measurements for the North Galactic Cap (NGC) are shown as solid circles, while the South Galactic Cap (SGC) data are displayed as open squares. The solid line represents the fit to the NGC, while the dashed line shows the result for the SGC. The best fitting models include the irregular $\mu$ distribution effect as explained in eq.~\ref{eq:binning}, which is more prominent in the SGC since the volume is smaller. The NGC and SGC power spectra are fitted simultaneously for $f\sigma_8$, $\alpha_\parallel$, and $\alpha_{\perp}$, while we marginalise over different NGC and SGC nuisance parameters ($b_1\sigma_8$, $b_2\sigma_8$, $N$ and $\sigma_v$). As a result, the best fit power spectra show different shapes for NGC and SGC, especially in the lowest redshift bin. The three lower panels show the residual for the three multipoles separately.}
\label{fig:best_fit}
\end{center}
\end{figure*}

\begin{figure}
\begin{center}
\includegraphics[height=10cm]{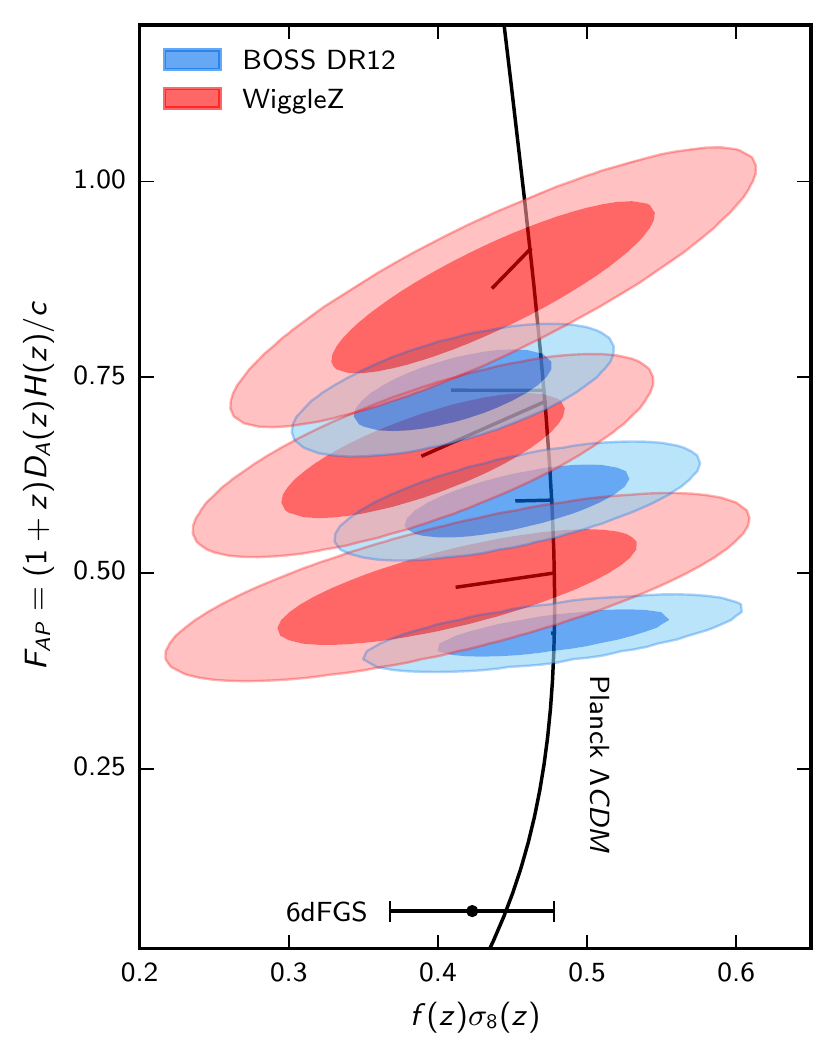}
\caption{The constraints on the Alcock-Paczynski parameter (y-axis) and the growth of structure (x-axis). The Planck prediction for these values is shown as the black solid line, where we used the best fit $\Lambda$CDM model for the Planck data to extrapolate from the redshift of decoupling to low redshift. The red contours represent the results of the WiggleZ survey~\citep{Blake:2012pj} at $\zeff = 0.44$, 0.6, and 0.73, and the black data point indicates the measurement in 6dFGS~\citep{Beutler:2012px} at $\zeff=0.067$. Both parameters, $F_{\rm AP}(z)$ and $f(z)\sigma_8(z)$, evolve with redshift and hence these contours at different $\zeff$ are not expected to overlap. Short black lines connect the best fitting values for each measurement with the Planck extrapolation for that particular redshift. The orientation of the degeneracy (i.e., the major axis of each contour ellipse) rotates with redshift, indicating that cosmological constraints can be improved by including measurements from many redshift bins.
}
\label{fig:F_fsig8}
\end{center}
\end{figure}

Here we will present the main results of our data analysis. The best fitting results of the BOSS DR12 data are summarised in Table~\ref{tab:results} and plotted in Figures~\ref{fig:best_fit},~\ref{fig:main_contours} and~\ref{fig:F_fsig8}. 

\subsection{Cosmological parameter constraints}

Marginalising over all other parameters produces the following constraints on the growth of structure parameter: $f(\zeff)\sigma_8 (\zeff)= 0.477\pm0.051$ at $z_{\rm eff}=0.38$, $0.453\pm0.050$ at $z_{\rm eff}=0.51$,  and $0.410\pm0.044$ at $z_{\rm eff}=0.61$ from the low, middle and high redshift bin, respectively.
For the Alcock-Paczynski parameter $F_{\rm AP} = (1+\zeff)D_A(\zeff)H(\zeff)/c = 0.424\pm0.020, 0.593\pm0.031$, and $0.732\pm0.034$; for the BAO scale parameter $D_Vr_s^{\rm fid}/r_s = 1490\pm33, 1913\pm44$, and $2134\pm46\,$Mpc at $z_{\rm eff}=0.38$, 0.51, and 0.61, respectively. These values are our default, final results.

\begin{figure*}
\begin{center}
\includegraphics[height=5.8cm]{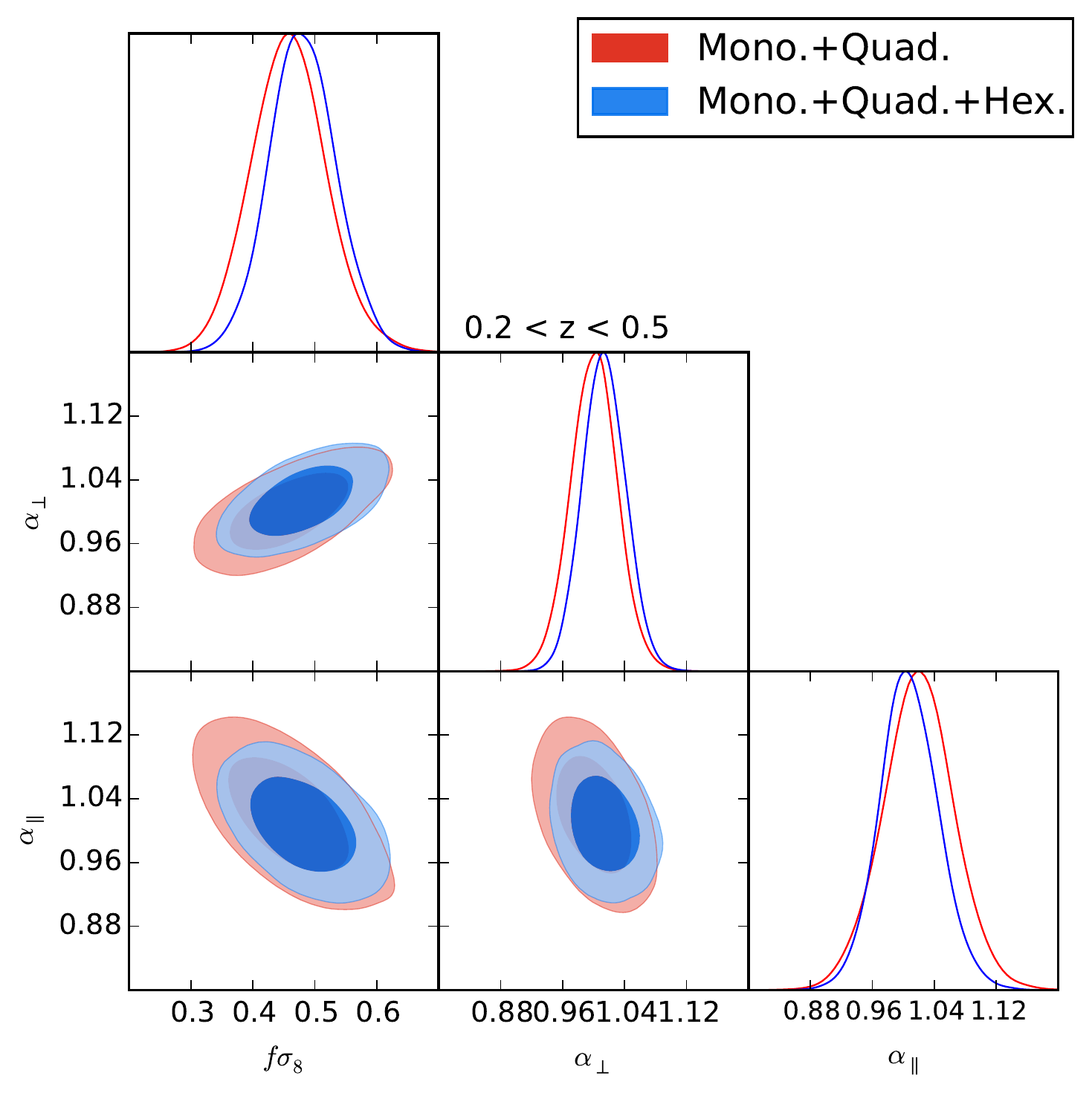}
\includegraphics[height=5.8cm]{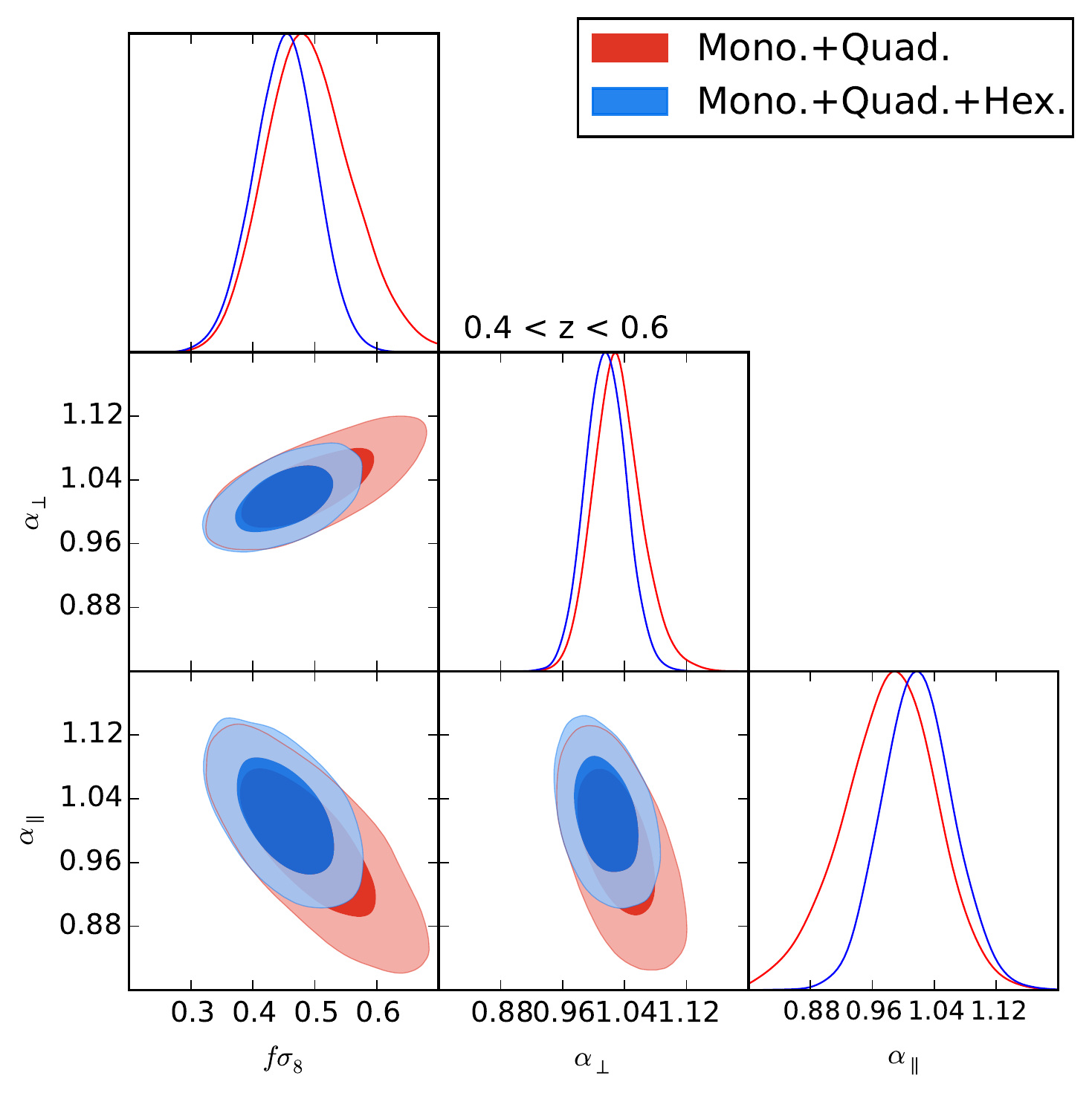}
\includegraphics[height=5.8cm]{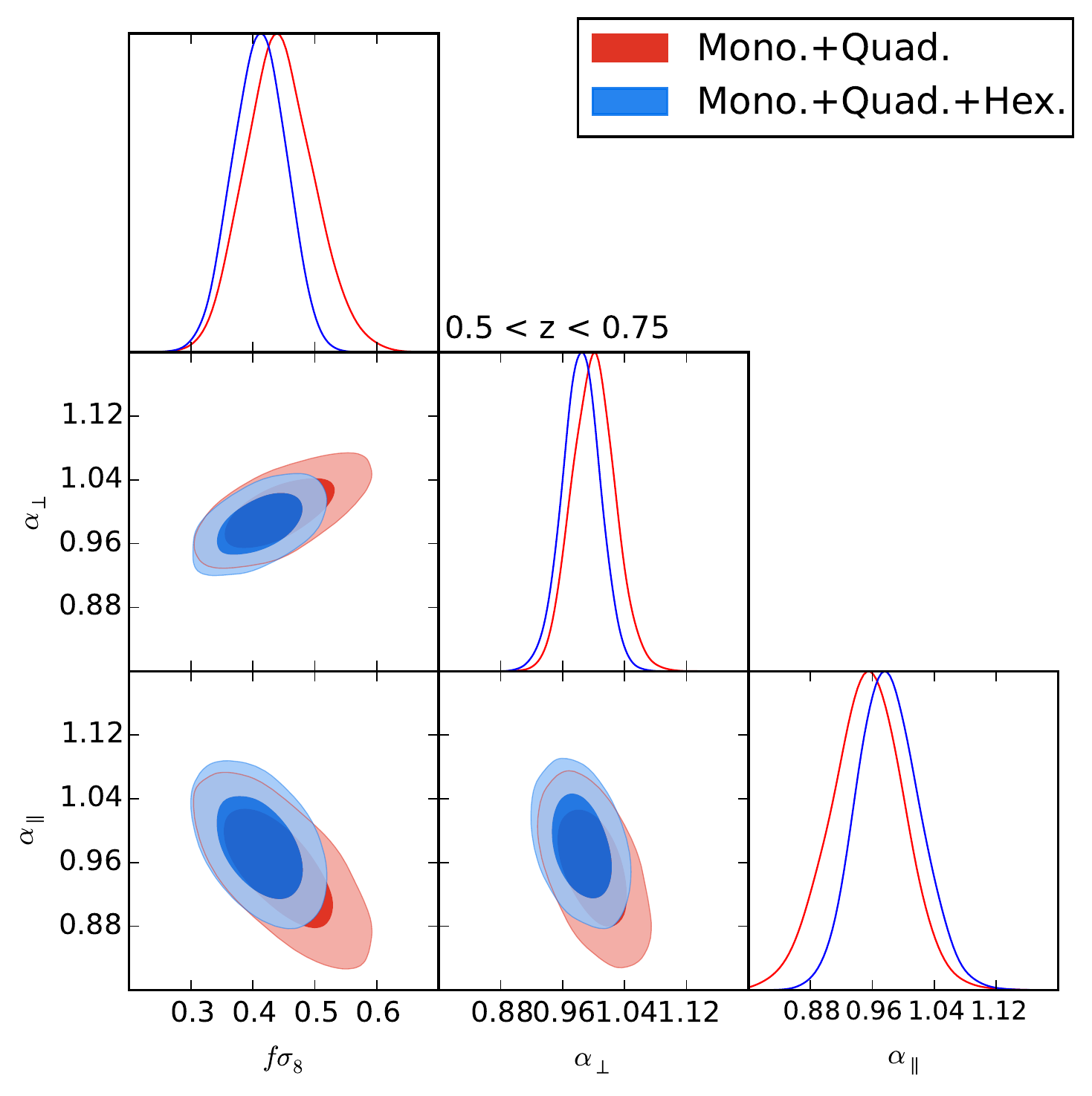}
\caption{Likelihood distributions for the three redshift bins of BOSS DR12. We show the results for the parameters $\alpha_{\perp}$, $\alpha_{\parallel}$, and $f\sigma_8$. The blue contours use the monopole, quadrupole and hexadecapole, while the red contours exclude the hexadecapole. The fitting range is $k = 0.01$ - $0.15\ihMpc$ for the monopole and quadrupole, and $k = 0.01$ - $0.10\ihMpc$ for the hexadecapole. The numerical values are summarised in Table~\ref{tab:results}.}
\label{fig:main_contours}
\end{center}
\end{figure*}

When excluding the hexadecapole, the best fitting values shift upwards in the high and middle redshift bin, while they shift downwards in the low redshift bin. In all cases the best fits agree with our default results within $1\sigma$.  Figure~\ref{fig:main_contours} shows the likelihood distributions for the fit with and without the hexadecapole. Including hexadecapoles reduces the uncertainties for all parameters in all redshift bins while maintaining the consistency in the constraints. This result agrees with our tests on mock catalogues as shown in Figure~\ref{fig:patchy_scatter}. 

The reduced $\chi^2$ for the three fits is $79.3/(74-11) =1.26$,  $74.1/(74-11)=1.18$ and $54.0/(74-11) =0.86$ from low to high redshift, and the probability of having a reduced $\chi^2$ value that exceeds this value is $Q = 8\%$, $16\%$ and $78\%$, respectively. The increase in $\chi^2$ for the lower redshift bins could arise because our model does not describe the low redshift measurements as well as the high redshift measurements due to the stronger nonlinearity at low redshift. To test the sensitivity of the low redshift result on our modelling of nonlinearity, we vary the choice of $k_{\rm max}$ and repeat the analysis using the fitting range $k = 0.01$ - $0.13\ihMpc$ for the monopole and quadrupole, while we keep the fitting range of $k = 0.01$ - $0.10\ihMpc$ for the hexadecapole. We obtain $f\sigma_8 = 0.470\pm0.066$, $F_{\rm AP} = 0.422\pm0.023$ and $D_Vr^{\rm fid}_s/r_s = 1482\pm38$Mpc, i.e., 15--30\% increase in the constraints by decreasing $k_{\rm max}$. The reduced $\chi^2$ of the best fit is $70.1/(66 - 11)$, i.e. the reduced $\chi^2$ increases slightly from $1.26$ to $1.27$. Moreover our best fitting constraints are in good agreement with our results for the larger fitting range. We therefore conclude that the fitting constraints from the low redshift bin are robust against the choice of $k_{\rm max}$. Given that the probability of exceeding this $\chi^2$ is still $8\%$, the most likely explanation is a statistical fluctuation.
 
\begin{table*}
\begin{center}
\caption{The best fitting values for the three redshift bins in BOSS DR12. The results are also shown in Figure~\ref{fig:best_fit}, \ref{fig:main_contours} and~\ref{fig:F_fsig8}. The first section of the table shows the fit to the monopole, quadrupole and hexadecapole, while the second part excludes the hexadecapole. The fitting range is $k = 0.01$ - $0.15\ihMpc$ for the monopole and quadrupole and $k = 0.01$ - $0.10\ihMpc$ for the hexadecapole. We use separate nuisance parameters for the North Galactic Cap (NGC) and South Galactic Cap (SGC). The parameters $F_{\rm AP} = (1+z)D_AH(z)/c$ and $D_Vr^{\rm fid}_s/r_s$ are derived from $\alpha_{\parallel}$ and $\alpha_{\perp}$ following eq.~\ref{eq:FAP} and~\ref{eq:DV}. The error-bars are obtained by marginalising over all other parameters.}
	\begin{tabular}{llllllll}
     		\hline
		 & & \multicolumn{6}{c}{Monopole + quadrupole + hexadecapole} \\
		 & & \multicolumn{2}{c}{$0.2 < z < 0.5$ ($z_{\rm eff} = 0.38$)} & \multicolumn{2}{c}{$0.4 < z < 0.6$ ($z_{\rm eff} = 0.51$)} & \multicolumn{2}{c}{$0.5 < z < 0.75$ ($z_{\rm eff} = 0.61$)}\\
		 & & max. like. & mean $\pm 1\sigma$\;\;\;\;\;\;\;\;\;\;\;  & max. like. & mean $\pm 1\sigma$\;\;\;\;\;\;\;\;\;\;\;  & max. like. & mean $\pm 1\sigma$\\
		\hline
		$\alpha_{\parallel}$ & & $1.001$ & $1.007\pm0.037$ & $1.007$ & $1.015\pm0.046$ & $0.977$ & $0.982\pm0.040$  \\
		$\alpha_{\perp}$ & & $1.008$ & $1.014\pm0.027$ & $1.014$ & $1.015\pm0.027$ & $0.982$ & $0.984\pm0.024$  \\
		$f(z)\sigma_{8}(z)$ & & $0.478$ & $0.482\pm0.053$ & $0.456$ & $0.455\pm0.050$ & $0.412$ & $0.410\pm0.042$  \\
		$\chi^2/$d.o.f. & & $79.3/(74-11)$ & --- & $74.1/(74-11)$ & --- & $54.0/(74-11)$ & ---  \\
		\hline
		$F_{\rm AP}$ & & $0.426$ & $0.427\pm0.022$ & $0.600$ & $0.594\pm0.035$ & $0.732$ & $0.736\pm0.040$  \\
		$D_V(z)r_s^{\rm fid}/r_s$ & [Mpc] & $1485$ & $1493\pm28$ & $1908$ & $1913\pm35$ & $2132$ & $2133\pm36$  \\
		$H(z)r_s/r_s^{\rm fid}$ & [km/s/Mpc] & $82.8$ & $82.4\pm3.0$ & $89.0$ & $88.5\pm4.0$ & $96.9$ & $97.1\pm3.9$  \\
		$D_A(z)r_s^{\rm fid}/r_s$ & [Mpc] & $1118$ & $1124\pm30$ & $1331$ & $1333\pm35$ & $1407$ & $1410\pm35$  \\
		\hline
		$b^{\rm NGC}_1\sigma_8$ & & $1.339$ & $1.336\pm0.040$ & $1.300$ & $1.303\pm0.040$ & $1.230$ & $1.235\pm0.041$ \\
		$b^{\rm SGC}_1\sigma_8$ & & $1.337$ & $1.332\pm0.057$ & $1.305$ & $1.305\pm0.046$ & $1.259$ & $1.247\pm0.043$ \\
		$b^{\rm NGC}_2\sigma_8$ & & $1.16$ & $1.11^{+0.77}_{-0.89}$ & $2.08$ & $1.95^{+0.58}_{-0.66}$ & $2.83$ & $2.70^{+0.47}_{-0.54}$ \\
		$b^{\rm SGC}_2\sigma_8$ & & $0.32$ & $0.52^{+0.64}_{-0.69}$ & $0.56$ & $0.61^{+0.60}_{-0.52}$ & $0.98$ & $0.71^{+0.55}_{-0.60}$ \\
		$N^{\rm NGC}$ & & $-1580$ & $-1100^{+1410}_{-780}$ & $-1710$ & $-1555^{+620}_{-570}$ & $-350$ & $-350^{+950}_{-740}$  \\
		$N^{\rm SGC}$ & & $-930$ & $-500^{+1880}_{-1400}$ & $-900$ & $790^{+1000}_{-970}$ & $-910$ & $-130^{+860}_{-650}$  \\
		$\sigma_v^{\rm NGC}$ & & $6.15$ & $6.10\pm0.69$ & $5.84$ & $5.84^{+0.70}_{-0.77}$ & $5.39$ & $5.35^{+0.76}_{-0.81}$ \\
		$\sigma_v^{\rm SGC}$ & & $6.80$ & $6.78\pm0.83$ & $6.39$ & $6.46\pm0.87$ & $5.08$ & $4.93^{+0.88}_{-0.95}$ \\
		\hline
		 & & \multicolumn{6}{c}{Monopole + quadrupole} \\
		 & & \multicolumn{2}{c}{$0.2 < z < 0.5$ ($z_{\rm eff} = 0.38$)} & \multicolumn{2}{c}{$0.4 < z < 0.6$ ($z_{\rm eff} = 0.51$)} & \multicolumn{2}{c}{$0.5 < z < 0.75$ ($z_{\rm eff} = 0.61$)}\\
		 & & max. like. & mean $\pm 1\sigma$\;\;\;\;\;\;\;\;\;\;\; & max. like. & mean $\pm 1\sigma$\;\;\;\;\;\;\;\;\;\;\; & max. like. & mean $\pm 1\sigma$\\
		\hline
		$\alpha_{\parallel}$ & & $1.017$ & $1.019\pm0.045$ & $0.982$ & $0.982\pm0.059$ & $0.951$ & $0.953\pm0.046$  \\
		$\alpha_{\perp}$ & & $0.999$ & $1.001\pm0.030$ & $1.025$ & $1.031\pm0.031$ & $0.996$ & $1.000\pm0.027$  \\
		$f(z)\sigma_{8}(z)$ & & $0.457$ & $0.459\pm0.060$ & $0.483$ & $0.494^{+0.071}_{-0.065}$ & $0.441$ & $0.443\pm0.054$  \\
		$\chi^2/$d.o.f. & & $54.5/(56-11)$ & --- & $40.8/(56-11)$ & --- & $34.9/(56-11)$ & ---  \\
		\hline
		F$_{\rm AP}$ & & $0.416$ & $0.417\pm0.027$ & $0.619$ & $0.626\pm0.051$ & $0.767$ & $0.771\pm0.052$  \\
		D$_V(z)r_s^{\rm fid}/r_s$ & [Mpc] & $1484$ & $1490\pm28$ & $1905$ & $1912\pm37$ & $2129$ & $2135\pm37$  \\
		$H(z)r_s/r_s^{\rm fid}$ & [km/s/Mpc] & $81.5$ & $81.5\pm3.6$ & $91.3$ & $91.6\pm5.6$ & $100.1$ & $100.1\pm4.9$  \\
		$D_A(z)r_s^{\rm fid}/r_s$ & [Mpc] & $1107.7$ & $1110\pm33$ & $1345.7$ & $1354\pm40$ & $1427$ & $1433\pm39$  \\
		\hline
		$b^{\rm NGC}_1\sigma_8$ & & $1.347$ & $1.332\pm0.047$ & $1.289$ & $1.283\pm0.050$ & $1.220$ & $1.225\pm0.045$ \\
		$b^{\rm SGC}_1\sigma_8$ & & $1.344$ & $1.335^{+0.55}_{-0.62}$ & $1.289$ & $1.271^{+0.049}_{-0.056}$ & $1.246$ & $1.229\pm0.046$ \\
		$b^{\rm NGC}_2\sigma_8$ & & $1.03$ & $0.84^{+0.78}_{-0.95}$ & $2.02$ & $1.48^{+0.64}_{-0.75}$ & $2.85$ & $2.66^{+0.45}_{-0.61}$ \\
		$b^{\rm SGC}_2\sigma_8$ & & $0.32$ & $0.6\pm1.2$ & $0.55$ & $0.46\pm0.63$ & $0.93$ & $0.55^{+0.50}_{-0.61}$ \\
		$N^{\rm NGC}$ & & $1460$ & $-600^{+1550}_{-750}$ & $-1760$ & $-1180\pm1000$ & $-380$ & $-440^{+940}_{-760}$  \\
		$N^{\rm SGC}$ & & $920$ & $-600^{+3370}_{-1650}$ & $-900$ & $-70^{+1290}_{-860}$ & $-910$ & $20^{+860}_{-690}$  \\
		$\sigma_v^{\rm NGC}$ & & $6.16$ & $6.03\pm0.73$ & $5.75$ & $5.67\pm0.74$ & $5.20$ & $5.19\pm0.74$ \\
		$\sigma_v^{\rm SGC}$ & & $6.89$ & $6.87\pm0.81$ & $6.20$ & $6.06\pm0.85$ & $4.94$ & $4.71\pm0.85$ \\
		\hline
		\hline
	  \end{tabular}
	  \label{tab:results}
\end{center}
\end{table*}

\section{Discussion}
\label{sec:dis}

\subsection{Power spectrum multipoles}

Figure~\ref{fig:best_fit} compares the best fitting power spectra models with the data, where we indicate the NGC with solid data points and the SGC with open squares. The corresponding best fitting models are indicated as a solid line for the NGC and a dashed line for the SGC. We use different nuisance parameters for the NGC and SGC, which makes the best fitting models appear quite different, even though the underlying cosmology is the same. The need to have separate nuisance parameters for NGC and SGC is limited to the lowest redshift bin, where the two power spectra have different amplitudes in the monopole. The source of this difference is connected to chunks 2-6 (in NGC) which have a different target selection from the rest of the survey, leading to a lower density at low redshift (see section~\ref{sec:data},~\citealt{Reid:2015gra} and~\citealt{Ross:2016}). The use of separate nuisance parameters for NGC and SGC does not degrade our parameter constraints and hence we used this approach for all redshift bins. The best fitting models include the correction for the irregular $\mu$-distribution as explained in section~\ref{sec:binning}. 

The lower panels in Figure~\ref{fig:best_fit} show the residual for the three multipoles. In the lowest redshift bin the monopole data seem to prefer a systematically larger amplitude at small $k$, which the model does not appear to be able to accommodate given the constraints on large $k$. This might contribute to the overall larger $\chi^2$ for this bin (see also the $k_{\rm max}$ test above). However, all fits result in reasonable reduced $\chi^2$, indicating that the model is adequate in describing the data.

\subsection{Parameter degeneracies and correlations}

Here we compare the correlation between different parameters with the theoretical expectation, with a focus on the second redshift bin. 
If we express the $\alpha$ value in $D_A(z)r^{\rm fid}_s/r_s$ and $H(z)r_s/r_s^{\rm fid}$ the following correlation matrix ($D_A(z)r^{\rm fid}_s/r_s$, $H(z)r_s/r_s^{\rm fid}$, $f\sigma_8$) is produced:
\begin{equation}
R^{D_A\text{-}H}_{\rm z2} = \begin{pmatrix} 1 & 0.257 & 0.503\\
			        0.257 & 1 & 0.547\\
			        0.503 & 0.547 & 1\end{pmatrix}.
\end{equation} 
The Fisher formalism~\citep{Seo:2003,Seo:2007ns,Shoji:2009} predicts that if we understand RSD perfectly, the pure AP limit will give a correlation coefficient between $D_A(z)$ and $H(z)$ of $1$ ($F_{\rm AP} \propto D_A(z)H(z)$). If we increase the free parameters for RSD, the coefficient decreases. If we marginalise over all RSD information and use the BAO alone, the expected correlation coefficient is $-0.4$. Therefore, for the BAO only analysis, we expect $r\sim -0.4$ (see our companion paper~\citealt{Beutler:2016}). Since we are using RSD as well as BAO information, we expect $r$ somewhere between $-0.4$ and $1$, depending on our freedom in RSD parameters. Our value of $r=0.257$ indicates a mixture of BAO and RSD information with a modest freedom in our RSD model. 
The most natural parametrisation is given by ($D_V(z)r^{\rm fid}_s/r_s$, $F_{\rm AP}$, $f\sigma_8$), which corresponds to the actual signals in the data. The correlation matrix is given by
\begin{equation}
R^{D_V\text{-}F_{\rm AP}}_{\rm z2} = \begin{pmatrix} 1 & -0.291 & -0.0562\\
			        -0.291 & 1 & 0.648\\
			        -0.0562 & 0.648 & 1\end{pmatrix}.
\end{equation} 
There is a clear correlation between the Alcock-Paczynski parameter ($F_{\rm AP}$) and growth rate $f\sigma_8$, while the BAO dilation parameter $D_Vr^{\rm fid}_s/r_s$ and $f\sigma_8$ are almost uncorrelated. We include the correlation matrices, covariance matrices and inverse covariance matrices for these three parameters in Appendix~\ref{app:covs}.

The correlation matrices indicate a correlation of about $60\%$ between $F_{\rm AP}$ and $f\sigma_8$. Therefore, if we hold $F_{\rm AP}$ fixed, i.e., if we assume that we know $F_{\rm AP}$ precisely, the constraints on $f\sigma_8$ can be significantly improved. This is an interesting case to consider, since when combining our results with the Planck constraints, we effectively fix the background cosmological model. Fixing $F_{\rm AP}$ to the best fit value yields $f\sigma_8 = 0.482\pm0.037$, $0.455\pm0.038$ and $0.410\pm0.034$ for the low ($z_{\rm eff}=0.38$), middle ($z_{\rm eff}=0.51$) and high redshift bin ($z_{\rm eff}=0.61$), respectively.

\subsection{Comparison to DR11 and other BOSS results}

We compare these new results with our DR11 analysis~\citep{Beutler:2013yhm}. Our DR11 study found a growth of structure constraint of $f(\zeff)\sigma_8(\zeff) = 0.419\pm0.043$ at $\zeff=0.57$, consistent with our high redshift measurement in this analysis of $f(\zeff)\sigma_8(\zeff) = 0.410\pm0.042$ at $\zeff=0.61$. Our new uncertainties are slightly larger compared to the DR11 result, which is caused by (1) the smaller redshift range given that our high redshift bin has a low redshift cutoff at $0.5$ compared to $0.43$ in the CMASS sample in~\citet{Beutler:2013yhm}, and (2) the fact that we use different mock catalogues compared to our DR11 analysis to generate the covariance matrix, which tend to result in larger uncertainties. 

In~\citet{Gil-Marin:2015sqa} the BOSS DR12 data have been analysed in Fourier-space using the LOWZ and CMASS samples. They found a growth of structure constraint of $f(\zeff)\sigma_8(\zeff) = 0.395\pm 0.064$ at $\zeff=0.32$ and $f(\zeff)\sigma_8(\zeff) = 0.442\pm 0.037$ at $\zeff=0.57$ for LOWZ and CMASS, respectively. The LOWZ result is significantly (more than $1\sigma$) smaller than our constraint in the low-redshift bin, which is $f(\zeff)\sigma_8(\zeff) = 0.482\pm 0.053$ at $\zeff=0.38$. There are many potential sources for this difference: (1) Our low redshift bin covers a redshift range of $z = 0.2$ - $0.5$, which is slighter higher compared to the redshift range of $z = 0.2$ - $0.43$ of LOWZ, (2) the additional data in our analysis (chunks 2-6) causes a difference in the target selection mainly in the low redshift bin, (3) \citet{Gil-Marin:2015sqa} fit the power spectrum monopole and quadrupole down to $k_{\rm max}=0.24\ihMpc$ compared to $k_{\rm max}=0.15\ihMpc$ in our analysis, which suggests that their constraint is dominated by high $k$ modes, and (4) we include the hexadecapole in our analysis, which is not used in~\citet{Gil-Marin:2015sqa}. 

The consistency between our results and our companion papers~\citet{Sanchez:2016},~\citet{Grieb:2016}, and~\citet{Satpathy2016} is discussed in~\citet{Alam2016}.

\subsection{Comparison to other galaxy survey}

Figure~\ref{fig:F_fsig8} compares our measurements of the AP parameter and $f\sigma_8$ with measurements from the 6-degree Field Galaxy Survey (6dFGS, black data point,~\citealt{Beutler:2012px}) at $\zeff=0.067$ and the WiggleZ survey (red contours,~\citealt{Blake:2012pj}) at $\zeff = 0.44$, 0.6 and 0.73. The 6dFGS measurement ignored the Alcock-Paczynski effect by assuming $D_A(z)$ and $H(z)$ are known, since the sensitivity to this signal becomes small at the 6dFGS redshift. The BOSS measurements cover a redshift range almost as wide as the WiggleZ measurement and with significantly reduced uncertainties. Given a smooth evolution of $f\sigma_8$ with redshift, the WiggleZ and BOSS measurements are consistent with each other\footnote{Note that there is a small level of correlation between these two surveys~\citep{Marin:2015ula,Beutler:2015tla}}. 

The only large-scale structure analysis in the literature we are aware of, which makes use of the hexadecapole, is the study by~\citet{Oka:2013cba}, which, however, ignores all window function effects. Our analysis suggests that the window function effects in the hexadecapole are indeed negligible when compared to the measurement uncertainties, while the effects of the window function on the monopole and quadrupole are significant; ignoring these effects would significantly bias our results.

\subsection{Comparison to Planck 2015}

Next we can compare our measurements to the Planck 2015 results~\citep{Ade:2015xua}. Using a $\Lambda$CDM model to extrapolate from the redshift of decoupling to the effective redshifts of our large-scale-structure measurements, Planck predicts values for the growth of structure of $f(\zeff)\sigma_8 (\zeff)= 0.4784\pm0.0077$, $0.4763\pm0.0060$ and $0.4707\pm0.0058$ for the low ($\zeff=0.38$), middle ($\zeff=0.51$) and high redshift bins ($\zeff=0.61$), respectively.
The largest deviation between our measurements and the Planck $\Lambda$CDM predictions occurs at the highest redshift bin, where our value is lower than Planck by about $1.4\sigma$. Figure~\ref{fig:F_fsig8} compares our 2-dimensional constraints on the growth of structure and the AP effect with the Planck $\Lambda$CDM predictions for these parameters. The deviation of the highest redshift datapoint is not statistically significant. Together with our low redshift measurements, which agree well with the Planck predictions, there is overall good consistency between the RSD constraints from BOSS and the Planck dataset. 

\section{Conclusion}
\label{sec:conclusion}

We measure the power spectrum multipoles from the final BOSS DR12 dataset in the redshift range $0.2 < z < 0.75$. We extract the Baryon Acoustic Oscillation, Alcock-Paczynski and redshift-space distortion signals using a model based on renormalised perturbation theory. For the first time we include the hexadecapole in our analysis, while appropriately accounting for the survey window function, which reduced the uncertainties on $f(z)\sigma_8(z)$ by about $20\%$. The main results of this analysis are:
\begin{enumerate}
\item A FFT based window function method, first suggested in~\citet{Wilson:2015lup} using the global plane parallel approximation, can be derived within the local plane parallel approximation, which makes it applicable to wide-angle surveys like BOSS. We present the detailed derivation in appendix~\ref{app:window}.
\item By fitting the monopole and quadrupole between $k = 0.01$ - $0.15\ihMpc$ and the hexadecapole between $k = 0.01$ - $0.10\ihMpc$, we were able to extract the constraints $f(\zeff)\sigma_8(\zeff) = 0.482\pm0.053$, $0.455\pm0.050$ and $0.410\pm0.042$ at the effective redshifts of $z_{\rm eff} = 0.38, 0.51$, and $0.61$, respectively. For the Alcock-Paczynski parameter $F_{\rm AP} = (1+\zeff)D_A(\zeff)H(\zeff)/c$ we find $0.427\pm0.022, 0.594\pm0.035$, and $0.736\pm0.040$ and the BAO scale parameter is $D_Vr_s^{\rm fid}/r_s = 1493\pm28, 1913\pm35$, and $2133\pm36\,$Mpc. Assuming Gaussian likelihood, we provide a covariance matrix which contains the parameter constraints as well as their correlations (see appendix~\ref{app:covs}). 
\item We demonstrated the accuracy of our analysis pipeline by participating in a mock challenge, which resulted in systematic uncertainties $\lesssim 10\%$ of the statistical error budget. The description of this mock challenge can be found in our companion paper~\citep{Tinker:2015}.
\item Our high redshift result on $f\sigma_8$ is in agreement with our DR11 analysis using the CMASS sample, and shows a small $1.4\sigma$ deviation from the Planck prediction. The low redshift results obtained in this analysis show good agreement with the Planck prediction. 
\end{enumerate} 
\citet{Alam2016} combines our measurements with the corresponding growth of structure measurements of~\citet{Grieb:2016},~\citet{Sanchez:2016} and~\citet{Satpathy2016} and the BAO-only measurements of~\citet{Beutler:2016} and~\citet{Ross:2016} into a final BOSS likelihood and investigates the cosmological implications. 

\section*{Acknowledgments}

FB acknowledges support from the UK Space Agency through grant ST/N00180X/1.

Funding for SDSS-III has been provided by the Alfred P. Sloan Foundation, the Participating Institutions, the National Science Foundation, and the U.S. Department of Energy Office of Science. The SDSS-III web site is http://www.sdss3.org/.

SDSS-III is managed by the Astrophysical Research Consortium for the Participating Institutions of the SDSS-III Collaboration including the University of Arizona, the Brazilian Participation Group, Brookhaven National Laboratory, Carnegie Mellon University, University of Florida, the French Participation Group, the German Participation Group, Harvard University, the Instituto de Astrofisica de Canarias, the Michigan State/Notre Dame/JINA Participation Group, Johns Hopkins University, Lawrence Berkeley National Laboratory, Max Planck Institute for Astrophysics, Max Planck Institute for Extraterrestrial Physics, New Mexico State University, New York University, Ohio State University, Pennsylvania State University, University of Portsmouth, Princeton University, the Spanish Participation Group, University of Tokyo, University of Utah, Vanderbilt University, University of Virginia, University of Washington, and Yale University.

This research used resources of the National Energy Research Scientific Computing Center, which is supported by the Office of Science of the U.S. Department of Energy under Contract No. DE-AC02-05CH11231. 

This work was supported by World Premier International Research Center Initiative (WPI Initiative), MEXT, Japan. Numerical computations were partly carried out on Cray XC30 at Center for Computational Astrophysics, National Astronomical Observatory of Japan.

H.-J. Seo is supported by the U.S. Department of Energy, Office of Science, Office of High Energy Physics under Award Number DE-SC0014329.
C. C. acknowledges support as a MultiDark Fellow. 
C. C. acknowledges support from the Spanish MICINNs Consolider-Ingenio 2010 Programme under grant MultiDark CSD2009-00064, MINECO Centro de Excelencia Severo Ochoa Programme under grant SEV-2012-0249, and grant AYA2014-60641-C2-1-P. 

\setlength{\bibhang}{2em}
\setlength{\labelwidth}{0pt}

\newpage

\appendix
\numberwithin{equation}{section}

\section{Window function}
\label{app:window}

Here we describe the inclusion of the survey window function effects in our power spectrum model. First we will discuss the convolution of the power spectrum model with the window function, followed by the integral constraint effect. We also show a re-derivation of the window function formalism of~\citet{Wilson:2015lup} using the local plain-parallel approximation (instead of the global plain-parallel approximation as in~\citealt{Wilson:2015lup}); this approach allows the application of this method to wide-angle surveys like BOSS.

\subsection{Derivation within the local plane parallel approximation}

The convolved correlation function multipoles can be expressed as
\begin{equation}
\hat{\xi}_{\ell}(s) = \frac{2\ell + 1}{2}\int d\mu_s \int \frac{d\phi}{2\pi} \xi(\vc{s}) W^2(\vc{s})\mathcal{L}_{\ell}(\hat{\vc{s}} \cdot \hat{\vc{x}}_h),
\label{eq:3Dxi}
\end{equation}
where $\mathcal{L}_{\ell}$ is the Legendre polynomial of order $\ell$ and $\xi(\vc{s})$ and $W^2(\vc{s})$ are the anisotropic correlation function and window function, respectively:
\begin{equation}
\xi(\vc{s}) = \sum_L \xi_L(s)\mathcal{L}_L(\mu_s)
\end{equation}
and 
\begin{equation}
W^2(\vc{s})=\int d\vc{x}_1 W(\vc{x}_1)W(\vc{x}_1+\vc{s}) = \sum_p W_{p}^2(s)\mathcal{L}_{p}(\mu_s).
\end{equation}
Here,  $\vc{s} = \vc{x}_2-\vc{x}_1$ is the pair separation vector,  and $\mu_s$ is the cosine of the angle of the separation vector relative to the line-of-sight, i.e., $\mu_s = \hat{\vc{s}} \cdot \hat{\vc{x}}_h$, where $\vc{x}_h =  (\vc{x}_1+\vc{x}_2)/2 = \vc{x}_1+\vc{s}/2$ is known as the local plane parallel approximation (see~\citealt{Beutler:2013yhm} section 3.1).
The window function multipoles $W_{p}^2(s)$ are given by 
\begin{equation}
W_{p}^2(s)  =\frac{2p+1}{2} \int d\mu_s  \int \frac{d\phi}{2\pi}\int d\vc{x}_1 W(\vc{x}_1)W(\vc{x}_1+\vc{s})\mathcal{L}_{\ell}(\mu_s).
\label{eq:Wp}
\end{equation}

Multipole expanding the correlation function and window function in eq.~\ref{eq:3Dxi} produces
\begin{equation}
	\begin{split}
		\hat{\xi}_{\ell}(s) &= \frac{2\ell + 1}{2}\int d\mu_s \sum_{L}\xi_{L}(s)\mathcal{L}_{L}(\mu_s)\\
		&\;\;\;\;\;\sum_pW_{p}^2(s)\mathcal{L}_{p}(\mu_s)\mathcal{L}_{\ell}(\mu_s), 
	\end{split}
\end{equation}
Using the relation
\begin{equation}
\mathcal{L}_{\ell}\mathcal{L}_{p} = \sum_t a^{\ell}_{p t}\mathcal{L}_{t},
\label{eq:multicol}
\end{equation}
leads to
\begin{equation}
	\begin{split}
		\hat{\xi}_{\ell}(s) &= \frac{2\ell + 1}{2}\int d\mu_s\int \frac{d\phi}{2\pi}\sum_{L}\xi_{L}(s) \sum_pW_{p}^2(s)\\
		&\;\;\;\;\;\sum_t a^{\ell}_{Lt}\mathcal{L}_{t}(\mu_s)\mathcal{L}_{p}(\mu_s).
	\end{split}
\end{equation}
We can further simplify this expression using the integral relation
\begin{equation}
\int d\mu_s \int \frac{d\phi}{2\pi}\mathcal{L}_{p}(\mu_s)\mathcal{L}_{t}(\mu_s) = \frac{2}{2p + 1}\delta_{pt}
\end{equation}
to find
\begin{equation}
\hat{\xi}_{\ell}(s) = (2\ell + 1)\sum_{L}\xi_{L}(s) \sum_p\frac{1}{2p + 1}W_{p}^2(s) a^{\ell}_{Lp}.
\label{eq:multixi}
\end{equation}
To determine the coefficients $a^{\ell}_{Lp}$ we use eq.~\ref{eq:multicol}, multiply the polynomial expressions for the Legendre polynomials on the left, and use 
\begin{equation}
\mu^n = \sum_{\ell = n,(n-1),...}\frac{(2\ell + 1)n!\mathcal{L}_{\ell}(\mu_s)}{2^{(n-\ell)/2}(\frac{1}{2}(n-\ell))!(\ell + n + 1)!!}.
\label{eq:mupower}
\end{equation}
Equation~\ref{eq:multixi} is fairly straightforward to evaluate for any model correlation function $\xi_{\ell}(s)$. The window function multipoles can be calculated from the random pair counts $RR(s, \mu_s)$ as 
\begin{equation}
W_{\ell}^2(s) \propto  \sum_{\mu_s} \sum_{\vec{x}_1} \sum_{\vec{x}_2}   RR(s,\mu_s)\mathcal{L}_{\ell}(\mu_s),
\end{equation}
where the normalisation is chosen as $W_{\ell}^2(s\rightarrow 0) = 1$ for $\ell = 0$.  

Now we want to include the simple window function treatment in configuration space into our Fourier-space model. The observed power spectrum is
\begin{align}
	\hat{P}(\vc{k}) 
	&= \int d\vc{x}_1 \int d\vc{x}_2 \llan \delta(\vc{x}_1) \delta(\vc{x}_2)W(\vc{x}_1)W(\vc{x}_2)\rran  e^{i\vc{k}\cdot \vc{x}_1}e^{-i\vc{k}\cdot \vc{x}_2}\notag\\
	\begin{split}
		&= \int d\vc{x}_1 \int d\vc{s} \llan \delta(\vc{x}_1) \delta(\vc{x}_1+\vc{s})\rran \\
		&\;\;\;\;W(\vc{x}_1)W(\vc{x}_1+\vc{s}) e^{i\vc{k}\cdot \vc{x}_1}e^{-i\vc{k}\cdot (\vc{x}_1+\vc{s})}
	\end{split}
	\label{eq:win1}\\
	&=  \int d\vc{x}_1 \int d\vc{s}\;\xi(\vc{s})  W(\vc{x}_1)W(\vc{x}_1+\vc{s})e^{-i\vc{k}\cdot \vc{s}} \\
	\begin{split}
		&=  \int d\vc{x}_1 \int d\vc{s}\;\left(\sum_L\xi_L(s) \mathcal{L}_{L}(\hat{\vc{x}}_h\cdot\hat{\vc{s}})  \right)\\
		&\;\;\;\;W(\vc{x}_1)W(\vc{x}_1+\vc{s}) e^{-i\vc{k}\cdot \vc{s}}
	\end{split}
\end{align}
where we used  
\begin{equation}
	\begin{split}
		&\llan \delta(\vc{x}_1) \delta(\vc{x}_1+\vc{s}) W(\vc{x}_1)W(\vc{x}_1+\vc{s}) \rran\\
		&=  \llan \delta(\vc{x}_1) \delta(\vc{x}_1+\vc{s})\rran W(\vc{x}_1)W(\vc{x}_1+\vc{s}).
	\end{split}
\end{equation}
The multipole moment power spectrum in the local plain-parallel approximation is then 
\begin{align}
	\begin{split}
		\hat{P}_\ell(k) &= \frac{2\ell+1}{2} \int d\mu_k\int \frac{d\phi}{2\pi} \int d\vc{x}_1 \int d\vc{x}_2e^{i\vc{k}\cdot \vc{x}_1}e^{-i\vc{k}\cdot \vc{x}_2}\\
		&\;\;\;\;\llan \delta(\vc{x}_1) \delta(\vc{x}_2)W(\vc{x}_1)W(\vc{x}_2)\rran    \mathcal{L}_{\ell}(\hat{\vc{k}}\cdot\hat{\vc{x}}_h) \notag
	\end{split}\\
	\begin{split}
		&=\frac{2\ell+1}{2} \int d\mu_k\int \frac{d\phi}{2\pi}\int d\vc{x}_1 \int d\vc{s} \;\times\\
		&\;\;\;\;\left(\sum_L\xi_L(s) \mathcal{L}_{L}(\hat{\vc{x}}_h\cdot\hat{\vc{s}})  \right) \;\times \\
		&\;\;\;\;W(\vc{x}_1)W(\vc{x}_1+\vc{s}) e^{-i\vc{k}\cdot \vc{s}} \mathcal{L}_{\ell}(\hat{\vc{k}}\cdot\hat{\vc{x}}_h),
	\end{split}
\end{align}
where  $\int d\mu_k$ represents the integration over all the possible cosine angles between $\hat{\vc{k}}$ and $\hat{\vc{x}}_h$.  Now we apply the relations
\begin{eqnarray}
e^{i \vc{k}\cdot \vc{s}} = \sum_p (-i)^p (2p+1)j_p(ks)\mathcal{L}_{p}(\hat{\vc{k}}\cdot \hat{\vc{s}}),
\end{eqnarray}
and
\begin{eqnarray}
\int d\mu_k\int \frac{d\phi}{2\pi} \mathcal{L}_{\ell}( \hat{\vc{k}}\cdot\hat{\vc{x}}_h ) \mathcal{L}_{p}(\hat{\vc{k}}\cdot \hat{\vc{s}}) = \frac{2}{2\ell+1}\mathcal{L}_{\ell}(\hat{\vc{s}}\cdot\hat{\vc{x}}_h)\delta_{\ell p},
\end{eqnarray}
which allows us to express the multipole power spectra as
\begin{align}
	\begin{split}
		\hat{P}_\ell(k) &= \frac{2\ell+1}{2} \int d\mu_k\int \frac{d\phi}{2\pi}\int d\vc{x}_1 \int d\vc{s} \;\times\\
		&\;\;\;\;\sum_L\xi_L(s) \sum_p i^p (2p+1)j_p(ks)\;\times\\
		&\;\;\;\;W(\vc{x}_1)W(\vc{x}_1+\vc{s}) \mathcal{L}_{\ell}(\hat{\vc{k}}\cdot\hat{\vc{x}}_h)\mathcal{L}_{p}(\hat{\vc{k}}\cdot \hat{\vc{s}}) \mathcal{L}_{L}(\hat{\vc{x}}_h\cdot\hat{\vc{s}})
	\end{split}\\
	\begin{split}
		&=\int d\vc{x}_1 \int d\vc{s} \; \sum_L\xi_L(s) i^\ell (2\ell+1)j_\ell(ks) \; \times\\
		&\;\;\;\;W(\vc{x}_1)W(\vc{x}_1+\vc{s}) \mathcal{L}_{\ell}(\hat{\vc{x}}_h\cdot\hat{\vc{s}})\mathcal{L}_{L}(\hat{\vc{x}}_h\cdot\hat{\vc{s}}). 
	\end{split}
\end{align}
Using $\mathcal{L}_{\ell}\mathcal{L}_{L} = \sum_t a^{\ell}_{L t}\mathcal{L}_{t}$ (from eq. \ref{eq:multicol}) leads to
\begin{align}
	\begin{split}
		\hat{P}_\ell(k) &=\int d\vc{s} \; \sum_L\xi_L(s) i^\ell (2\ell+1)j_\ell(ks) \;\times\\
		&\;\;\;\;\int d\vc{x}_1 W(\vc{x}_1)W(\vc{x}_1+\vc{s})  \sum_t a^{\ell}_{L t}\mathcal{L}_{t}(\hat{\vc{x}}_h\cdot\hat{\vc{s}})
	\end{split}\\
	\begin{split}
		&= \int 2\pi s^2ds \; \sum_L\xi_L(s) i^\ell (2\ell+1)j_\ell(ks) \;\times\\
		&\;\;\;\; \sum_t a^{\ell}_{L t}\int d\mu_s \int \frac{d\phi}{2\pi}\int d\vc{x}_1\;\times\\
		&\;\;\;\; W(\vc{x}_1)W(\vc{x}_1+\vc{s}) \mathcal{L}_{t}(\hat{\vc{x}}_h\cdot\hat{\vc{s}}). 
	\end{split}
\end{align}
Using the definition of the window function multipoles of eq.~\ref{eq:Wp} we can write the equation above as
\begin{equation}
	\begin{split}
		\hat{P}_\ell(k) &= i^\ell (2\ell+1)\int 2\pi s^2ds\;  j_\ell(ks)\;\times\\
		&\;\;\;\;\sum_L\sum_t \frac{2}{2t+1} a^{\ell}_{L t} \xi_L(s)  W_{t}^2(s).
	\end{split}
\end{equation}
Substituting with eq.~\ref{eq:multixi} the convolved power spectrum multipoles are given by 
\begin{equation}
\hat{P}_{\ell}(k) = 4\pi i^\ell \int ds\,s^2\hat{\xi}_{\ell}(s) j_{\ell}(sk).
\end{equation}
For our analysis we need to calculate the convolved monopole, quadrupole, and hexadecapole power spectra. Therefore, the convolved correlation function multipoles in eq.~\ref{eq:PWell}, relevant for our analysis, are given by  
\begin{align}
	\hat{\xi}_{0}(s) &= \xi_{0}W_{0}^2  + \frac{1}{5}\xi_2W^2_2 + \frac{1}{9}\xi_4W^2_4 +...\\
	\begin{split}
		\hat{\xi}_{2}(s) &= \xi_{0} W_{2}^2  + \xi_2\left[W^2_0 + \frac{2}{7}W^2_2 + \frac{2}{7}W^2_4\right]\\
		&\;\;\;\;\;\;\;\;\;\;\;\;\;\,+\xi_4\left[\frac{2}{7}W^2_2 + \frac{100}{693}W^2_4 + \frac{25}{143}W^2_6\right]\\
		&\;\;\;\;\;\;\;\;\;\;\;\;\;\,+...
	\end{split}\\
	\begin{split}
		\hat{\xi}_{4}(s) &= \xi_{0} W_{4}^2  + \xi_2\left[\frac{18}{35}W^2_2 + \frac{20}{77}W^2_4 + \frac{45}{143}W^2_6\right]\\
		&\;\;\;\;\;\;\;\;\;\;\;\;\;\,+\xi_4\bigg[W^2_0 + \frac{20}{77}W^2_2 + \frac{162}{1001}W^2_4\\
		&\;\;\;\;\;\;\;\;\;\;\;\;\;\,+ \frac{20}{143}W^2_6 + \frac{490}{2431}W^2_8 \bigg]\\
		&\;\;\;\;\;\;\;\;\;\;\;\;\;\,+...
	\end{split}
\end{align}
We truncate the formula after the hexadecapole contribution of the correlation function, but use all window function multipoles up to $\ell = 8$.

\subsection{The integral constraint correction}

Whenever we estimate a power spectrum we must make an assumption of the mean density of the Universe, so that we can properly define an over-density field. The standard assumption is that the mean density of the Universe is equivalent to the mean density of the survey. The non-zero sample variance expected at the wavelengths that correspond to the size of the survey invalidates this assumption. In general, this assumption affects only the mean density, i.e., forcing the power spectrum near $k=0$ to be zero, which is known as the integral constraint. However, the window function correlates various modes with $k=0$ and therefore propagates the incorrect estimation of $k=0$ to other scales that are relevant for cosmological measurements.  Eq~\ref{eq:conv1} and \ref{eq:conv2} demonstrate how the integral constraint affects the correlation function. Without the window function effect, i.e., $W_0=1$ with $W_{\ell \ge 2}=0$, the integral constraint will simply introduce a constant offset to $\xi_0(s)$ and therefore to $\hat{\xi}_0(s)$. With scale-dependent nonzero $W_0$ and $W_2$, however, a constant offset in $\xi_0(s)$ becomes scale-dependent in $\hat{\xi}_0(s)$ and $\hat{\xi}_2(s)$, affecting the shape of the correlation function.

We can account for the integral constraint bias by correcting the model power spectrum as 
\begin{equation}
P^{\rm ic-corrected}_{\ell}(k) = \hat{P}_{\ell}(k) - P_{0}W_{\ell}^2(k),
\end{equation}
where the window functions $W^2_\ell(k)$ can be obtained from $W^2(s)$ defined in eq.~\ref{eq:Well} as
\begin{equation}
W^2_{\ell}(k) = 4\pi\int ds\; s^2W_{\ell}^2(s)j_{\ell}(sk).
\end{equation}
The integral constraint correction in BOSS only affects modes $\lesssim 0.005\ihMpc$ and does not affect any of our results.

\section{Correlation, covariance and inverse covariance matrices}
\label{app:covs}

We determine the correlation between the three cosmological parameter constraints ($D_V(z)r^{\rm fid}_s/r_s$, $F_{\rm AP}$, $f\sigma_8$) using the MultiDark-Patchy mock catalogues. This approach leads to the following correlation matrix for the first redshift bin
\begin{equation}
R_{\rm z1} = \begin{pmatrix} 1 & -0.206 & 0.0490\\
			        -0.206 & 1 & 0.652\\
			        0.0490 & 0.652 & 1\end{pmatrix},
\label{eq:cor_matrix_z1}
\end{equation} 
which leads to a covariance matrix of 
\begin{equation}
C_{\rm z1} = \begin{pmatrix} 1090 & -0.136 & 0.0825\\
			        -0.136 & 0.000400 & 0.000665\\
			        0.0825 & 0.000665 & 0.00260\end{pmatrix}
\label{eq:cov_matrix_z1}
\end{equation} 
and the inverse of this matrix is
\begin{equation}
C^{-1}_{\rm z1} = \begin{pmatrix} 0.00102 & 0.697 & -0.211\\
			        0.697 & 4830 & -1260\\
			        -0.211 & -1260 & 713\end{pmatrix}.
\label{eq:invcov_matrix_z1}
\end{equation} 
For the second redshift bin we find
\begin{equation}
R_{\rm z2} = \begin{pmatrix} 1 & -0.291 & -0.056\\
			        -0.291 & 1 & 0.648\\
			        -0.056 & 0.648 & 1\end{pmatrix}
\label{eq:cor_matrix_z2}
\end{equation} 
and the covariance matrix is 
\begin{equation}
C_{\rm z2} = \begin{pmatrix} 1940 & -0.397 & -0.124\\
			        -0.397 & 0.000961 & 0.00100\\
			        -0.124 & 0.00100 & 0.00250\end{pmatrix}.
\label{eq:cov_matrix_z2}
\end{equation} 
Inverting this matrix yields
\begin{equation}
C^{-1}_{\rm z2} = \begin{pmatrix} 0.000582 & 0.360 & -0.115\\
			        0.360 & 2010 & -784\\
			        -0.115 & -784 & 708\end{pmatrix}.
\label{eq:invcov_matrix_z2}
\end{equation} 
Finally, for the third redshift bin, we have
\begin{equation}
R_{\rm z3} = \begin{pmatrix} 1 & -0.126 & 0.101\\
			        -0.126 & 1 & 0.619\\
			        0.101 & 0.619 & 1\end{pmatrix}
\label{eq:cor_matrix_z3}
\end{equation} 
and
\begin{equation}
C_{\rm z3} = \begin{pmatrix} 2120 & -0.197 & 0.205\\
			        -0.197 & 0.00116 & 0.000927\\
			        0.205 & 0.000927 & 0.00194\end{pmatrix}.
\label{eq:cov_matrix_z3}
\end{equation} 
The inverse matrix is
\begin{equation}
C^{-1}_{\rm z3} = \begin{pmatrix} 0.000506 & 0.208 & -0.153\\
			        0.208 & 1480 & -729\\
			        -0.153 & -729 & 880\end{pmatrix}.
\label{eq:invcov_matrix_z3}
\end{equation} 
Only the first and second redshift bins are independent, while the middle redshift bin is correlated with the other two. These results can be used together with the data vector for any likelihood analysis, e.g. for the first redshift bin the data vector is $D_{\rm z1} = (1485, 0.426, 0.478)$ and the corresponding likelihood is $\mathcal{L}_{\rm z1} = D_{\rm z1}^TC^{-1}_{\rm z1}D_{\rm z1}$.

\label{lastpage}


\begin{thebibliography}{99}

\bibitem[\protect\citeauthoryear{Ade et al.}{2015}]{Ade:2015xua} 
  Ade~P.~A.~R. {\it et al.} [Planck Collaboration],
  arXiv:1502.01589 [astro-ph.CO].
  
\bibitem[\protect\citeauthoryear{Anderson et al.}{2013b}]{Anderson2.0} 
  Anderson~L. {\it et al.},
  arXiv:1312.4877 [astro-ph.CO].
  
\bibitem[\protect\citeauthoryear{Alam et al.}{2015}]{Alam:2015mbd} 
  Alam~S. {\it et al.} [SDSS-III Collaboration],
  Astrophys.\ J.\ Suppl.\  {\bf 219}, no. 1, 12 (2015)
  doi:10.1088/0067-0049/219/1/12
  [arXiv:1501.00963 [astro-ph.IM]].
  
\bibitem[\protect\citeauthoryear{Alam et al.}{2016}]{Alam2016}
 Alam et al. 2016, 

\bibitem[\protect\citeauthoryear{Alcock \& Paczynski}{1979}]{Alcock:1979mp}
  Alcock~C. and Paczynski~B.,
  Nature {\bf 281} (1979) 358.
  
\bibitem[\protect\citeauthoryear{Baldauf et al.}{2012}]{Baldauf:2012hs} 
  Baldauf~T., Seljak~U., Desjacques~V. and McDonald~P.,
  Phys.\ Rev.\ D {\bf 86}, 083540 (2012)
  [arXiv:1201.4827 [astro-ph.CO]].

\bibitem[\protect\citeauthoryear{Ballinger, Peacock \& Heavens}{1996}]{Ballinger:1996cd} 
  Ballinger~W.~E., Peacock~J.~A. and Heavens~A.~F.,
  Mon.\ Not.\ Roy.\ Astron.\ Soc.\  {\bf 282}, 877 (1996)
  [astro-ph/9605017].
  
\bibitem[\protect\citeauthoryear{Beutler et al.}{2011}]{Beutler:2011hx} 
  Beutler~F. {\it et al.},
  Mon.\ Not.\ Roy.\ Astron.\ Soc.\  {\bf 416}, 3017 (2011)
  [arXiv:1106.3366 [astro-ph.CO]].
  
\bibitem[\protect\citeauthoryear{Beutler et al.}{2012}]{Beutler:2012px} 
  Beutler~F. {\it et al.},
  Mon.\ Not.\ Roy.\ Astron.\ Soc.\  {\bf 423}, 3430 (2012)
  
\bibitem[\protect\citeauthoryear{Beutler et al.}{2013}]{Beutler:2013yhm} 
  Beutler~F. {\it et al.}  [BOSS Collaboration],
  Mon.\ Not.\ Roy.\ Astron.\ Soc.\  {\bf 443}, 1065 (2014)
  
\bibitem[\protect\citeauthoryear{Beutler et al.}{2014}]{Beutler:2014} Beutler F., et al., 2014, MNRAS, 444, 3501   

\bibitem[\protect\citeauthoryear{Beutler et al.}{2015}]{Beutler:2015tla}
  Beutler~F., Blake~C., Koda~J., Marin~F., Seo~H.~J., Cuesta~A.~J. and Schneider~D.~P.,
  Mon.\ Not.\ Roy.\ Astron.\ Soc.\  {\bf 455} (2016) no.3,  3230
  doi:10.1093/mnras/stv1943
  [arXiv:1506.03900 [astro-ph.CO]].
  
\bibitem[\protect\citeauthoryear{Beutler et al.}{2016}]{Beutler:2016} 
  Beutler~F. {\it et al.} 
  
\bibitem[\protect\citeauthoryear{Biagetti et al.}{2014}]{Biagetti:2013hfa} 
  Biagetti~M., Chan~K.~C., Desjacques~V. and Paranjape~A.,
  Mon.\ Not.\ Roy.\ Astron.\ Soc.\  {\bf 441}, no. 2, 1457 (2014)
  doi:10.1093/mnras/stu680
  [arXiv:1310.1401 [astro-ph.CO]].
  
\bibitem[\protect\citeauthoryear{Bianchi et al.}{2015}]{Bianchi:2015oia}
  Bianchi~D., Gil-Mar'n~H., Ruggeri~R. and Percival~W.~J.,
  arXiv:1505.05341 [astro-ph.CO].

\bibitem[\protect\citeauthoryear{Blake et al.}{2011a}]{Blake:2011rj} 
  Blake~C. {\it et al.},
  Mon.\ Not.\ Roy.\ Astron.\ Soc.\  {\bf 415}, 2876 (2011a)
  [arXiv:1104.2948 [astro-ph.CO]].
  
\bibitem[\protect\citeauthoryear{Blake et al.}{2012}]{Blake:2012pj} 
  Blake~C. {\it et al.},
  Mon.\ Not.\ Roy.\ Astron.\ Soc.\  {\bf 425}, 405 (2012)
  [arXiv:1204.3674 [astro-ph.CO]].
  
\bibitem[\protect\citeauthoryear{Bolton et al.}{2012}]{Bolton:2012hz} 
  Bolton~A.~S. {\it et al.}  [Cutler Group, LP Collaboration],
  2012 {\bf 144}, 144
  [arXiv:1207.7326 [astro-ph.CO]].
  
\bibitem[\protect\citeauthoryear{Bundy et al.}{2015}]{Bundy:2014vpa} 
  Bundy~K. {\it et al.},
  Astrophys.\ J.\  {\bf 798}, no. 1, 7 (2015)
  doi:10.1088/0004-637X/798/1/7
  [arXiv:1412.1482 [astro-ph.GA]].
  
\bibitem[\protect\citeauthoryear{Chan, Scoccimarro \& Sheth}{2012}]{Chan:2012jj} 
  Chan~K.~C., Scoccimarro~R. and Sheth~R.~K.,
  Phys.\ Rev.\ D {\bf 85}, 083509 (2012)
  [arXiv:1201.3614 [astro-ph.CO]].
  
\bibitem[\protect\citeauthoryear{Chuang et al.}{2013a}]{Chuang:2013hya} 
  Chuang~C.~-H. {\it et al.},
  arXiv:1303.4486 [astro-ph.CO].
  
\bibitem[\protect\citeauthoryear{Dawson et al.}{2013}]{Dawson:2012va} 
  Dawson~K.~S. {\it et al.}  [BOSS Collaboration],
  Astrophys.\ J.\  {\bf 145}, 10 (2013)
  
\bibitem[\protect\citeauthoryear{Doi et al.}{2010}]{Doi:2010rf}
  Doi~M. {\it et al.},
  Astron.\ J.\  {\bf 139} (2010) 1628
  [arXiv:1002.3701 [astro-ph.IM]].
  
\bibitem[\protect\citeauthoryear{Eisenstein et al.}{2007}]{Eisenstein:2006nk} 
  Eisenstein~D.~J., Seo~H.~J., Sirko~E. and Spergel~D.,
  Astrophys.\ J.\  {\bf 664}, 675 (2007)
  [astro-ph/0604362].
  
\bibitem[\protect\citeauthoryear{Eisenstein et al.}{2011}]{Eisenstein:2011sa} 
  Eisenstein~D.~J. {\it et al.}  [SDSS Collaboration],
  Astron.\ J.\  {\bf 142}, 72 (2011)
  [arXiv:1101.1529 [astro-ph.IM]].

\bibitem[\protect\citeauthoryear{Feldman, Kaiser \& Peacock}{1993}]{Feldman:1993ky} 
  Feldman~H.~A., Kaiser~N. and Peacock~J.~A.,
  Astrophys.\ J.\  {\bf 426}, 23 (1994)
  [astro-ph/9304022].
  
\bibitem[\protect\citeauthoryear{Foreman-Mackey et al.}{2013}]{ForemanMackey:2012ig} 
  Foreman-Mackey~D., Hogg~D.~W., Lang~D. and Goodman~J.,
  Publ.\ Astron.\ Soc.\ Pac.\  {\bf 125}, 306 (2013)
  doi:10.1086/670067
  [arXiv:1202.3665 [astro-ph.IM]].
  
\bibitem[\protect\citeauthoryear{Fukugita et al.}{1996}]{Fukugita:1996qt} 
  Fukugita~M., Ichikawa~T., Gunn~J.~E., Doi~M., Shimasaku~K. and Schneider~D.~P.,
  Astron.\ J.\  {\bf 111}, 1748 (1996).
  
\bibitem[\protect\citeauthoryear{Garilli et al.}{2014}]{Garilli:2013eoa} 
  Garilli~B. {\it et al.},
  Astron.\ Astrophys.\  {\bf 562}, A23 (2014)
  doi:10.1051/0004-6361/201322790
  [arXiv:1310.1008 [astro-ph.CO]].
  
\bibitem[\protect\citeauthoryear{Gelman \& Rubin}{1992}]{Gelman:1992zz} 
  Gelman~A. and Rubin~D.~B.,
  Statist.\ Sci.\  {\bf 7}, 457 (1992).
  doi:10.1214/ss/1177011136
  
\bibitem[\protect\citeauthoryear{Gil-Marin et al.}{2015}]{Gil-Marin:2015sqa} 
  Gil-Marin~H. {\it et al.},
  arXiv:1509.06386 [astro-ph.CO].

\bibitem[\protect\citeauthoryear{Grieb et al.}{2016}]{Grieb:2016} 
  Grieb~J.~N. {\it et al.} 
  
\bibitem[\protect\citeauthoryear{Gunn et al.}{1998}]{Gunn:1998vh} 
  Gunn~J.~E. {\it et al.}  [SDSS Collaboration],
  Astron.\ J.\  {\bf 116}, 3040 (1998)
  [astro-ph/9809085].
  
\bibitem[\protect\citeauthoryear{Gunn et al.}{2006}]{Gunn:2006tw} 
  Gunn~J.~E. {\it et al.}  [SDSS Collaboration],
  Astron.\ J.\  {\bf 131}, 2332 (2006)
  [astro-ph/0602326].
  
\bibitem[\protect\citeauthoryear{Guo, Zehavi \& Zheng}{2012}]{Guo:2011ai} 
  Guo~H., Zehavi~I. and Zheng~Z.,
  Astrophys.\ J.\  {\bf 756}, 127 (2012)
  [arXiv:1111.6598 [astro-ph.CO]].

\bibitem[\protect\citeauthoryear{Guzzo et al.}{2008}]{Guzzo:2008ac}
  Guzzo~L. {\it et al.},
  Nature {\bf 451} (2008) 541
  [arXiv:0802.1944 [astro-ph]].
  
\bibitem[\protect\citeauthoryear{Hartlap et al.}{2007}]{Hartlap:2006kj} 
  Hartlap~J., Simon~P. and Schneider~P.,
  [astro-ph/0608064].
  
\bibitem[\protect\citeauthoryear{Hawkins et al.}{2003}]{Hawkins:2002sg} 
  Hawkins~E. {\it et al.},
  Mon.\ Not.\ Roy.\ Astron.\ Soc.\  {\bf 346}, 78 (2003)
  [astro-ph/0212375].
  
\bibitem[\protect\citeauthoryear{Jing}{2005}]{Jing:2004fq} 
  Jing~Y.~P.,
  Astrophys.\ J.\  {\bf 620}, 559 (2005)
  doi:10.1086/427087
  [astro-ph/0409240].
  
\bibitem[\protect\citeauthoryear{Kaiser}{1987}]{Kaiser:1987qv} 
  Kaiser~N.,
  Mon.\ Not.\ Roy.\ Astron.\ Soc.\  {\bf 227}, 1 (1987).
  
\bibitem[\protect\citeauthoryear{Kitaura et al.}{2015}]{Kitaura:2015uqa} 
  Kitaura~F.~S. {\it et al.},
  arXiv:1509.06400 [astro-ph.CO].
  
\bibitem[\protect\citeauthoryear{Klypin et al.}{2014}]{Klypin:2014kpa} 
  Klypin~A., Yepes~G., Gottlober~S., Prada~S. and Hess~S.,
  arXiv:1411.4001 [astro-ph.CO].
  
\bibitem[\protect\citeauthoryear{Lazeyras et al.}{2016}]{Lazeyras:2015lgp}
  Lazeyras~T., Wagner~C., Baldauf~T. and Schmidt~F.,
  JCAP {\bf 1602} (2016) no.02,  018
  doi:10.1088/1475-7516/2016/02/018
  [arXiv:1511.01096 [astro-ph.CO]].
  
\bibitem[\protect\citeauthoryear{Leauthaud et al.}{2016}]{Leauthaud:2015fva} 
  Leauthaud~A. {\it et al.},
  doi:10.1093/mnras/stw117
  arXiv:1507.04752 [astro-ph.GA].

\bibitem[\protect\citeauthoryear{Lesgourgues \& Pastor}{2006}]{Lesgourgues:2006nd} 
  Lesgourgues~J. and Pastor~S.,
  Phys.\ Rept.\  {\bf 429}, 307 (2006)
  doi:10.1016/j.physrep.2006.04.001
  [astro-ph/0603494].
  
\bibitem[\protect\citeauthoryear{Lewandowski et al.}{2015}]{Lewandowski:2015ziq} 
  Lewandowski~M., Senatore~L., Prada~F., Zhao~C. and Chuang~C.~H.,
  arXiv:1512.06831 [astro-ph.CO].
  
\bibitem[\protect\citeauthoryear{Li, Hu \& Takada}{2016}]{Li:2015jsz} 
  Li~Y., Hu~W. and Takada~M.,
  Phys.\ Rev.\ D {\bf 93}, no. 6, 063507 (2016)
  doi:10.1103/PhysRevD.93.063507
  [arXiv:1511.01454 [astro-ph.CO]].
  
\bibitem[\protect\citeauthoryear{Matsubara \& Suto}{1996}]{Matsubara:1996nf} 
  Matsubara~T. and Suto~Y.,
  Astrophys.\ J.\  {\bf 470}, L1 (1996)
  [astro-ph/9604142].
  
\bibitem[\protect\citeauthoryear{McDonald \& Roy}{2009}]{McDonald:2009dh} 
  McDonald~P. and Roy~A.,
  JCAP {\bf 0908}, 020 (2009)
  [arXiv:0902.0991 [astro-ph.CO]].
  
\bibitem[\protect\citeauthoryear{Nishimichi \& Oka}{2013}]{Nishimichi:2013aba} 
  Nishimichi~T. and Oka~A.,
  arXiv:1310.2672 [astro-ph.CO].
  
\bibitem[\protect\citeauthoryear{Mar'in et al.}{2015}]{Marin:2015ula}
  Mar'in~F.~A., Beutler~F., Blake~C., Koda~J., Kazin~E. and Schneider~D.~P.,
  Mon.\ Not.\ Roy.\ Astron.\ Soc.\  {\bf 455} (2016) no.4,  4046
  doi:10.1093/mnras/stv2502
  [arXiv:1506.03901 [astro-ph.CO]].
   
\bibitem[\protect\citeauthoryear{Oka et al.}{2014}]{Oka:2013cba} 
  Oka~A., Saito~S., Nishimichi~T., Taruya~A. and Yamamoto~K.,
  Mon.\ Not.\ Roy.\ Astron.\ Soc.\  {\bf 439}, 2515 (2014)
  doi:10.1093/mnras/stu111
  [arXiv:1310.2820 [astro-ph.CO]].
  
\bibitem[\protect\citeauthoryear{Okumura et al.}{2008}]{Okumura:2007br} 
  Okumura~T., Matsubara~T., Eisenstein~D.~J., Kayo~I., Hikage~C., Szalay~A.~S. and Schneider~D.~P.,
  Astrophys.\ J.\  {\bf 676}, 889 (2008)
  [arXiv:0711.3640 [astro-ph]].
  
\bibitem[\protect\citeauthoryear{Okumura et al.}{2015}]{Okumura:2015fga} 
  Okumura~T., Hand~N., Seljak~U., Vlah~Z. and Desjacques~V.,
  Phys.\ Rev.\ D {\bf 92}, no. 10, 103516 (2015)
  doi:10.1103/PhysRevD.92.103516
  [arXiv:1506.05814 [astro-ph.CO]].
  
\bibitem[\protect\citeauthoryear{Padmanabhan \& White}{2008}]{Padmanabhan:2008ag} 
  Padmanabhan~N. and White~M.~J.,
  Phys.\ Rev.\ D {\bf 77}, 123540 (2008)
  doi:10.1103/PhysRevD.77.123540
  [arXiv:0804.0799 [astro-ph]].
  
\bibitem[\protect\citeauthoryear{Peacock et al.}{2001}]{Peacock:2001gs} 
  Peacock~J.~A. {\it et al.},
  Nature {\bf 410}, 169 (2001)
  [astro-ph/0103143].
  
\bibitem[\protect\citeauthoryear{Percival et al.}{2013}]{Percival:2013} 
  Percival~W.~J. {\it et al.},
  arXiv:1312.4841 [astro-ph.CO].
  
\bibitem[\protect\citeauthoryear{Reid et al.}{2012}]{Reid:2012sw} 
  Reid~B.~A. {\it et al.},
  arXiv:1203.6641 [astro-ph.CO].
  
\bibitem[\protect\citeauthoryear{Reid et al.}{2015}]{Reid:2015gra} 
  Reid~B. {\it et al.},
  Mon.\ Not.\ Roy.\ Astron.\ Soc.\  {\bf 455}, 1553 (2016)
  doi:10.1093/mnras/stv2382
  [arXiv:1509.06529 [astro-ph.CO]].
  
\bibitem[\protect\citeauthoryear{Rodr'guez-Torres et al.}{2015}]{Rodriguez-Torres:2015vqa} 
  Rodr'guez-Torres~S.~A. {\it et al.},
  arXiv:1509.06404 [astro-ph.CO].
  
\bibitem[\protect\citeauthoryear{Ross et al.}{2012a}]{Ross:2012qm} 
  Ross~A.~J. {\it et al.}  [BOSS Collaboration],
  Mon.\ Not.\ Roy.\ Astron.\ Soc.\  {\bf 424}, 564 (2012)
  [arXiv:1203.6499 [astro-ph.CO]].
  
\bibitem[\protect\citeauthoryear{Ross et al.}{2016}]{Ross:2016} 
  Ross~A.~J. {\it et al.},
  
\bibitem[\protect\citeauthoryear{Saito et al.}{2014}]{Saito2013} 
  Saito~S., Baldauf~T., Vlah~Z., Seljak~U., Okumura~T. and McDonald~P.,
  arXiv:1405.1447 [astro-ph.CO].
  
\bibitem[\protect\citeauthoryear{Saito et al.}{2015}]{Saito:2015eka} 
  Saito~S. {\it et al.},
  arXiv:1509.00482 [astro-ph.CO].
  
\bibitem[\protect\citeauthoryear{Samushia, Percival \& Raccanelli}{2011}]{Samushia:2011cs} 
  Samushia~L., Percival~W.~J. and Raccanelli~A.,
  Mon.\ Not.\ Roy.\ Astron.\ Soc.\  {\bf 420}, 2102 (2012)
  [arXiv:1102.1014 [astro-ph.CO]].
  
\bibitem[\protect\citeauthoryear{Samushia et al.}{2013}]{Samushia:2012iq} 
  Samushia~L. {\it et al.},
  Mon.\ Not.\ Roy.\ Astron.\ Soc.\  {\bf 429}, 1514 (2013)
  [arXiv:1206.5309 [astro-ph.CO]].

\bibitem[\protect\citeauthoryear{Sanchez et al.}{2016}]{Sanchez:2016} 
  Sanchez~A.~G. {\it et al.},
  
\bibitem[\protect\citeauthoryear{Satpathy et al.}{2016}]{Satpathy2016} 
  Satpathy et al. (2016)
  
\bibitem[\protect\citeauthoryear{Schmidt}{2016}]{Schmidt:2015gwz} 
  Schmidt~F.,
  Phys.\ Rev.\ D {\bf 93}, no. 6, 063512 (2016)
  doi:10.1103/PhysRevD.93.063512
  [arXiv:1511.02231 [astro-ph.CO]].
  
\bibitem[\protect\citeauthoryear{Scoccimarro}{2004}]{Scoccimarro:2004tg} 
  Scoccimarro~R.,
  Phys.\ Rev.\ D {\bf 70}, 083007 (2004)
  [astro-ph/0407214].
  
\bibitem[\protect\citeauthoryear{Scoccimarro}{2015}]{Scoccimarro:2015bla} 
  Scoccimarro~R.,
  arXiv:1506.02729 [astro-ph.CO].
  
\bibitem[\protect\citeauthoryear{Sefusatti et al.}{2015}]{Sefusatti:2015aex} 
  Sefusatti~E., Crocce~M., Scoccimarro~R. and Couchman~H.,
  arXiv:1512.07295 [astro-ph.CO].
  
\bibitem[\protect\citeauthoryear{Seo 
\& Eisenstein}{2003}]{Seo:2003} Seo H.-J., Eisenstein D.~J., 2003, ApJ, 598, 720 

\bibitem[\protect\citeauthoryear{Seo \& Eisenstein}{2007}]{Seo:2007ns} 
  Seo~H.~J. and Eisenstein~D.~J.,
  Astrophys.\ J.\  {\bf 665}, 14 (2007)
  [astro-ph/0701079].
  
  \bibitem[\protect\citeauthoryear{Shoji, Jeong, 
\& Komatsu}{2009}]{Shoji:2009} Shoji M., Jeong D., Komatsu E., 2009, ApJ, 693, 1404 
  
\bibitem[\protect\citeauthoryear{Smee et al.}{2013}]{Smee:2012wd} 
  Smee~S. {\it et al.},
  Astron.\ J.\  {\bf 126}, 32 (2013)
  arXiv:1208.2233 [astro-ph.IM].
  
\bibitem[\protect\citeauthoryear{Smith et al.}{2002}]{Smith:2002pca} 
  Smith~J.~A. {\it et al.}  [SDSS Collaboration],
  Astron.\ J.\  {\bf 123}, 2121 (2002)
  [astro-ph/0201143].
  
\bibitem[\protect\citeauthoryear{Taruya, Nishimichi \& Saito}{2010}]{Taruya:2010mx}
  Taruya~A., Nishimichi~T. and Saito~S.,
  Phys.\ Rev.\ D {\bf 82} (2010) 063522
  [arXiv:1006.0699 [astro-ph.CO]].
  
\bibitem[\protect\citeauthoryear{Taruya et al.}{2012}]{Taruya:2012ut} 
  Taruya~A., Bernardeau~F., Nishimichi~T. and Codis~S.,
  Phys.\ Rev.\ D {\bf 86}, 103528 (2012)
  [arXiv:1208.1191 [astro-ph.CO]].
  
\bibitem[\protect\citeauthoryear{Tegmark et al.}{2006}]{Tegmark:2006az} 
  Tegmark~M. {\it et al.}  [SDSS Collaboration],
  Phys.\ Rev.\ D {\bf 74}, 123507 (2006)
  [astro-ph/0608632].
  
\bibitem[\protect\citeauthoryear{Tinker et al. }{2016}]{Tinker:2015} 
  Tinker~J.~L. {\it et al.},
  
\bibitem[\protect\citeauthoryear{Vargas-Magana et al.}{2016}]{Vargas-Magana2016} 
  Vargas-Magana et al. (2016)
  
\bibitem[\protect\citeauthoryear{Wilson}{2015}]{Wilson:2015lup} 
  Wilson~M.~J., Peacock~J.~A., Taylor~A.~N. and de la Torre~S.,
  arXiv:1511.07799 [astro-ph.CO].
  
\bibitem[\protect\citeauthoryear{Yamamoto et al.}{2006}]{Yamamoto:2005dz} 
  Yamamoto~K., Nakamichi~M., Kamino~A., Bassett~B.~A. and Nishioka~H.,
  Publ.\ Astron.\ Soc.\ Jap.\  {\bf 58}, 93 (2006)
  [astro-ph/0505115].
  
\bibitem[\protect\citeauthoryear{Yamamoto, Sato \& Huetsi}{2008}]{Yamamoto:2008gr} 
  Yamamoto~K., Sato~T. and Huetsi~G.,
  Prog.\ Theor.\ Phys.\  {\bf 120}, 609 (2008)
  [arXiv:0805.4789 [astro-ph]].
  
\bibitem[\protect\citeauthoryear{Yoo \& Seljak}{2013}]{Yoo:2013zga} 
  Yoo~J. and Seljak~U.,
  arXiv:1308.1093 [astro-ph.CO].
  
\bibitem[\protect\citeauthoryear{Zheng \& Song}{2016}]{Zheng:2016zxc}
  Zheng~Y. and Song~Y.~S.,
  arXiv:1603.00101 [astro-ph.CO].

\end{thebibliography}
\end{document}